\documentclass[aps,showpacs,eqsecnum,twocolumn,superscriptaddress]{revtex4}

\usepackage{amsmath}
\usepackage{latexsym}
\usepackage{graphicx}
\usepackage{times}

\begin{document}


\title{Three-dimensional relativistic simulations of rotating neutron-star
collapse to a Kerr black hole}


\author{Luca~Baiotti} 
\affiliation{SISSA, International School for
Advanced Studies and INFN, Via Beirut 2, 34014 Trieste, Italy}

\author{Ian~Hawke}
\affiliation{Max-Planck-Institut f\"ur Gravitationsphysik,
Albert-Einstein-Institut, 14476 Golm, Germany}

\author{Pedro~J.~Montero}
\affiliation{SISSA, International School for
Advanced Studies and INFN, Via Beirut 2, 34014 Trieste, Italy}

\author{Frank~L\"offler}
\affiliation{Max-Planck-Institut f\"ur Gravitationsphysik,
Albert-Einstein-Institut, 14476 Golm, Germany}

\author{Luciano~Rezzolla}
\affiliation{SISSA, International School for
Advanced Studies and INFN, Via Beirut 2, 34014 Trieste, Italy}
\affiliation{Department of Physics, Louisiana State University, Baton
Rouge, LA 70803 USA}

\author{Nikolaos~Stergioulas}
\affiliation{Department of Physics, Aristotle University of Thessaloniki,
Thessaloniki 54124, Greece}

\author{Jos\'e~A.~Font}
\affiliation{Departamento de Astronom\'{\i}a y Astrof\'{\i}sica,
Universidad de Valencia, Dr. Moliner 50, 46100 Burjassot, Spain}

\author{Ed~Seidel}
\affiliation{Center for Computation and Technology,
Louisiana State University, Baton Rouge, LA 70803 USA}
\affiliation{Department of Physics, Louisiana State University, Baton
Rouge, LA 70803 USA}
\affiliation{Max-Planck-Institut f\"ur Gravitationsphysik,
Albert-Einstein-Institut, 14476 Golm, Germany}


\date{\today}


\begin{abstract}
        We  present a  new three-dimensional  fully  general-relativistic
        hydrodynamics   code    using   high-resolution   shock-capturing
        techniques and a  conformal traceless formulation of the Einstein
        equations.  Besides presenting a  thorough set of tests which the
        code passes  with very high accuracy, we  discuss its application
        to the study of  the gravitational collapse of uniformly rotating
        neutron stars  to Kerr black  holes.  The initial  stellar models
        are  modelled   as  relativistic  polytropes   which  are  either
        secularly  or dynamically  unstable and  with  angular velocities
        which range  from slow rotation  to the mass-shedding  limit.  We
        investigate the gravitational  collapse by carefully studying not
        only the  dynamics of  the matter, but  also that of  the trapped
        surfaces,  {\it  i.e.}  of  both  the  apparent  {\it and}  event
        horizons formed during the  collapse.  The use of these surfaces,
        together  with  the dynamical  horizon  framework,  allows for  a
        precise measurement of the black-hole mass and spin.  The ability
        to successfully  perform these simulations  for sufficiently long
        times  relies on excising  a region  of the  computational domain
        which  includes  the  singularity  and  is  within  the  apparent
        horizon.   The  dynamics of  the  collapsing  matter is  strongly
        influenced  by the  initial  amount of  angular  momentum in  the
        progenitor star  and, for  initial models with  sufficiently high
        angular velocities, the collapse can  lead to the formation of an
        unstable disc  in differential rotation.  All  of the simulations
        performed with uniformly rotating initial data and a
        polytropic or ideal-fluid equation of state show no evidence
        of shocks or of the presence of matter on stable orbits outside
        the black hole. 
\end{abstract}




\maketitle


\section{Introduction}
\label{sec:introduction}


        The numerical investigation of gravitational collapse of rotating
stellar configurations leading to black-hole formation is a long standing
problem in numerical relativity. However, it is through numerical
simulations in general relativity that one can hope to improve our
knowledge of fundamental aspects of Einstein's theory such as the cosmic
censorship hypothesis and black-hole no-hair theorems, along with that of
current open issues in relativistic astrophysics research, such as the
mechanism responsible for gamma-ray bursts. Furthermore, numerical
simulations of stellar gravitational collapse to black holes provide a
unique mean of computing the gravitational waveforms emitted in such
events, believed to be among the most important sources of detectable
gravitational radiation.

        However, the modelling of black-hole spacetimes with collapsing matter-sources in
multidimensions is one of the most formidable efforts of numerical relativity. This is due, on one
hand, to the inherent difficulties and complexities of the system of equations which is to be
integrated, the Einstein field equations coupled to the general-relativistic hydrodynamics
equations, and, on the other hand, to the immense computational resources needed to integrate the
equations in the case of three-dimensional (3D) evolutions. In addition to the practical
difficulties encountered in the accurate treatment of the hydrodynamics involved in the
gravitational collapse of a rotating neutron star to a black hole, the precise numerical computation
of the gravitational radiation emitted in the process is particularly challenging as the energy
released in gravitational waves is much smaller than the total rest-mass energy of the system.


The ability to perform long-term numerical simulations of
self-gravitating systems in general relativity strongly depends on the
formulation adopted for the Einstein equations. The covariant nature
of these equations (the ``many-fingered time'' of relativity) leads to
difficulties in constructing an appropriate coordinate representation
which would allow for stable and accurate simulations.  Over the
years, the standard approach has been mainly based upon the
unconstrained solution of the $3+1$ ADM formulation of the field
equations, which, despite large-scale and dedicated
collaborations~\cite{Cook97a, Abrahams97a, Gomez98a} has gradually
been shown to lack the stability properties necessary for long-term
numerical simulations. In recent years, however, a considerable effort
has been invested in extending the set of ADM equations solved by
including at some level the solution of the constraint equations on
each spatial hypersurface~\cite{AM:03,Schnetter03c,Bonazzola03a}, or
by reformulating the ADM approach in order to achieve long-term
stability (see, {\it e.g.,}~\cite{Lehner01a} and references therein).
Building on the experience developed with lower-dimensional
formulations, Nakamura, Oohara and Kojima~\cite{Nakamura87} presented
in 1987 a conformal traceless reformulation of the ADM system which
subsequent authors (see, {\it e.g.,}~\cite{Shibata95, Baumgarte99,
  Shibata99e, Shibata03,Alcubierre99d, Font02c, Yoneda02a,
  Kawamura03}) gradually showed to be robust enough to accomplish such
a goal for different classes of spacetimes, including black holes and
neutron stars (both isolated and in coalescing binary systems). The
most widespread version developed from this formalism, which we refer
to here as the NOK formulation, was given
by~\cite{Shibata95,Baumgarte99} and is commonly referred to as the
BSSN formulation.


        In addition to the improvements achieved in the formulation of the field equations,
successful long-term 3D evolutions of black holes in vacuum have been obtained in the last few years
using excision techniques (see, {\it e.g.,}~\cite{Seidel92a, Brandt00, Alcubierre00a, Kidder00a,
Kidder01a,Alcubierre01a, Yo02a, Laguna02, Choptuik:2003ac, Sperhake:2003fc}), although the original
idea is much older~\cite{MayWhite67}. In this approach, the spacetime region within the black-hole
horizon, and so causally disconnected, can be safely ignored without affecting the evolution outside
the horizon as long as suitable boundary conditions are specified at the excision surface. The
simulations presented here show the applicability of excision techniques {\it also} in non-vacuum
spacetimes, namely during the collapse of rotating neutron stars to Kerr black holes. The excision
technique, which is applied once the black-hole apparent horizon is found, permits to extend
considerably the lifetime of the simulations, at least with the resolutions used here. This, in
turn, allows for an accurate investigation of the dynamics of the trapped surfaces formed during the
collapse, from which important information on the mass and spin of the black hole, as well as on the
amount of energy which is lost to gravitational radiation, can be extracted. While our study was
nearing completion, we have learnt that a similar approach has also been implemented with
success~\cite{Duez04}.


        The presence of rotation in the collapsing stellar models
requires multidimensional investigations, either in axisymmetry or in
full 3D. The numerical investigations of black-hole formation (beyond
spherical symmetry) started in the early 1980's with the pioneering
work of Nakamura~\cite{Nakamura81b}. He adopted the (2+1)+1
formulation of the Einstein equations in cylindrical coordinates and
introduced regularity conditions to avoid divergences at coordinate
singularities. Nakamura used a ``hypergeometric" slicing condition
which prevents the grid points from reaching the singularity when a
black hole forms. The simulations could track the evolution of the
collapse of a $ 10 M_\odot $ ``core'' of a massive star with different
amounts of rotational energy, up to the formation of a rotating black
hole. However, the numerical scheme employed was not accurate enough
to compute the emitted gravitational radiation. In subsequent works,
Nakamura \cite{Nakamura83} (see also \cite{Nakamura87}) considered a
configuration consisting of a neutron star with an accreting envelope,
which was thought to mimic mass fall-back in a supernova
explosion. Rotation and infall velocity were added to such a
configuration, simulating an evolution dependent on the prescribed
rotation rates and rotation laws.

        Later on, in a series of papers~\cite{Bardeen83,Stark85,stark86,stark87}, Bardeen, Stark and
Piran studied the collapse of rotating relativistic polytropes to black holes, succeeding in
computing the associated gravitational radiation. The gravitational field and hydrodynamics
equations were formulated using the $3+1$ formalism in two spatial dimensions, using the radial
gauge and a mixture of singularity-avoiding polar and maximal slicings. The initial model was a
spherically symmetric relativistic polytrope of mass $M$ in equilibrium. The gravitational collapse
was induced by lowering the pressure in the initial model by a prescribed (and often very large)
fraction and by simultaneously adding an angular momentum distribution approximating rigid-body
rotation. Their simulations showed that Kerr black-hole formation occurs only for angular momenta
less than a critical value.  Furthermore, the energy $\Delta E$ carried away through gravitational
waves from the collapse to a Kerr black hole was found to be $\Delta E/Mc^2<7\times 10^{-4}$, the
shape of the waveforms being nearly independent of the details of the collapse.

        The axisymmetric codes employed in the aforementioned works
adopted curvilinear coordinate systems which may lead to long-term
numerical instabilities at coordinate singularities. These coordinate
problems have deterred researchers from building 3D codes in spherical
coordinates. Recently, a general-purpose method (called ``cartoon''),
has been proposed to enforce axisymmetry in numerical codes based on
Cartesian coordinates and which does not suffer from stability
problems at coordinate singularities~\cite{alcubierre00b}. It should
be noted, however, that the stability properties of the cartoon method
are not fully understood yet, as discussed
by~\cite{Frauendiener02}. Using this method, Shibata~\cite{Shibata00a}
investigated the effects of rotation on the criterion for prompt
adiabatic collapse of rigidly and differentially rotating polytropes
to a black hole, finding that the criterion for black-hole formation
depends strongly on the amount of angular momentum, but only weakly on
its (initial) distribution. The effects of shock heating when using a
non-isentropic equation of state (EOS hereafter) are important in
preventing prompt collapse to black holes in the case of large
rotation rates.

        More recently, Shibata~\cite{Shibata03,Shibata03b} has performed
axisymmetric simulations of the collapse of rotating supramassive neutron
stars to black holes for a wide range of polytropic EOSs and with an
improved implementation of the hydrodynamics solver (based on approximate
Riemann solvers) with respect to the original implementation used
in~\cite{Shibata00a}. Parameterizing the ``stiffness" of the EOS through
the polytropic index $N$, the final state of the collapse is a Kerr black
hole without any noticeable disc formation, when the polytropic index $N$
is in the range $2/3 \le N \le 2$. Based on the specific angular momentum
distribution in the initial star, Shibata has estimated an upper limit to
the mass of a possible disc as being less than $10^{-3}$ of the initial
stellar mass~\cite{Shibata03b}.  Unfortunately, such small masses cannot
currently be confirmed with the presently-available resolutions in
3D simulations on uniform grids.


        Three-dimensional, fully relativistic simulations of the collapse
of supramassive uniformly rotating neutron stars to rotating black holes
were presented in~\cite{Shibata99e}. The simulations focused on $N=1$
polytropes and showed no evidence of massive disc formation or
outflows. These results are in agreement with those obtained in
axisymmetry~\cite{Shibata03,Shibata03b} and with the new simulations
reported by \cite{Duez04} (both in axisymmetry and in 3D) which show that
for a rapidly rotating polytrope with $J/M^2<0.9$ ($J$ being the angular
momentum) all the mass falls promptly into the black hole, with no disc
being formed. Hence, all existing simulations agree that massive disc
formation from the collapse of neutron stars requires differential
rotation, at least for a polytropic EOS with $1 \le N \le 2$.

        Here, we present the results of new, fully 3D simulations of
gravitational collapse of uniformly rotating neutron stars, both
secularly and dynamically unstable, which we model as relativistic
polytropes.  The angular velocities of our sample of initial models range
from slow rotation to the mass-shedding limit. For the first time in such
3D simulations, we have detected the {\it event} horizon of the forming
black hole and showed that it can be used to achieve a more accurate
determination of the black-hole mass and spin than it would be otherwise
possible using the area of the {\it apparent} horizon. We have also
considered several other approaches to measure the properties of
the newly formed Kerr black hole, including the recently proposed {\it
isolated} and {\it dynamical-horizon} frameworks. A comparison among the
different methods has indicated that the dynamical-horizon approach is
simple to implement and yields estimates which are accurate and more
robust than those of the equivalent methods.

        The simulations are performed with a new general relativistic
hydrodynamics code, the {\tt Whisky} code, in which the Einstein and
hydrodynamics equations are finite-differenced on a Cartesian grid and
solved using state-of-the-art numerical schemes (a first description of
the code was given in \cite{Baiotti03a}). The code incorporates the
expertise developed over the past few years in the numerical solution of
the Einstein equations and of the hydrodynamics equations in a curved
spacetime (see~\cite{Alcubierre99d,Font02c}, but also~\cite{Font03} and
references therein) and is the result of a collaboration among several
European Institutes~\cite{eunetworkweb}.

        As mentioned before, we have implemented in the {\tt Whisky}
code a robust excision algorithm which warrants the extension of the
lifetime of the simulations far beyond the evolution times when the
black holes first form. Our calculations are starting from initially
axisymmetric stellar models but are performed in full 3D to allow for
departures from the initial axial symmetry. Overall, our results show
that the dynamics of the collapsing matter is strongly influenced by
the initial amount of angular momentum in the progenitor neutron star,
which, when sufficiently high, leads to the formation of an unstable
flattened object. All of the simulations performed for realistic
initial data and a polytropic equation of state show no evidence of
shock formation preventing a prompt collapse to a black hole, nor the
presence of matter on stable orbits outside the black hole. It should
be remarked, however, that both of these conclusions may change if
the initial stellar models are rotating differentially.

        The use of numerical grids with uniform spacing and the present
computational resources have placed the outer boundary of our
computational box in regions of the spacetime where the gravitational
waves have not yet reached their asymptotic form. As a result, the
information on the gravitational waveforms that we extract through
perturbative techniques~\cite{Rupright98,Rezzolla99a} does not provide
interesting information besides the obvious change in the stellar
quadrupole moment. Work is now in progress to use mesh refinement
techniques~\cite{Schnetter-etal-03b} to move the outer boundary
sufficiently far from the source so that important information can be
extracted on the gravitational wave emission produced during the
collapse. The results of these investigations will be presented in a
companion paper~\cite{Baiotti04}.


        The paper is organized as follows: Section~\ref{equations}
describes the formulation we adopt for the Einstein and hydrodynamics
equations, the way they are implemented in the code and a brief
discussion of how the excision techniques can be employed within a
hydrodynamical treatment making use of high-resolution shock-capturing
(HRSC) schemes. To avoid detracting the reader's attention from the
physical problem considered here, we have confined most of the technical
details concerning the numerical implementation of the hydrodynamical
equations to Appendix~\ref{sec:numerical-methods}.
Section~\ref{sec:model} is therefore devoted to describing the various
properties of the initial stellar models. The following two Sections,
~\ref{sec:dynamics-fluid} and~\ref{sec:dynamics-horizons}, present our
results regarding the dynamics of the collapsing stars and of the trapped
surfaces, respectively. In both cases we will be considering and
comparing the dynamics of slowly and of rapidly rotating stellar
models. The paper ends with Section~\ref{sec:conclusion}, which contains
a summary of the results obtained and the perspectives for further
investigations. Finally, Appendix~\ref{sec:numerical-tests} is devoted to
presenting some of the tests the code passes with very high accuracy.

        We here use a spacelike signature $(-,+,+,+)$ and a system of
units in which $c=G=M_\odot=1$ (unless explicitly shown otherwise for
convenience). Greek indices are taken to run from 0 to 3, Latin indices
from 1 to 3 and we adopt the standard convention for the summation over
repeated indices. 


\section{Basic equations and their implementation}
\label{equations}

        The {\tt Whisky} code solves the general relativistic
hydrodynamics equations on a 3D numerical grid with
Cartesian coordinates. The code has been constructed within the framework
of the {\tt Cactus} Computational Toolkit (see \cite{Cactusweb} for
details), developed at the Albert Einstein Institute (Golm) and at the
Louisiana State University (Baton Rouge). This public domain code
provides high-level facilities such as parallelization, input/output,
portability on different platforms and several evolution schemes to solve
general systems of partial differential equations. Clearly, special
attention is dedicated to the solution of the Einstein equations, whose
matter-terms in non-vacuum spacetimes are handled by the {\tt Whisky}
code. While the {\tt Whisky} code is entirely new, its initial
development has benefitted in part from the release of a public version
of the general relativistic hydrodynamics code described
in~\cite{Font98b,Font02c}, and developed mostly by the group at the
Washington University (St. Louis).

        The {\tt Whisky} code, however, incorporates important recent
developments regarding, in particular, new numerical methods for the
solution of the hydrodynamics equations that have been described in
detail in~\cite{Baiotti03a} and will be briefly reviewed in
Appendix~\ref{sec:numerical-methods}. These include: {\it (i)} the
Piecewise Parabolic Method (PPM) \cite{Colella84} and the Essentially
Non-Oscillatory (ENO) methods \cite{Harten87} for the cell-reconstruction
procedure; {\it (ii)} the Harten-Lax-van Leer-Einfeldt (HLLE)
\cite{Harten83} approximate Riemann solver, the Marquina flux formula
\cite{Aloy99b}; {\it (iii)} the analytic expression for the left
eigenvectors \cite{Ibanez01} and the compact flux formulae \cite{Aloy99a}
for a Roe-type Riemann solver and a Marquina flux formula; {\it (iv)} the
use of a ``method of lines'' (MoL) approach for the implementation of
high-order time evolution schemes; {\it (v)} the possibility to couple
the general relativistic hydrodynamics equations with a conformally
decomposed three-metric. The incorporation of these new numerical techniques
in the code has led to a much improved ability to simulate relativistic
stars, as shown in Appendix~\ref{sec:numerical-tests} which is devoted to
code tests.

        While the {\tt Cactus} code provides at each time step a solution
of the Einstein equations~\cite{Alcubierre99d}
\begin{equation}
\label{efes}
G_{\mu \nu}=8\pi T_{\mu \nu}\ , 
\end{equation}
where $G_{\mu \nu}$ is the Einstein tensor and $T_{\mu \nu}$ is the
stress-energy tensor, the {\tt Whisky} code provides the time evolution
of the hydrodynamics equations, expressed through the conservation
equations for the stress-energy tensor $T^{\mu\nu}$ and for the matter
current density $J^\mu$
\begin{equation}
\label{hydro eqs}
\nabla_\mu T^{\mu\nu} = 0\;,\;\;\;\;\;\;
\nabla_\mu J^\mu = 0.
\end{equation}

        In what follows we briefly discuss how both the right and the
left-hand side of equations (\ref{efes}) are computed within the coupled
{\tt Cactus/Whisky} codes.


\subsection{Evolution of the field equations}   
\label{feqs}

        We here give only a brief overview of the system of equations for
the evolution of the field equations, but refer the reader
to~\cite{Alcubierre99d} for more details. Many different formulations of
the equations have been proposed throughout the years, starting with the
ADM formulation in 1962~\cite{Arnowitt62}. As mentioned in the
Introduction, we use the NOK \cite{Nakamura87} formulation, which is
based on the ADM construction and has been further developed
in~\cite{Shibata95}.

        In the ADM formulation~\cite{Arnowitt62}, the spacetime is
foliated with a set of non-intersecting spacelike hypersurfaces. Two
kinematic variables relate the hypersurfaces: the lapse function $\alpha$,
which describes the rate of advance of time along a timelike unit vector
$n^\mu$ normal to a spacelike hypersurface, and the shift three-vector
$\beta^i$ that relates the coordinates of two spacelike hypersurfaces.
In this construction the line element reads
\begin{equation}
ds^2 = -(\alpha^{2} -\beta _{i}\beta ^{i}) dt^2 + 
        2 \beta_{i} dx^{i} dt +\gamma_{ij} dx^{i} dx^{j}
        \ .
\end{equation}

        The original ADM formulation casts the Einstein equations into a
first-order (in time) quasi-linear~\cite{Richtmyer67} system of
equations. The dependent variables are the three-metric $\gamma_{ij}$ and the
extrinsic curvature $K_{ij}$, with first-order evolution equations given
by
\begin{eqnarray}
\partial_t \gamma_{ij} &=& - 2 \alpha K_{ij}+\nabla_i
        \beta_j + \nabla_j \beta_i, 
\label{dtgij} \\
        \partial_t K_{ij} &=& -\nabla_i \nabla_j \alpha + \alpha \Biggl[
        R_{ij}+K\ K_{ij} -2 K_{im} K^m_j  \nonumber \\
        &\ & - 8 \pi \left( S_{ij} - \frac{1}{2}\gamma_{ij}S \right)
        - 4 \pi {\rho}_{_{\rm ADM}} \gamma_{ij}
        \Biggr] \nonumber \\ 
        &\ & + \beta^m \nabla_m K_{ij}+K_{im} 
        \nabla_j \beta^m+K_{mj} \nabla_i \beta^m .
        \nonumber \\
\label{dtkij}
\end{eqnarray}
Here, $\nabla_i$ denotes the covariant derivative with respect to the
three-metric $\gamma_{ij}$, $R_{ij}$ is the Ricci curvature of the
three-metric, $K\equiv\gamma^{ij}K_{ij}$ is the trace of the extrinsic
curvature, $S_{ij}$ is the projection of the stress-energy tensor onto
the spacelike hypersurfaces and $S \equiv \gamma^{ij} S_{ij}$ (for a
more detailed discussion, see~\cite{York79}). In addition to the
evolution equations, the Einstein equations also provide four
constraint equations to be satisfied on each spacelike
hypersurface. These are the Hamiltonian constraint equation
\begin{equation}
\label{ham_constr}
{}^{(3)}R + K^2 - K_{ij} K^{ij} - 16 \pi
        {\rho}_{_{\rm ADM}} = 0 \ ,
\end{equation}
and the momentum constraint equations
\begin{equation}
\label{mom_constr}
\nabla_j K^{ij} - \gamma^{ij} \nabla_j K - 8 \pi j^i = 0 \ .
\end{equation}
In equations (\ref{dtgij})--(\ref{mom_constr}), $ {\rho}_{_{\rm ADM}}$
and $j^i$ are the energy density and the momentum density as measured by
an observer moving orthogonally to the spacelike hypersurfaces.

Details of our particular implementation of the conformal traceless
reformulation of the ADM system as proposed
by~\cite{Nakamura87,Shibata95,Baumgarte99} are extensively described
in~\cite{Alcubierre99d,Alcubierre02a} and will not be repeated here. We
only mention, however, that this formulation makes use of a conformal
decomposition of the three-metric, \hbox{$\tilde \gamma_{ij} = e^{- 4 \phi}
\gamma_{ij}$}, and the trace-free part of the extrinsic curvature,
\hbox{$A_{ij} = K_{ij} - \gamma_{ij} K/3$}, with the conformal factor
$\phi$ chosen to satisfy $e^{4 \phi} = \gamma^{1/3}$, where $\gamma$ is
the determinant of the spatial three-metric $\gamma_{ij}$. In this
formulation, in addition to the evolution equations for the conformal
three-metric $\tilde \gamma_{ij}$ and the conformal traceless extrinsic
curvature $\tilde A_{ij}$, there are evolution equations for
the conformal factor $\phi$, for the trace of the extrinsic curvature $K$
and for the ``conformal connection functions'' $\tilde \Gamma^i \equiv
\tilde \gamma^{ij}{}_{,j}$. We note that although the final mixed, first
and second-order, evolution system for $\left\{ \phi, K, \tilde
\gamma_{ij}, {\tilde A_{ij}}, {\tilde \Gamma^i} \right\}$ is not in any
immediate sense hyperbolic, there is evidence showing that the
formulation is at least equivalent to a hyperbolic
system~\cite{Sarbach02a,Bona:2003qn,Nagy04}. In the formulation
of~\cite{Shibata95}, the auxiliary variables ${\tilde F}_i = -\sum_j
{\tilde \gamma_{ij,j}}$ were used instead of the ${\tilde \Gamma^i}$.

        In Refs.~\cite{Alcubierre99d,Alcubierre99e} the improved
properties of this conformal traceless formulation of the Einstein
equations were compared to the ADM system. In particular,
in~\cite{Alcubierre99d} a number of strongly gravitating systems were
analysed numerically with {\em convergent} HRSC methods with {\it
total-variation-diminishing} (TVD) schemes using the equations
described in \cite{Font98b}. These included weak and strong
gravitational waves, black holes, boson stars and relativistic
stars. The results showed that this treatment leads to numerical
evolutions of the various strongly gravitating systems which did not
show signs of numerical instabilities for sufficiently long
times. However, it was also found that the conformal traceless
formulation requires grid resolutions higher than the ones needed in
the ADM formulation to achieve the same accuracy, when the foliation
is made using the ``$K$-driver'' approach discussed
in~\cite{Balakrishna96a}. Because in long-term evolutions a small
error-growth rate is the most desirable property, we have adopted the
conformal traceless formulation as our standard form for the evolution
of the field equations.

\subsubsection{Gauge choices}

        The code is designed to handle arbitrary shift and lapse
conditions, which can be chosen as appropriate for a given spacetime
simulation.  More information about the possible families of spacetime
slicings which have been tested and used with the present code can be
found in~\cite{Alcubierre99d,Alcubierre01a}. Here, we limit ourselves to
recalling details about the specific foliations used in the present
evolutions. In particular, we have used hyperbolic $K$-driver slicing
conditions of the form
\begin{equation}
(\partial_t - \beta^i\partial_i) \alpha = - f(\alpha) \;
\alpha^2 (K-K_0),
\label{eq:BMslicing}
\end{equation}
with $f(\alpha)>0$ and $K_0 \equiv K(t=0)$. This is a generalization of
many well known slicing conditions.  For example, setting $f=1$ we
recover the ``harmonic'' slicing condition, while, by setting
\mbox{$f=q/\alpha$}, with $q$ an integer, we recover the generalized
``$1+$log'' slicing condition~\cite{Bona94b}.  In particular, all of the
simulations discussed in this paper are done using condition
(\ref{eq:BMslicing}) with $f=2/\alpha$. This choice has been made mostly because
of its computational efficiency, but we are aware that ``gauge
pathologies'' could develop with the ``$1+$log''
slicings~\cite{Alcubierre97a,Alcubierre97b}.


        As for the spatial gauge, we use one of the ``Gamma-driver''
shift conditions proposed in~\cite{Alcubierre01a} (see also 
\cite{Alcubierre02a}), that essentially
act so as to drive the $\tilde{\Gamma}^{i}$ to be constant. In this
respect, the ``Gamma-driver'' shift conditions are similar to the
``Gamma-freezing'' condition $\partial_t \tilde\Gamma^k=0$, which, in
turn, is closely related to the well-known minimal distortion shift
condition~\cite{Smarr78b}. The differences between these two conditions
involve the Christoffel symbols and are basically due to the fact that the
minimal distortion condition is covariant, while the Gamma-freezing
condition is not.

        In particular, all of the results reported here have been
obtained using the hyperbolic Gamma-driver condition,
\begin{equation}
\partial^2_t \beta^i = F \, \partial_t \tilde\Gamma^i - \eta \,
\partial_t \beta^i,
\label{eq:hyperbolicGammadriver}
\end{equation}
where $F$ and $\eta$ are, in general, positive functions of space and
time. For the hyperbolic Gamma-driver conditions it is crucial to add
a dissipation term with coefficient $\eta$ to avoid strong
oscillations in the shift. Experience has shown that by tuning the
value of this dissipation coefficient it is possible to almost freeze
the evolution of the system at late times. We typically choose
$F=\frac{3}{4}$ and $\eta=3$ and do not vary them in time.

\subsection{Evolution of the hydrodynamics equations}    

        An important feature of the {\tt Whisky} code is the
implementation of a \textit{conservative formulation} of the
hydrodynamics equations \cite{Marti91,Banyuls97,Ibanez01}, in which the
set of equations (\ref{hydro eqs}) is written in a hyperbolic,
first-order and flux-conservative form of the type
\begin{equation}
\label{eq:consform1}
\partial_t {\mathbf q} + 
        \partial_i {\mathbf f}^{(i)} ({\mathbf q}) = 
        {\mathbf s} ({\mathbf q})\ ,
\end{equation}
where ${\mathbf f}^{(i)} ({\mathbf q})$ and ${\mathbf s}({\mathbf q})$
are the flux-vectors and source terms, respectively~\cite{Font03}.  Note
that the right-hand side (the source terms) depends only on the metric,
and its first derivatives, and on the stress-energy tensor. Furthermore,
while the system (\ref{eq:consform1}) is not strictly hyperbolic,
strong hyperbolicity is recovered in a flat spacetime, where ${\mathbf s}
({\mathbf q})=0$.

        As shown by \cite{Banyuls97}, in order to write system
(\ref{hydro eqs}) in the form of system (\ref{eq:consform1}), the
\textit{primitive} hydrodynamical variables ({\it i.e.} the
rest-mass density $\rho$ and the pressure $p$ (measured in the
rest-frame of the fluid), the fluid three-velocity $v^i$ (measured by a local
zero-angular momentum observer), the specific internal energy $\epsilon$
and the Lorentz factor $W$) are mapped to the so called \textit{conserved}
variables \mbox{${\mathbf q} \equiv (D, S^i, \tau)$} via the relations

\vspace{-0.2 cm}
\begin{eqnarray}
  \label{eq:prim2con}
   D &\equiv& \sqrt{\gamma}W\rho\ , \nonumber\\
   S^i &\equiv& \sqrt{\gamma} \rho h W^2 v^i\ ,  \\
   \tau &\equiv& \sqrt{\gamma}\left( \rho h W^2 - p\right) - D\ , \nonumber
\end{eqnarray}
where $h \equiv 1 + \epsilon + p/\rho$ is the specific enthalpy and
\hbox{$W \equiv (1-\gamma_{ij}v^i v^j)^{-1/2}$}. Note that only five of
the seven primitive variables are independent.

        In order to close the system of equations for the hydrodynamics
an EOS which relates the pressure to the rest-mass density and to the
energy density must be specified. The code has been written to use any
EOS, but all of the tests so far have been performed using either an
(isentropic) polytropic EOS
\begin{eqnarray}
\label{poly}
p &=& K \rho^{\Gamma}\ , \\
e &=& \rho + \frac{p}{\Gamma-1}\ ,
\end{eqnarray}
or an ``ideal fluid'' EOS
\begin{equation}
\label{id fluid}
p = (\Gamma-1) \rho\, \epsilon \ . 
\end{equation}
Here, $e$ is the energy density in the rest-frame of the fluid, $K$ the polytropic constant (not to
be confused with the trace of the extrinsic curvature defined earlier) and $\Gamma$ the adiabatic
exponent. In the case of the polytropic EOS (\ref{poly}), $\Gamma=1+1/N$, where $N$ is the
polytropic index and the evolution equation for $\tau$ needs not be solved. In the case of the
ideal-fluid EOS (\ref{id fluid}), on the other hand, non-isentropic changes can take place in the
fluid and the evolution equation for $\tau$ needs to be solved. In addition to the EOSs (\ref{poly})
and (\ref{id fluid}), a ``hybrid" EOS (suitable for core-collapse simulations), as described
in~\cite{Zwerger95,Zwerger97}, has also been implemented.

Note that polytropic EOSs \eqref{poly} do not allow any transfer of kinetic energy to thermal
energy, a process which occurs in physical shocks (shock heating). However, we have verified, by
performing simulations with the more general EOS \eqref{id fluid} on some selected cases, that for
the physical systems treated here, shock heating is not important (no shocks form during our
simulations). Since in our numerical scheme using e general EOS like \eqref{id fluid} is more
expensive than using a polytropic EOS, the systematic investigations presented here have been
obtained using the latter.

        Additional details of the formulation we use for the
hydrodynamics equations can be found in \cite{Font03}. We stress that an
important feature of this formulation is that it has allowed to extend to
a general relativistic context the powerful numerical methods developed
in classical hydrodynamics, in particular HRSC schemes
based on linearized Riemann solvers (see~\cite{Font03}). Such schemes are
essential for a correct representation of shocks, whose presence is
expected in several astrophysical scenarios. Two important results
corroborate this view. The first one, by Lax and Wendroff~\cite{Lax60},
states that a stable scheme converges to a weak solution of the
hydrodynamical equations. The second one, by Hou and
LeFloch~\cite{Hou94}, states that, in general, a non-conservative scheme
will converge to the wrong weak solution in the presence of a shock,
hence underlining the importance of flux-conservative formulations. For a
full introduction to HRSC methods the reader is also referred
to~\cite{Laney98,Toro99,Leveque98}.

\subsection{Hydrodynamical excision}
\label{sec:hydr-excis}

        Excision boundaries are usually based on the principle that a
region of spacetime that is causally disconnected can be ignored without
this affecting the solution in the remaining portion of the
spacetime. Although this is true for signals and perturbations travelling
at physical speeds, numerical calculations may violate this assumption
and disturbances, such as gauge waves, may travel at larger speeds thus
leaving the physically disconnected regions. 

        Note that, in non-vacuum spacetimes, the excision boundaries for
the hydrodynamical and the metric fields need not be the same. For the
fluid quantities, in fact, all characteristics emanating from an event in
spacetime will propagate within the sound-cone at that event, and, for
physically realistic EOSs, this sound-cone will always be contained
within the light-cone at that event. As a result, if a region of
spacetime contains trapped surfaces, both the hydrodynamical and the
metric fields are causally disconnected and both can be excised
there. On the other hand, there may be situations, {\it e.g.,} when the
bulk flow is locally supersonic but no trapped surfaces are present, in
which it is possible (at least in principle) to excise the hydrodynamical
fields without having to do the same for the metric fields. We have not
used this option here and the hydrodynamical excision implemented in our
simulations has always been made within regions of the spacetime
contained inside trapped surfaces.

        A first naive implementation of an excision algorithm within a
HRSC method could ensure that the data used to construct the flux
at the excision boundary is extrapolated from data outside the excision
region. This may appear to be a good idea since HRSC methods naturally
change the stencils depending on the data locally. In general, however,
this approach is not guaranteed to reduce the total variation of the
solution and simple examples may be produced that fail with this
boundary condition.

        An effective solution, however, is not much more complicated and
can be obtained by applying at the excision boundary the simplest outflow
boundary condition (here, by outflow we mean flow into the excision
region). In practice, we apply a zeroth-order extrapolation to all
variables at the boundary, {\it i.e.}\ a simple copy of the hydrodynamical
variables across the excision boundary (note that setting the
hydrodynamical fields inside the excised region to zero would still yield
an outflow boundary condition, but leads to incorrect outflow speeds). If
the reconstruction method requires more cells inside the excision region,
we force the stencil to only consider the data in the exterior and the
first interior cell. Although the actual implementation of this excision
technique may depend on the reconstruction method used, the working
principle is always the same.

        The location of the excision boundary itself is based on the
determination of the apparent horizon which, within the {\tt Cactus}
code, is obtained using the fast finder of Thornburg
~\cite{Thornburg2003:AH-finding}. The excision boundary is placed a few
gridpoints (typically 4), within a surface which is 0.6 times the size of
the apparent horizon. This may not be a suitable outflow boundary on a
Cartesian grid, as pointed out by~\cite{Kidder00b,
Calabrese:2003a}. However, similar or larger excision regions show no
problems in vacuum evolutions and since the sound-cones are always
contained within the light-cones, we expect no additional problems to
arise from the hydrodynamics.

        More details on how the hydrodynamical excision is applied in
practice, as well as tests showing that this method is stable, consistent
and converges to the expected order will be published in a separate
paper~\cite{Hawke04}.


\section{Initial stellar models}
\label{sec:model}

        As mentioned earlier, this paper is specially dedicated to study
the gravitational collapse of slowly and rapidly rotating supramassive
relativistic stars, in uniform rotation, that have become unstable to
axisymmetric perturbations.  Given equilibrium models of gravitational
mass $M$ and central energy density $e_c$ along a sequence of fixed
angular momentum or fixed rest mass, the Friedman, Ipser \& Sorkin
criterion \hbox{$\partial M/\partial e_c= 0$} \cite{Friedman88} can be
used to locate the exact onset of the {\it secular} instability to
axisymmetric collapse. The onset of the {\it dynamical} instability to
collapse is located near that of the secular instability but at somewhat
larger central energy densities. Unfortunately, no simple criterion
exists to determine this location, but the expectation mentioned above
has been confirmed by the simulations performed here and by those
discussed in~\cite{Shibata99e}. Note that in the absence of viscosity or
strong magnetic fields, the star is not constrained to rotate uniformly
after the onset of the secular instability and could develop differential
rotation. In a realistic neutron star, however, viscosity or intense
magnetic fields are likely to enforce a uniform rotation and cause the
star to collapse soon after it passes the secular instability limit.

        The initial data for our simulations are constructed using a 2D
numerical code, that computes accurate stationary equilibrium solutions
for axisymmetric and rapidly rotating relativistic stars in polar
coordinates \cite{Stergioulas95}. The data are then transformed to
Cartesian coordinates using standard coordinate transformations.  The
same initial data routines have been used in previous 3D simulations
\cite{Alcubierre99d,Font02c,Stergioulas01} and details on the accuracy of
the code can be found in~\cite{Stergioulas03}.

\begin{figure}
\centering
\includegraphics[angle=0,width=8.5cm]{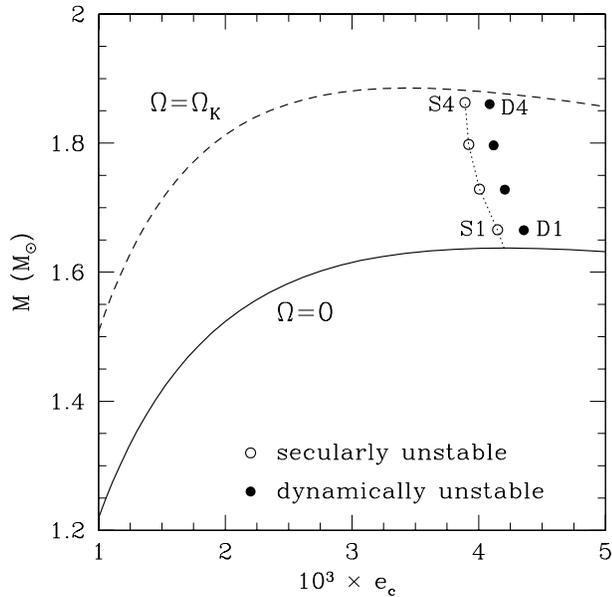} 
\caption{Gravitational mass shown as a function of the central energy
        density for equilibrium models constructed with the polytropic
        EOS, for $\Gamma=2$ and polytropic constant $K_{\rm ID}=100$. The
        solid, dashed and dotted lines correspond to the sequence of
        non-rotating models, the sequence of models rotating at the
        mass-shedding limit and the sequence of models that are at the
        onset of the secular instability to axisymmetric
        perturbations. Also shown are the secularly (open circles) and
        dynamically unstable (filled circles) initial models used in the
        collapse simulations.}
\label{figInitial} 
\end{figure}

        For simplicity, we have focused on initial models constructed
with the polytropic EOS (\ref{poly}), choosing $\Gamma=2$ and polytropic
constant $K_{\rm ID}=100$ to produce stellar models that are, at least
qualitatively, representative of what is expected from observations of
neutron stars. More specifically, we have selected four models located on
the line defining the onset of the secular instability and having
polar-to-equatorial axes ratio of roughly 0.95, 0.85, 0.75 and 0.65
(these models are indicated as S1--S4 in Fig.\ref{figInitial}), respectively. Four
additional models were defined by increasing the central energy density
of the secularly unstable models by $5\%$, keeping the same axis
ratio. These models (indicated as D1--D4 in Fig.\ref{figInitial}) were
expected and have been found to be dynamically unstable.

        Figure~\ref{figInitial} shows the gravitational mass as a
function of the central energy density for equilibrium models constructed
with the chosen polytropic EOS. The solid, dashed and dotted lines
correspond respectively to: the sequence of non-rotating models, the
sequence of models rotating at the mass-shedding limit and the sequence
of models that are at the onset of the secular instability to
axisymmetric perturbations. Furthermore, the secularly and dynamically
unstable initial models used in the collapse simulations are shown as
open and filled circles, respectively.

\begin{table}
\caption{Equilibrium properties of the initial stellar models. The
        different columns refer respectively to: the central rest-mass
        density $\rho_c$, the ratio of the polar to equatorial coordinate
        radii $r_p/r_e$, the gravitational mass $M$, the circumferential
        equatorial radius $R_e$, the angular velocity $\Omega$, the ratio
        $J/M^2$ where $J$ is the angular momentum, the ratio of
        rotational kinetic energy to gravitational binding energy
        $T/|W|$, and the ``height'' of the corotating ISCO $h_+$. All
        models have been computed with a polytropic EOS with $K_{\rm
        ID}=100$ and $\Gamma=2$.}
\medskip
\begin{tabular}{*{9}{c}}
\hline
Model & $\rho_c^{\dagger}$   & $r_p/r_e$ & $M$ & $R_e$ &
        $\Omega^{\ddagger}$  &$J/M^2$
        & $T/|W|^{\ddagger}$            & $h_+$ \\ 
\hline 
S1 & 3.154 & 0.95 & 1.666 & 7.82 & 1.69 & 0.207 & 1.16 & 1.18 \\
S2 & 3.066 & 0.85 & 1.729 & 8.30 & 2.83 & 0.363 & 3.53 & 0.51 \\
S3 & 3.013 & 0.75 & 1.798 & 8.90 & 3.49 & 0.470 & 5.82 & 0.04 \\
S4 & 2.995 & 0.65 & 1.863 & 9.76 & 3.88 & 0.545 & 7.72 &  --  \\
\hline 
D1 & 3.280 & 0.95 & 1.665 & 7.74 & 1.73 & 0.206 & 1.16 & 1.26 \\
D2 & 3.189 & 0.85 & 1.728 & 8.21 & 2.88 & 0.362 & 3.52 & 0.58 \\
D3 & 3.134 & 0.75 & 1.797 & 8.80 & 3.55 & 0.468 & 5.79 & 0.10 \\
D4 & 3.116 & 0.65 & 1.861 & 9.65 & 3.95 & 0.543 & 7.67 &  --  \\
\hline \\
${}^{\dagger} \  \times 10^{-3}$ \\
${}^{\ddagger} \ \times 10^{-2}$ \\
\end{tabular}
\label{tableInitial}
\end{table}

        Table~\ref{tableInitial} summarizes the main equilibrium
properties of the initial models. The circumferential equatorial radius
is denoted as $R_e$, while $\Omega$ is the angular velocity with respect
to an inertial observer at infinity, and $r_p/r_e$ is the ratio of the
polar to equatorial coordinate radii. The height of the corotating
innermost stable circular orbit (ISCO) is defined as $h_+=R_+ - R_e$,
where $R_+$ is the circumferential radius for a corotating ISCO observer. Note
that in those models for which a value of $h_+$ is not reported, all
circular geodesic orbits outside the stellar surface are stable. Other
quantities shown are the central rest-mass density $\rho_c$, the angular
momentum $J$, and the ratio of rotational kinetic energy to gravitational
binding energy $T/|W|$.


\section{Dynamics of the matter}
\label{sec:dynamics-fluid}

        This Section discusses the dynamics of the matter during the
collapse of the initial stellar models described in the preceding
section. All of the simulations reported here have been computed using a
uniformly spaced computational grid for which symmetry conditions are
imposed across the equatorial plane. Different spatial
resolutions have been used to check convergence and improve the accuracy
of the results, with the finest resolution having been obtained using
$288^2\times 144$ cells. 
While the precise numbers depend on the resolution used and on the model
simulated, as a general rule we have used $\sim 50\%$ of the cells in the
$x$-direction to cover the star in case D1 and $\sim 66\%$ of the
cells in the $x$-direction to cover the star in D4. As a result, the outer
boundary is set at $\sim 2.0$ times the stellar equatorial radius for D1 and
at $\sim 1.4$ times the stellar equatorial radius for D4.

The hydrodynamics equations have been solved employing the Marquina flux formula and a third-order
PPM reconstruction, which was shown in \cite{Font99} to be superior to other methods in maintaining
a highly-accurate angular-velocity profile (see also \cite{Stergioulas01,Stergioulas03b} for recent
3D evolutions of rotating relativistic stars with a third-order order PPM reconstruction). The
Einstein field equations, on the other hand, have been evolved using the ICN evolution scheme, the
``$1+$log'' slicing condition and the ``Gamma-driver'' shift
conditions~\cite{Alcubierre01a}. Finally, both polytropic and ideal-fluid EOSs have been used,
although no significant difference has been found in the dynamics between the two cases. This is
most probably related to the small $J/M^2$ of the uniformly rotating initial models considered
here. This implies a relatively rapid collapse and as a result we do not see any shock formation (see
below for a more complete discussion). Hereafter
we will restrict our attention to a polytropic EOS only.

\begin{figure}
\centering
\includegraphics[angle=0,width=8.5cm]{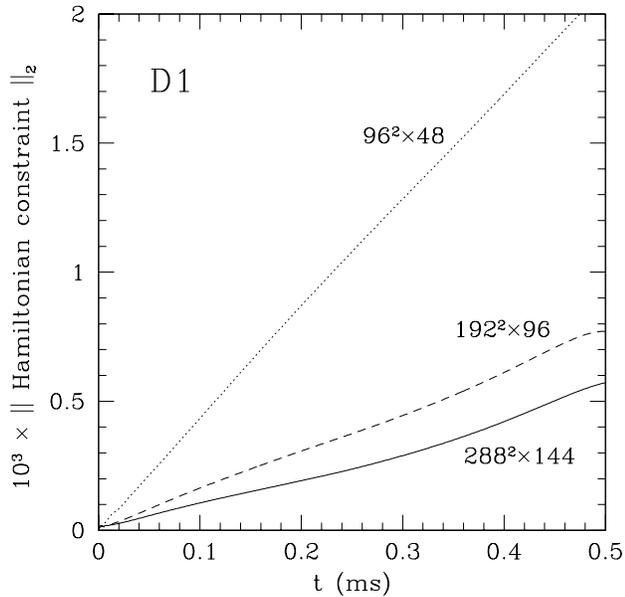}
\caption{$L_2$ norm of the Hamiltonian constraint violation for the initial
model D1 shown as a function of time. The different lines refer to
different grid resolutions, but in all cases the IVP was solved after
the pressure was uniformly decreased to trigger the collapse.}
\label{fig_ham_n2_conv} 
\end{figure}

\begin{figure*}
\centering
\includegraphics[angle=0,width=6.2cm]{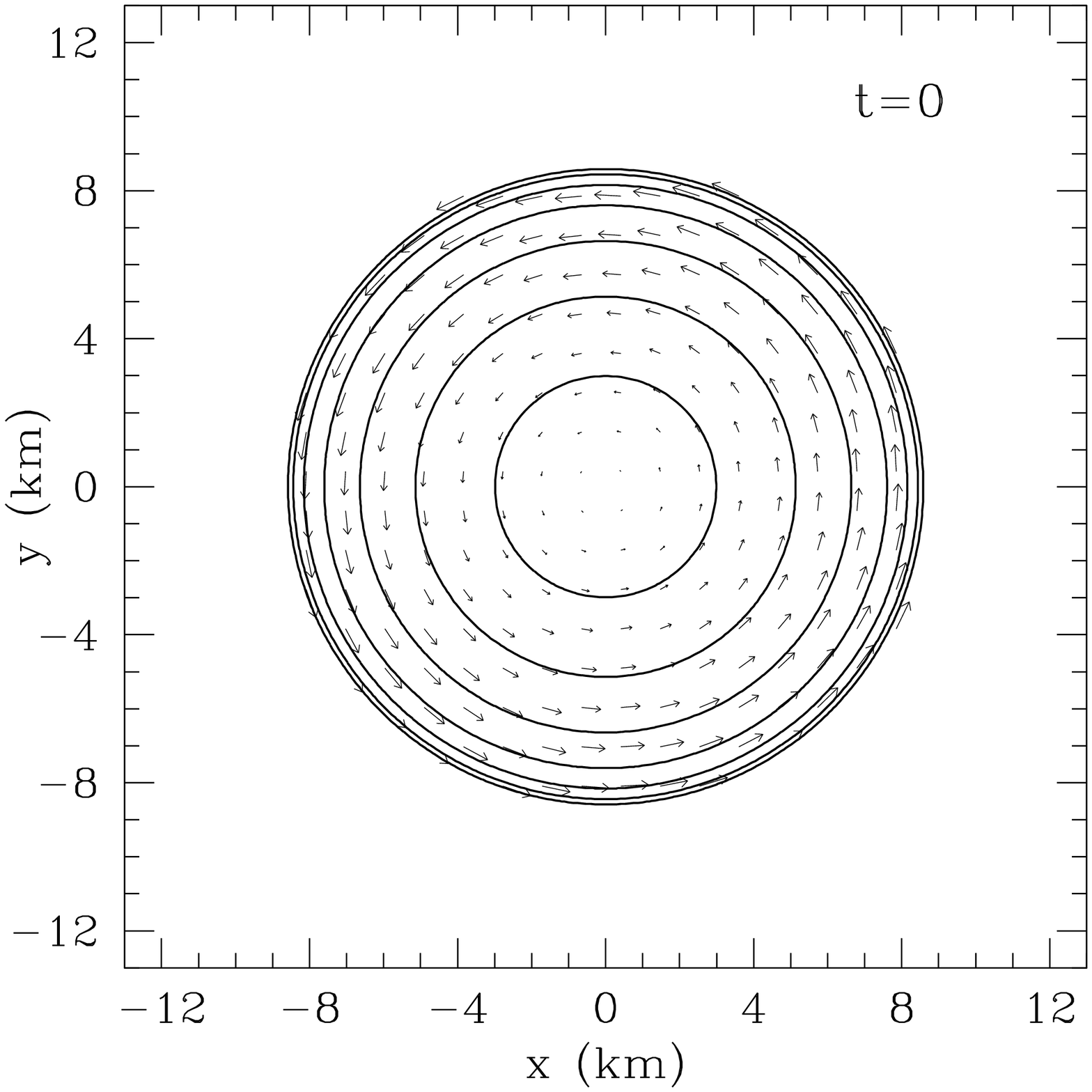} 
\hskip 1.0cm
\includegraphics[angle=0,width=6.2cm]{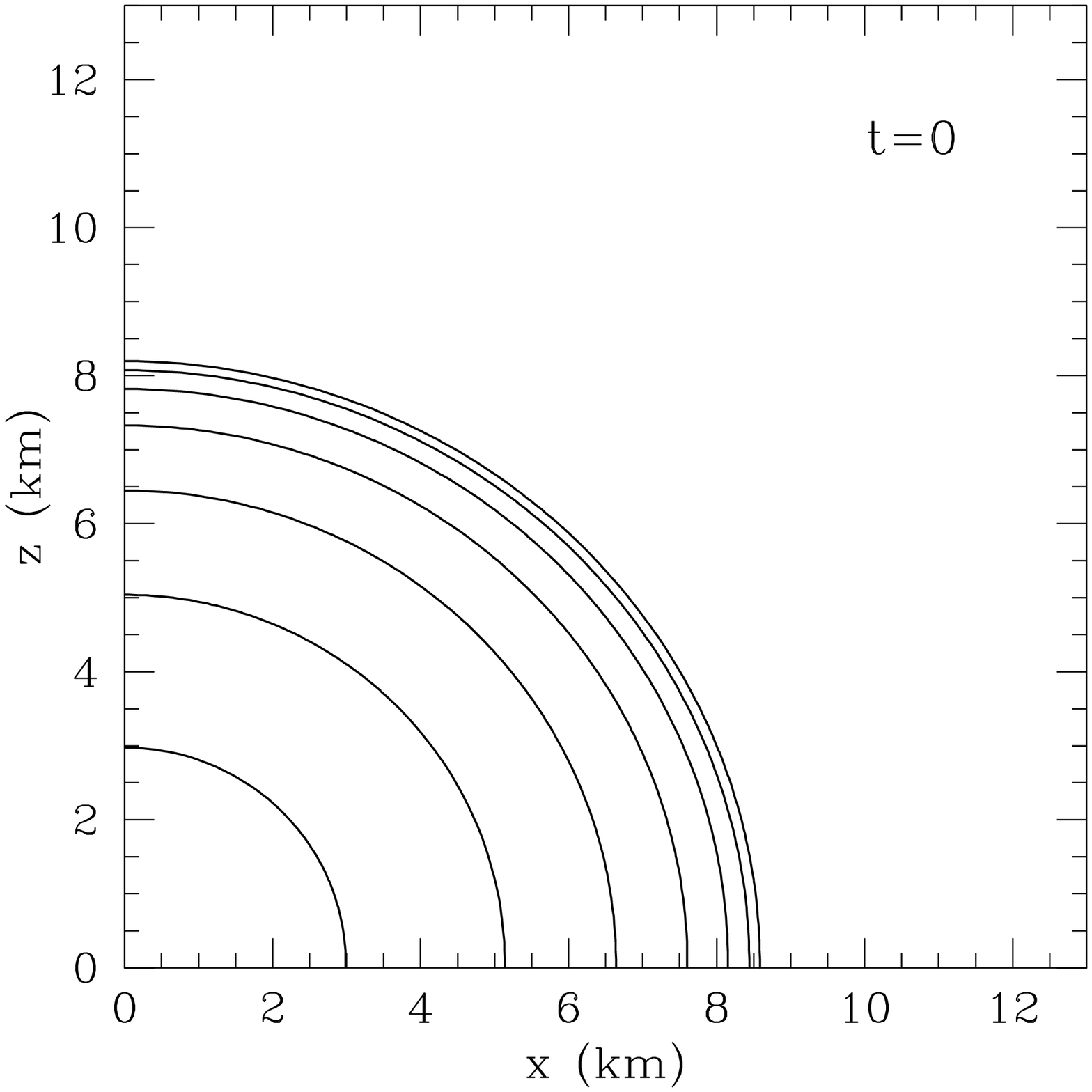} 
\vskip 0.25cm
\includegraphics[angle=0,width=6.2cm]{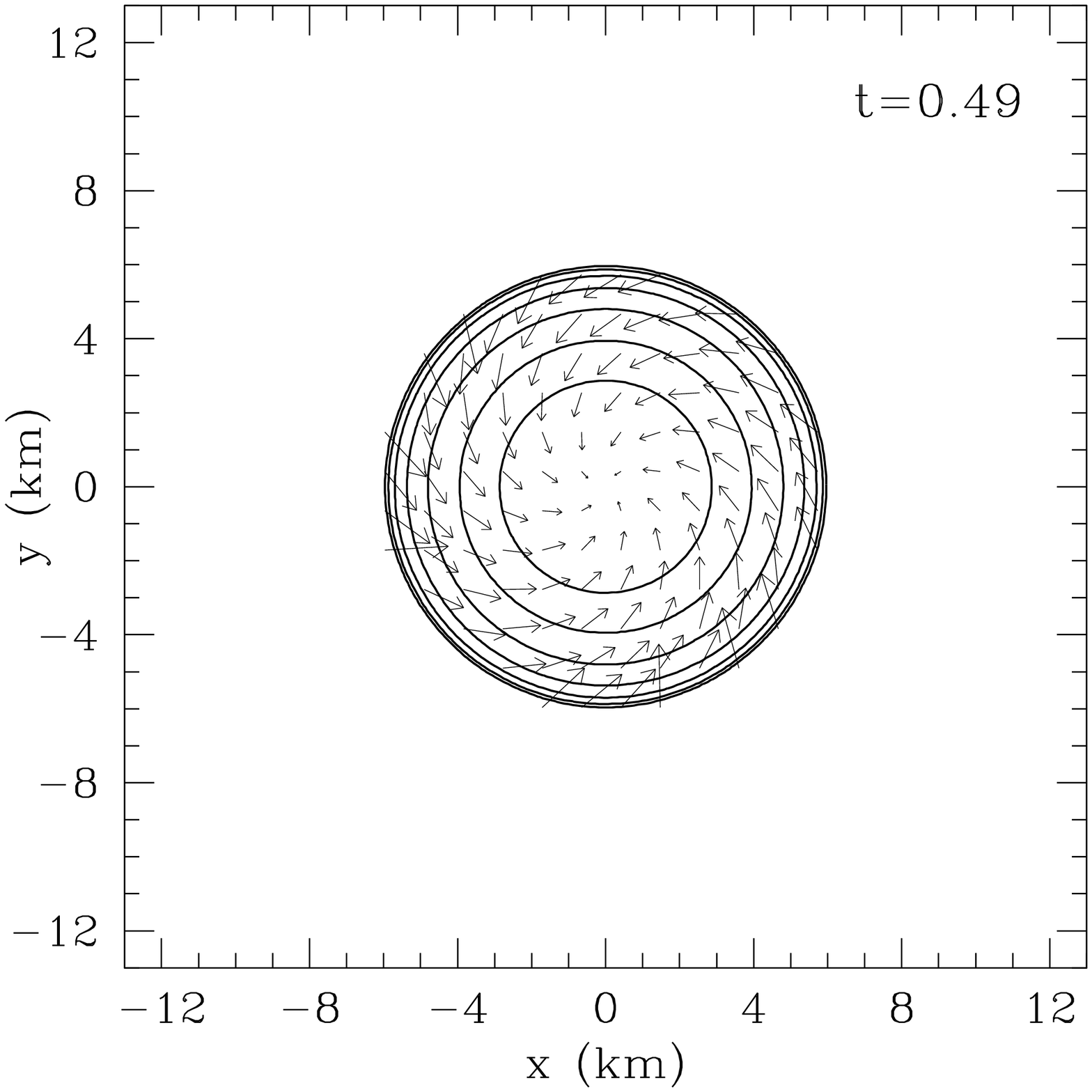} 
\hskip 1.0cm
\includegraphics[angle=0,width=6.2cm]{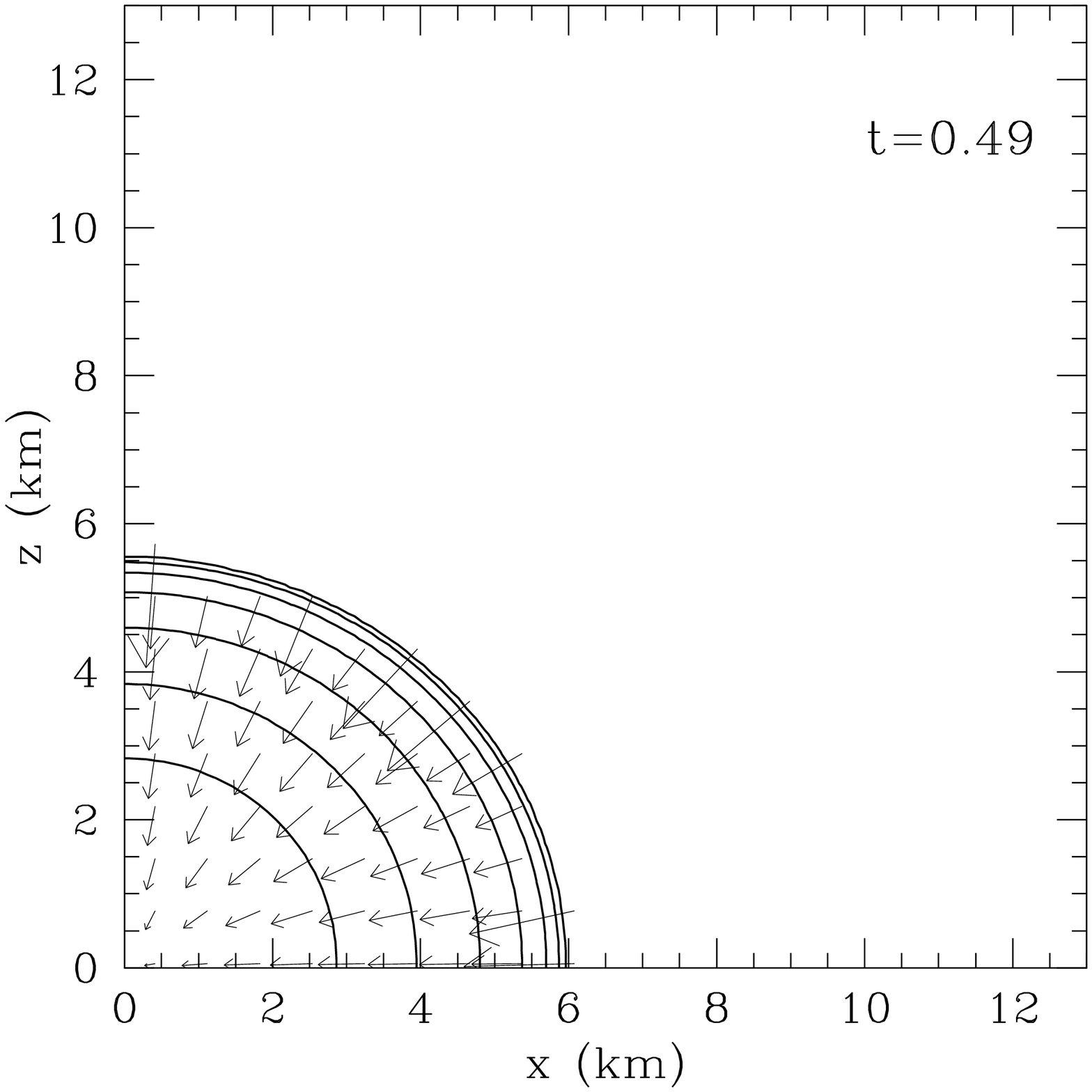} 
\vskip 0.25cm
\includegraphics[angle=0,width=6.2cm]{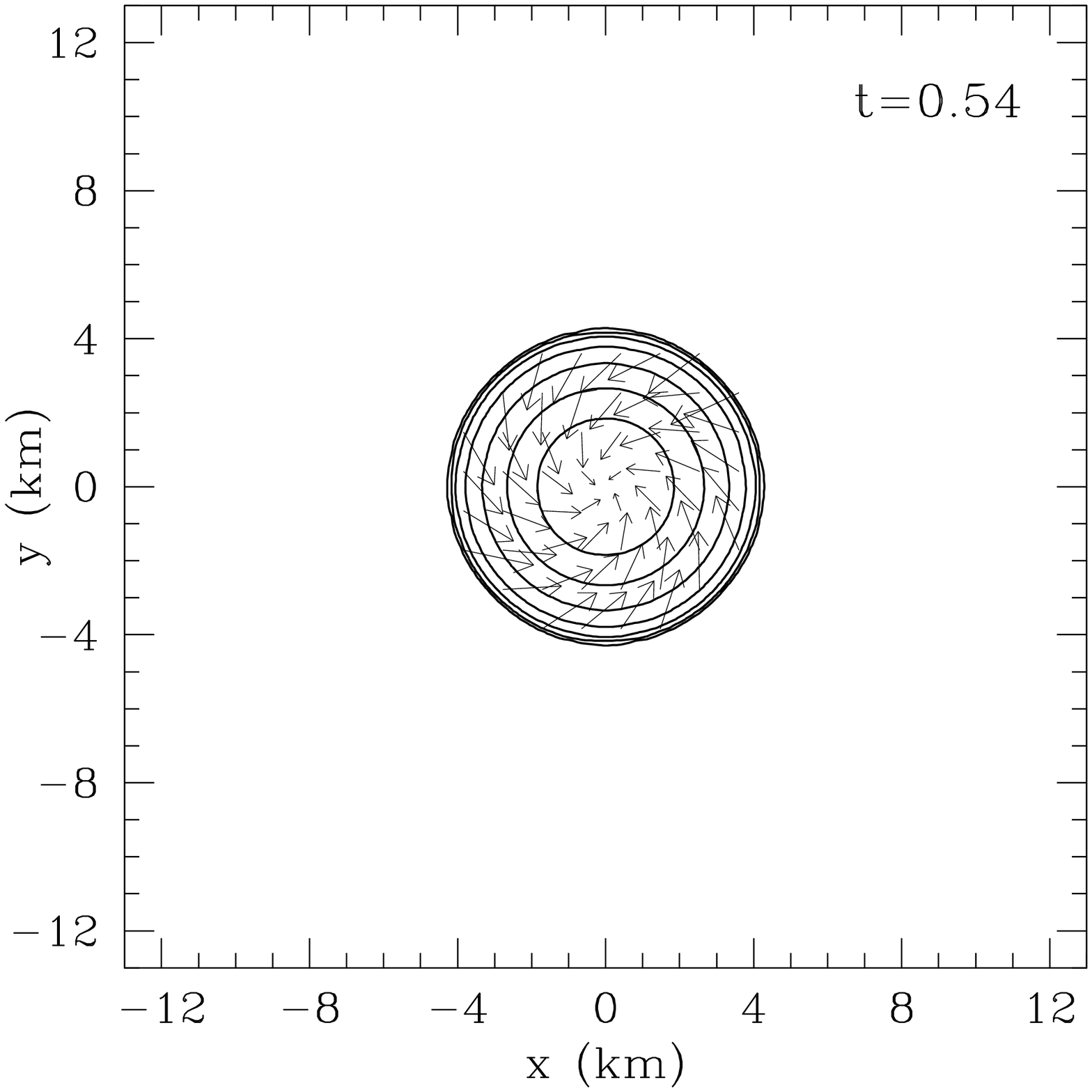} 
\hskip 1.0cm
\includegraphics[angle=0,width=6.2cm]{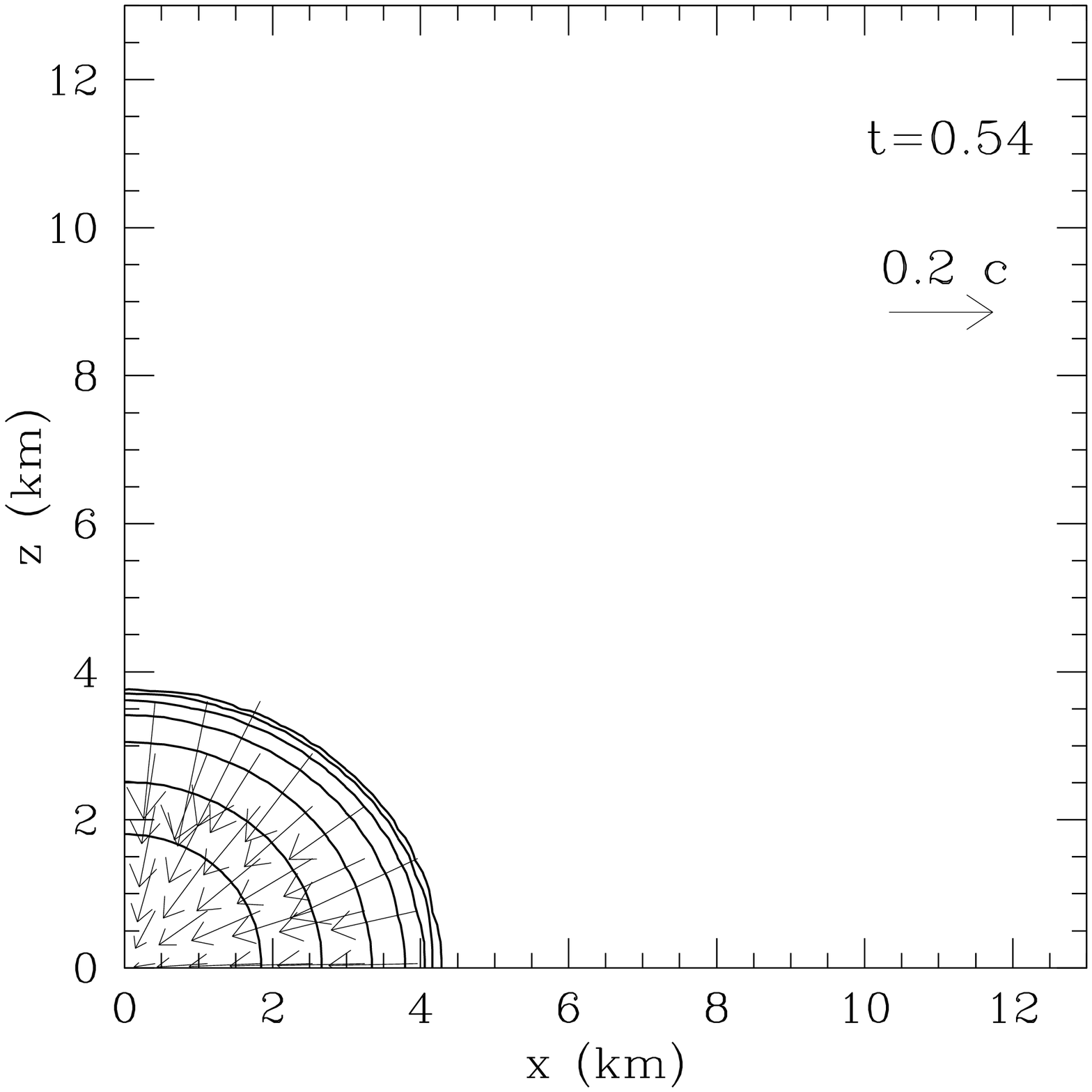} 
\caption{Collapse sequence for the slowly rotating model D1. Different
        panels refer to different snapshots during the collapse and show
        the isocontours of the rest-mass density and velocity field in
        the $(x,y)$ plane (left column) and in the $(x,z)$ plane (right
        column), respectively. The isobaric surfaces are logarithmically
        spaced and a reference length for the vector field is shown in the
        lower right panel for a velocity of $0.2\ c$. The time of the
        different snapshots appears in the top right corner of each
        panel and it is given in ms, while the units on both axes are
        expressed in km. See main text for a discussion and
        Fig.~\ref{figcollseq_D4} for a comparison with the collapse of a
        rapidly rotating model.}
\label{figcollseq_D1} 
\end{figure*}

        Given an initial stellar model which is dynamically unstable,
simple round-off errors would be sufficient to produce an evolution
leading either to the gravitational collapse to a black hole or to the
migration to the stable branch of the equilibrium
configurations~\cite{Font02c} (we recall that both evolutions are
equally probable mathematically, although it is easier to imagine that
it would collapse to a black hole). In general, however, leaving the
onset of the dynamical instability to the cumulative effect of the
numerical truncation error is not a good idea, since this produces
instability-growth times that are dependent on the grid-resolution
used.

        For this reason, we induce the collapse by slightly reducing the
pressure in the initial configuration. This is done uniformly throughout
the star by using a polytropic constant for the evolution $K$ that is
smaller than the one used to calculate the initial data $K_{\rm ID}$.
The accuracy of the code is such that only very small perturbations
are sufficient to produce the collapse and we have usually adopted
\mbox{$(K_{\rm{ID}} - K)/ K_{\rm{ID}} \lesssim 2\%$}.

        After imposing the pressure reduction, the Hamiltonian and
momentum constraints are solved to enforce that the constraint violation
is at the truncation-error level. We refer to this procedure as to
solving the initial value problem (IVP), which ensures that second order
convergence holds from the start of the simulations, as shown in
Fig.~\ref{fig_ham_n2_conv} for the $L_2$ norm of the Hamiltonian
constraint. Strict second-order convergence is lost when the
excision is introduced, although the code remains convergent at a lower
rate while the norms of the Hamiltonian constraint start to grow
exponentially (this is not shown in Fig.~\ref{fig_ham_n2_conv}).  We are
presently investigating the origin of the deterioration of the
convergence rate at the time of excision, although this is somewhat
unavoidable when excising a spherical region in a Cartesian rectangular
grid in the course of the evolution.

        Details on how we solve the IVP implementing the York-Lichnerowicz conformal
transverse-traceless decomposition can be found in \cite{Cook00a}. If, on the other hand, the IVP is
not solved after the pressure change, the constraints violation increases twice as fast and
converges to second order only after an initial period of about $20\ M \sim 0.17$ ms. To assess the
validity of our procedure to trigger the collapse, we also perform the pressure change after the
evolution has started and without solving the IVP. In this case, after the system has recovered from
the perturbation, the violation of the constraints is only a few percent different from the case in
which the IVP is solved. Furthermore, other dynamical features of the collapse, such as the instant
at which the apparent horizon is first formed (see Section~\ref{sec:dynamics-horizons} for a
detailed discussion), do not vary by more than 1\%.

        The dynamics resulting from the collapse of models S1--S4 and
D1--D4 are extremely similar and no qualitative differences have been
detected. However, as one would expect, models D1--D4 collapse more
rapidly to a black hole (the formation of the apparent horizon appears
about 5\% earlier in coordinate time), are computationally less
expensive and therefore better suited for a detailed investigation. As a
result, in what follows we will restrict our discussion to the collapse
of the dynamically unstable models and distinguish the dynamics of case
D1, in Section~\ref{sec:d1}, from that of model D4, in
Section~\ref{sec:d4}.

\begin{figure*}
\centering
\includegraphics[angle=0,width=7.5cm]{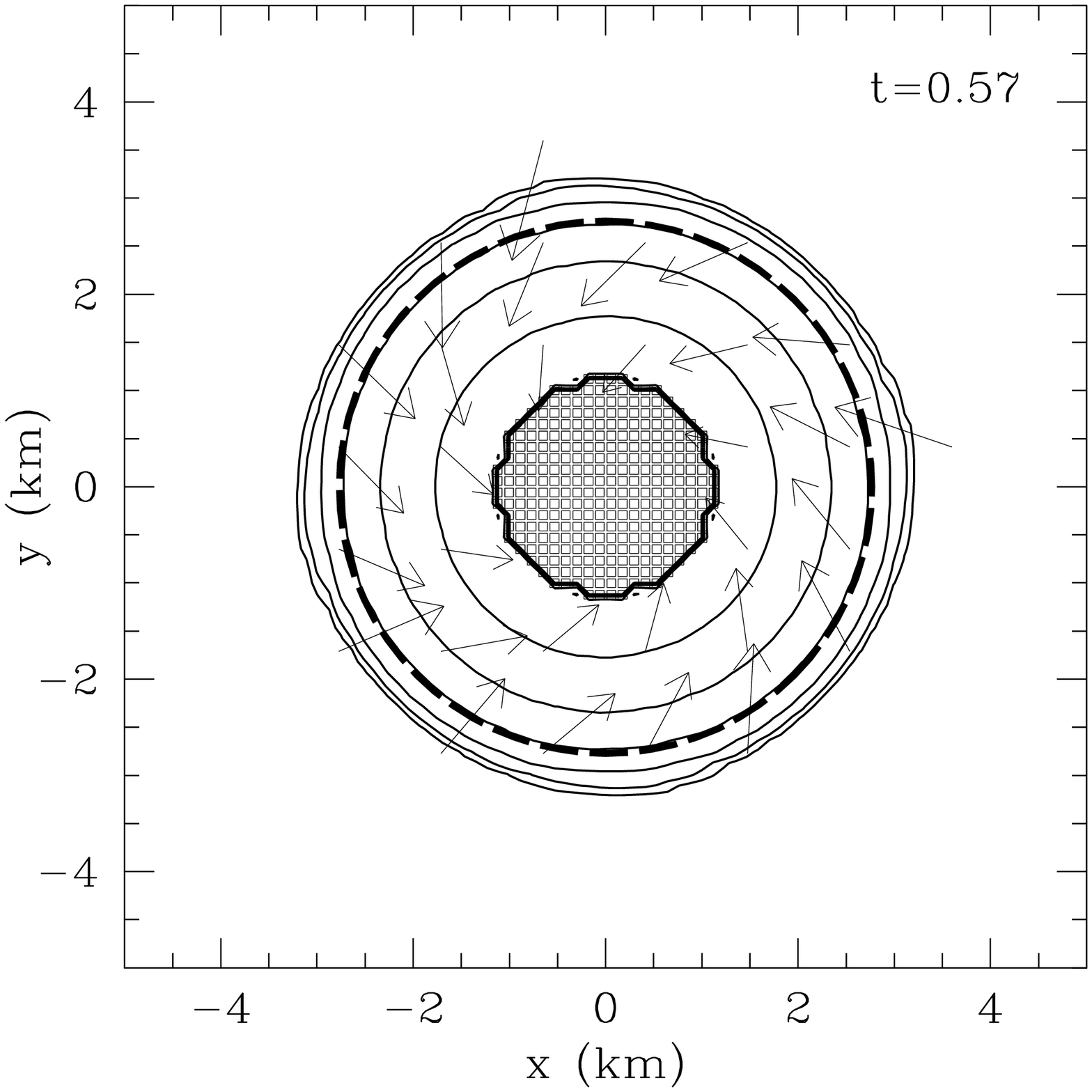} 
\hskip 1.0cm
\includegraphics[angle=0,width=7.5cm]{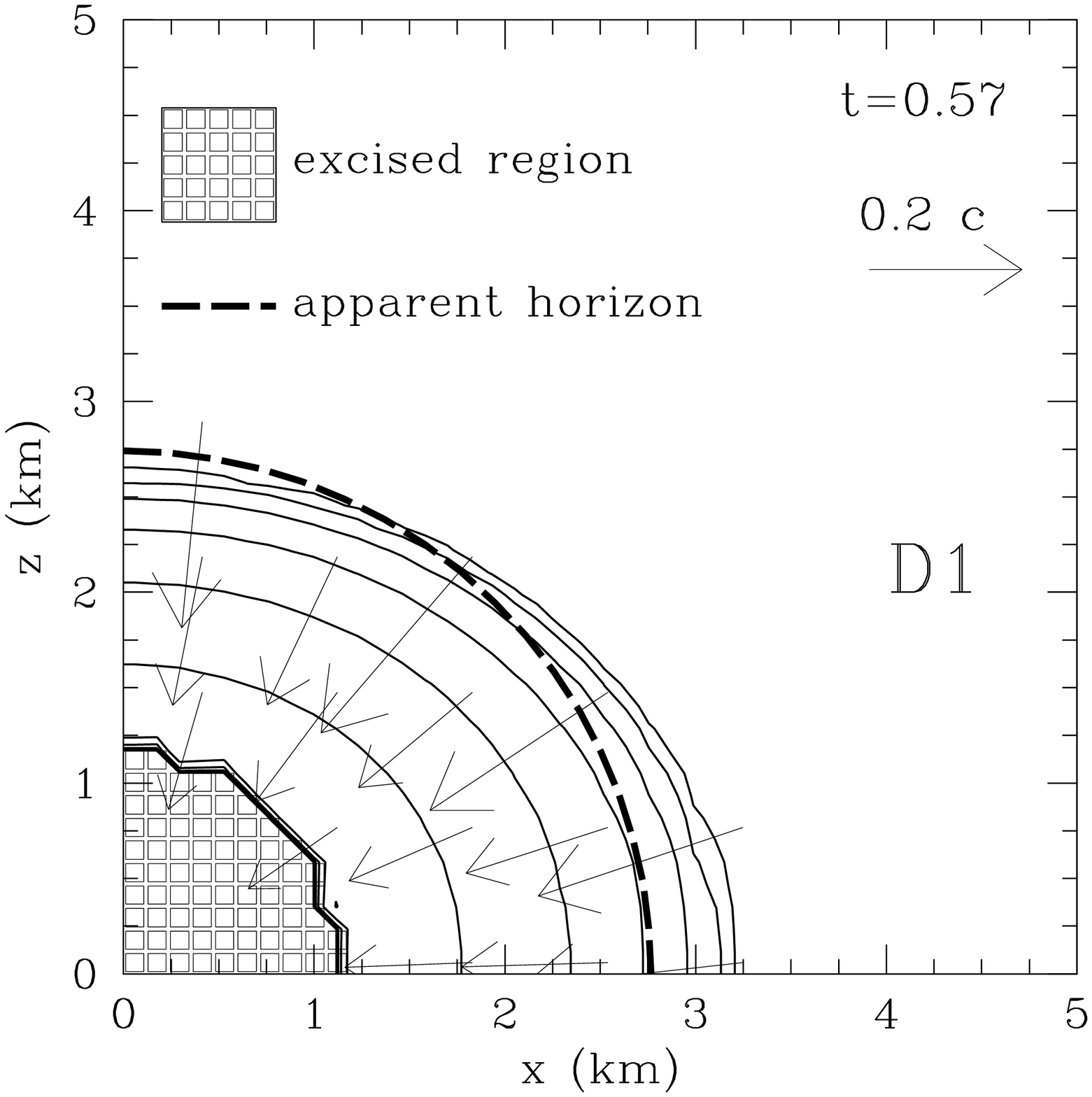} 
\caption{Magnified view of the final stages of the collapse of model D1. A region around the
        singularity that has formed is excised from the computational domain and is shown as an area
        filled with squares. Also shown with a thick dashed line is the coordinate location of the
        apparent horizon. Note that because of rotation this surface is not a two-sphere, although the
        departures are not significant and cannot be appreciated from the Figure ({\it cf.}
        Fig.~\ref{surfcs_spctm} for a clearer view).}
\label{figcollseq_D1z} 
\end{figure*}

\subsection{Slowly rotating stellar models}
\label{sec:d1}

        We start by discussing the dynamics of the matter by looking at
the evolution of the initial stellar model D1 which is slowly
rotating (thus almost spherical, with $r_p/r_e =0.95$) and has the
largest central density ({\it cf.}\ Fig.~\ref{figInitial} and
Table~\ref{tableInitial}).

        We show in Figs.~\ref{figcollseq_D1}--\ref{figcollseq_D1z} some
representative snapshots of the evolution of this initial model.  The
different panels of Fig.~\ref{figcollseq_D1}, in particular, refer to the
initial and intermediate stages of the collapse and show the isocontours
of the rest-mass density and velocity field in the $(x,y)$ plane (left
column) and in the $(x,z)$ plane (right column), respectively. The
isobaric surfaces are logarithmically spaced starting from
$\rho=2.0\times 10^{-5}$ and going up to $\rho=2.0\times 10^{-3}$ at the
stellar interior. The velocity vector field is expressed in units of $c$
and the length for a velocity of $0.2\,c$ is shown in the lower-right
panel. Panels on the same row refer to the same instant in time and this
is indicated in ms in the top-right corner of each panel.  The units on
both axes are expressed in km and refer to coordinate
lengths. This sequence has been obtained with a grid of $288^2\times 144$
zones but the data for the velocity field has been down-sampled to
produce clearer figures. The data have also been restricted to a single octant in the $(x,z)$
plane to provide a magnified view.
In all cases, however, the whole extent of the numerical grid is reported in the
figures.

        Overall, the sequence shown in Fig.~\ref{figcollseq_D1} is simple to illustrate. 
During the collapse of model D1 spherical symmetry is almost preserved;
as the star increases its compactness and the matter is compressed to
larger pressures, the velocity field acquires a radial component (which
was zero initially) that grows to relativistic values. This is clearly
shown in the panels at $t=0.49$ ms and $t=0.54$ ms, which indicate that
the star roughly preserves the ratio of its polar $r_p$ and equatorial
$r_e$ radii (see also Fig.~\ref{figRadii_ratio}), while radial velocities
in excess of $\sim 0.28\ c$ can be easily reached. The behaviour of the
angular velocity during this collapse will be analysed in more detail in
Section~\ref{sec:discf_dr}, but we can here anticipate that it does not
show appreciable departures from a profile which is uniform inside the
star.

        Soon after $t=0.54$ ms, ({\it i.e.} at $t=0.546\ {\rm ms}=66.6\
M$ in the high-resolution run), an apparent horizon is found and when
this has grown to a sufficiently large area, the portion of the
computational domain containing the singularity is excised. A discussion
of how the trapped surfaces are studied in practice will be presented in
Section~\ref{sec:dynamics-horizons}, while details on the hydrodynamical
excision have been given in Section~\ref{sec:hydr-excis}. Here, we just
remark that the use of an excised region and the removal of the
singularity from the computational domain is essential for extending the
calculations significantly past this point in time.
Figure~\ref{figcollseq_D1z} shows a magnified view of the final stages of
the collapse of model D1. Indicated as an area filled with squares is the
excised region of the computational domain, which is an approximation of
a sphere on the uniform Cartesian grid, {\it i.e.}\ a ``lego-sphere''.
Also shown with a thick dashed line is the coordinate location of the apparent horizon and it should
be remarked that, because of rotation, this surface is not a coordinate two-sphere, although the
departures are not significant and cannot be appreciated in Fig.~\ref{figcollseq_D1z} (see
Section~\ref{sec:dynamics-horizons} and Table~\ref{tab:eh_oblateness} for details). At $t=0.57$ ms,
the time which Fig.~\ref{figcollseq_D1z} refers to, most of the matter has already fallen within the
apparent horizon and has assumed an oblate shape.

        The numerical calculations were carried out up to \hbox{$t\simeq
0.73 \ {\rm ms}\sim 89\ M$}, thus using an excised region in a dynamical
spacetime for more than $26$\% of the total computing time. By this
point, all of the stellar matter has collapsed well within the event
horizon and the Hamiltonian constraint violation has become very large.

\begin{figure*}
\centering
\includegraphics[angle=0,width=6.2cm]{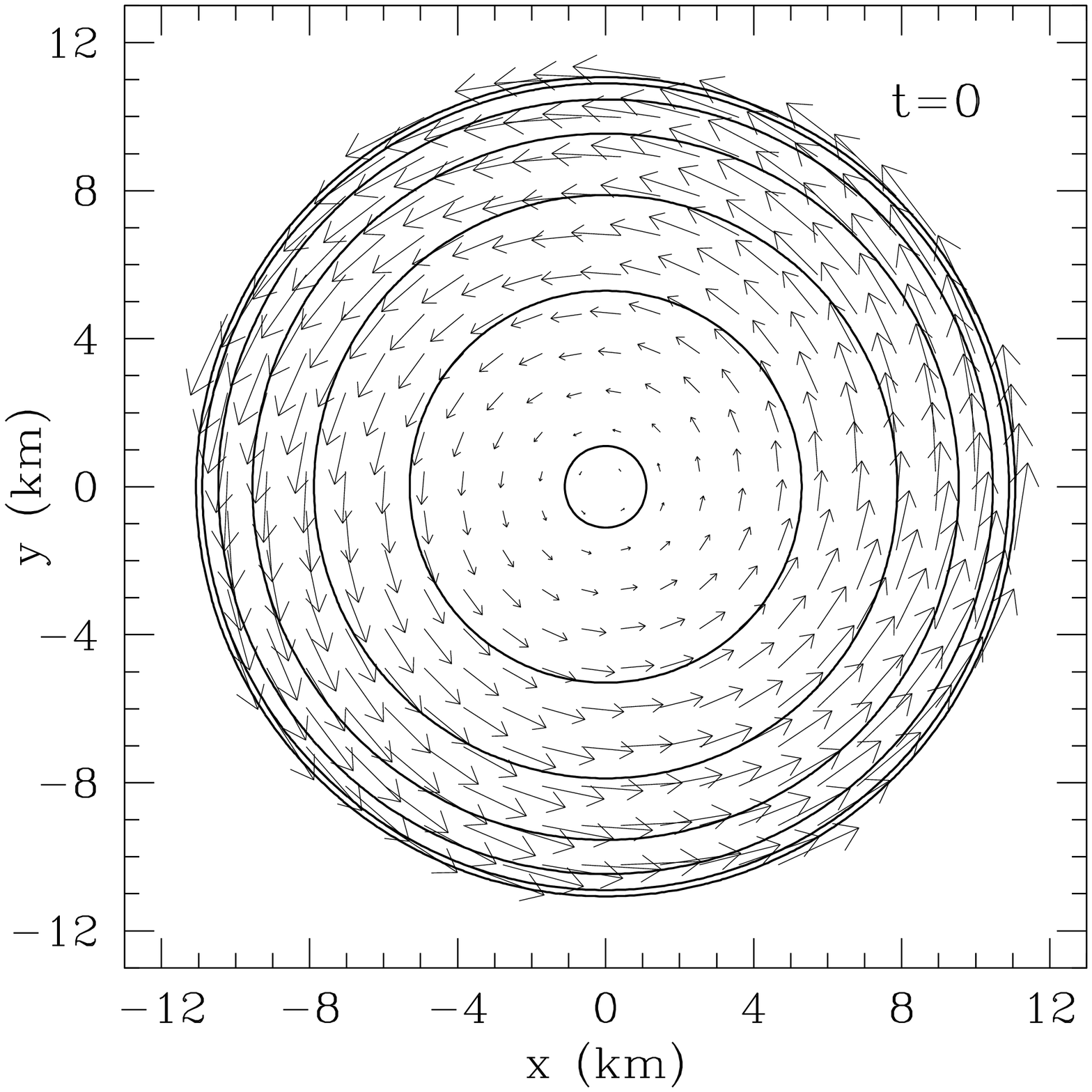} 
\hskip 1.0cm
\includegraphics[angle=0,width=6.2cm]{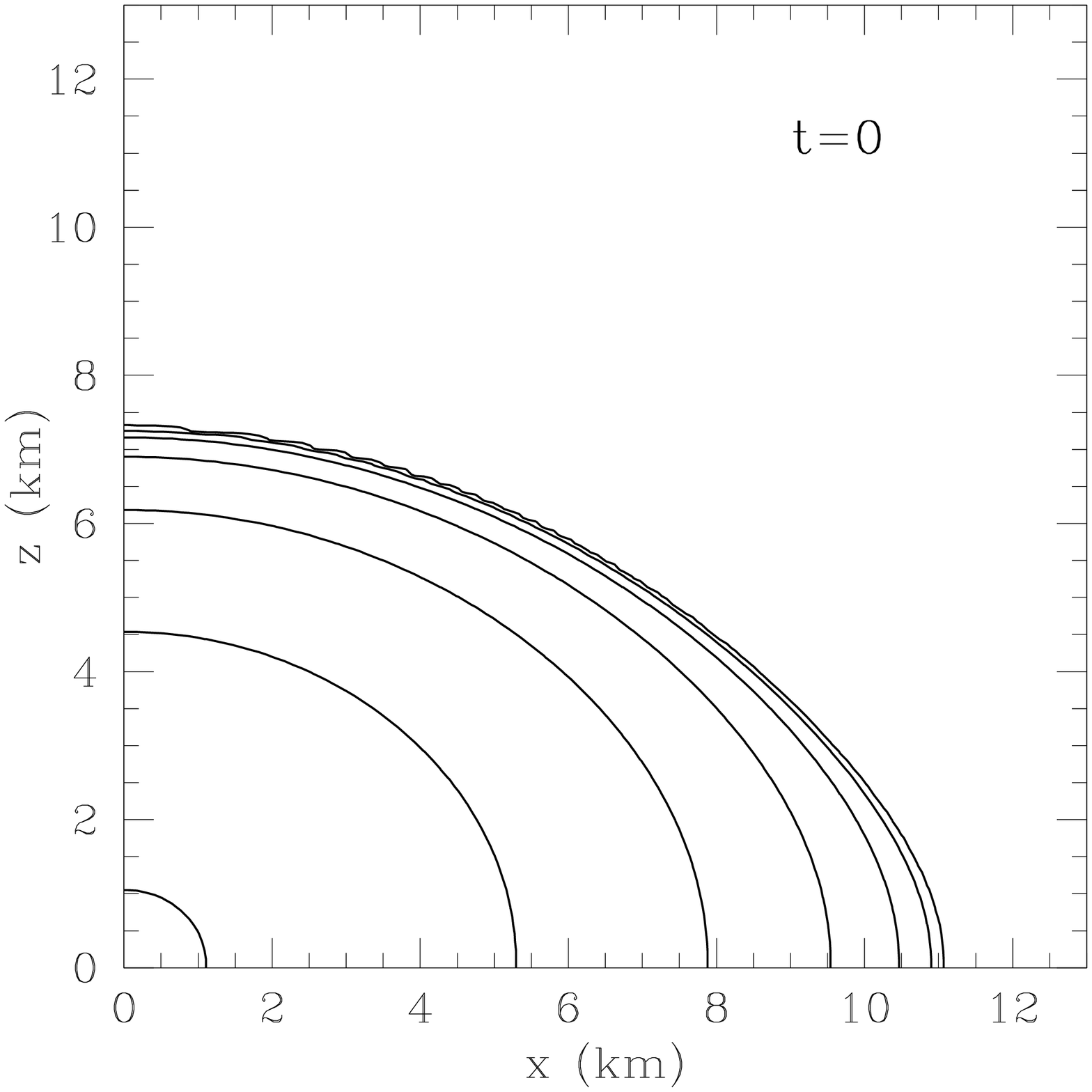} 
\vskip 0.25cm
\includegraphics[angle=0,width=6.2cm]{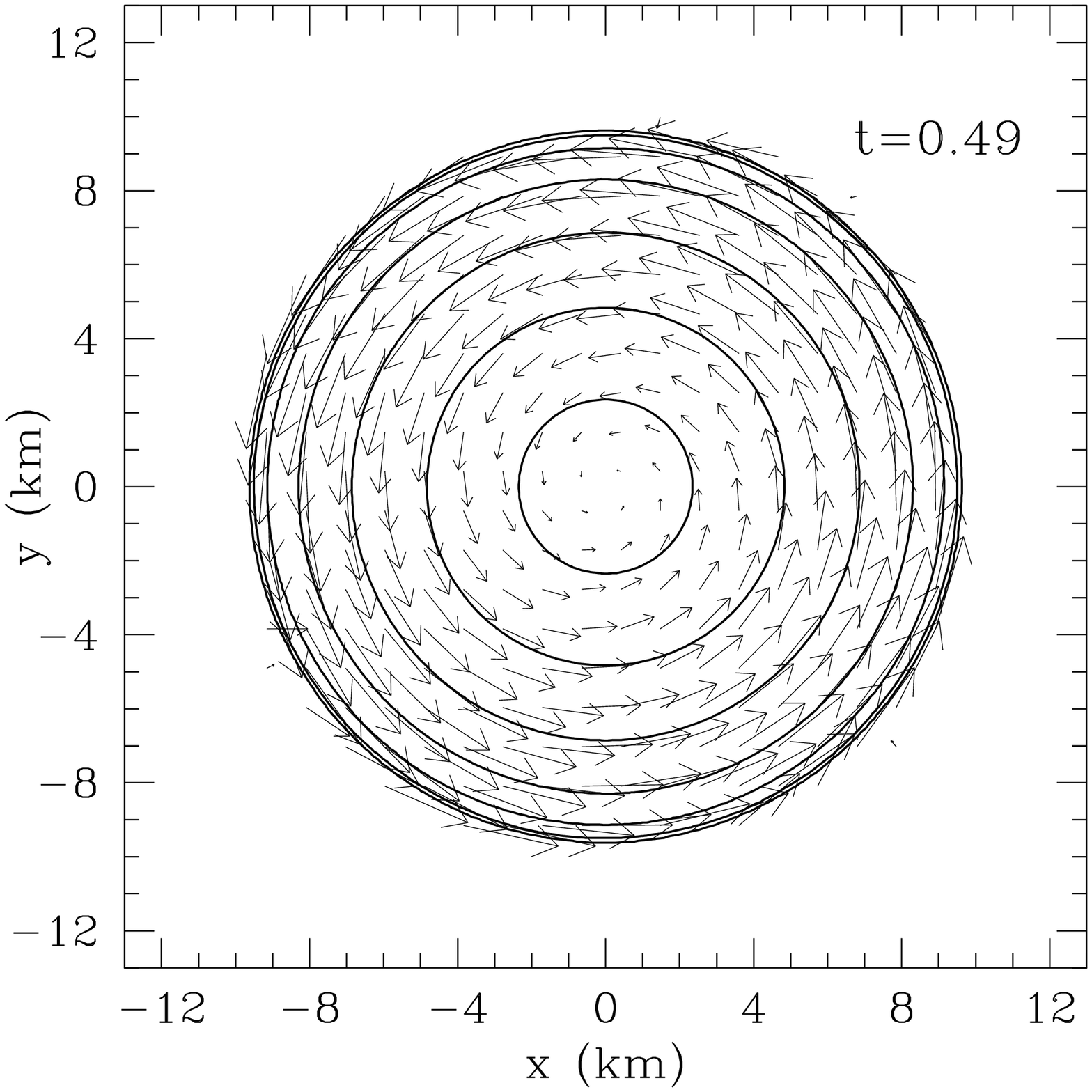} 
\hskip 1.0cm
\includegraphics[angle=0,width=6.2cm]{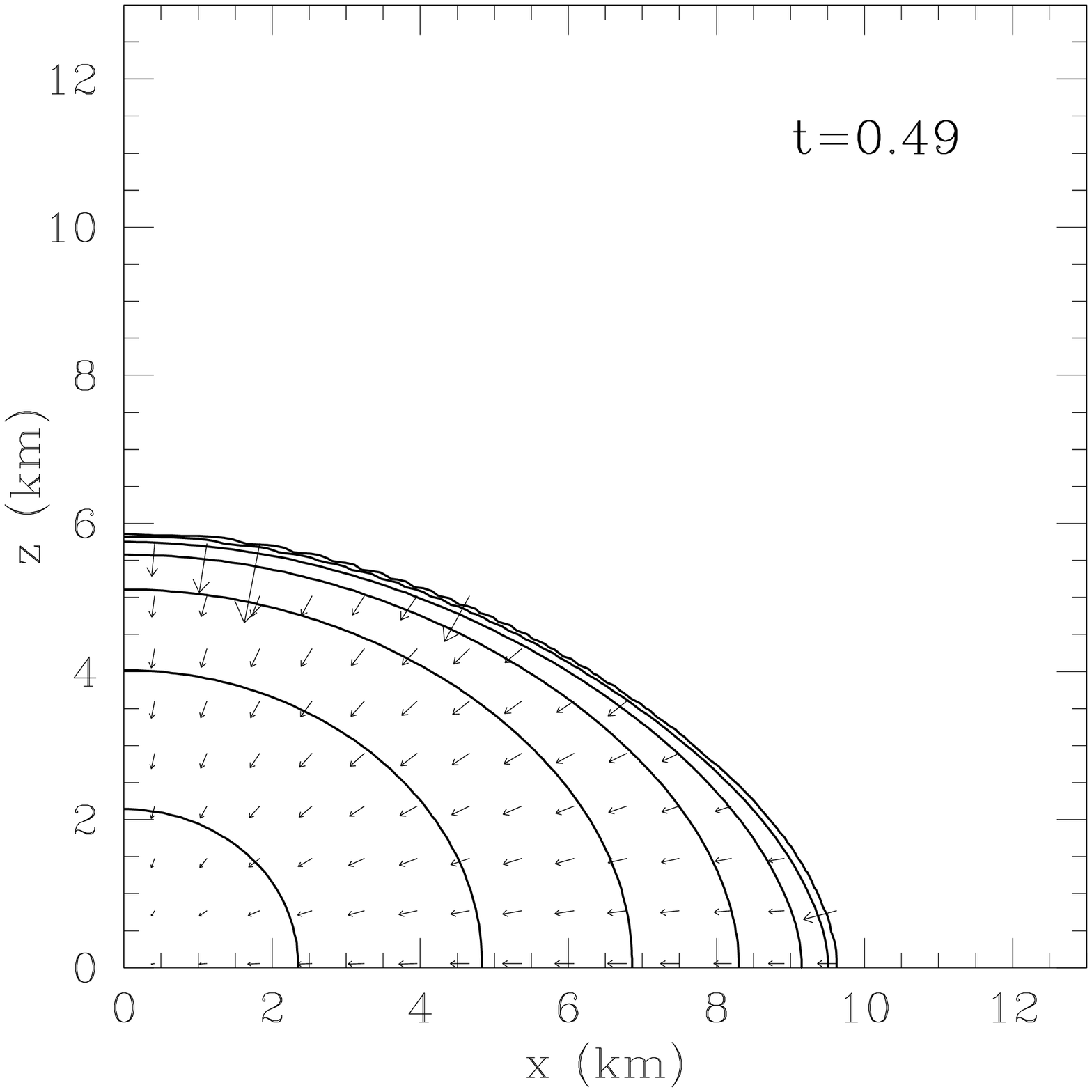} 
\vskip 0.25cm
\includegraphics[angle=0,width=6.2cm]{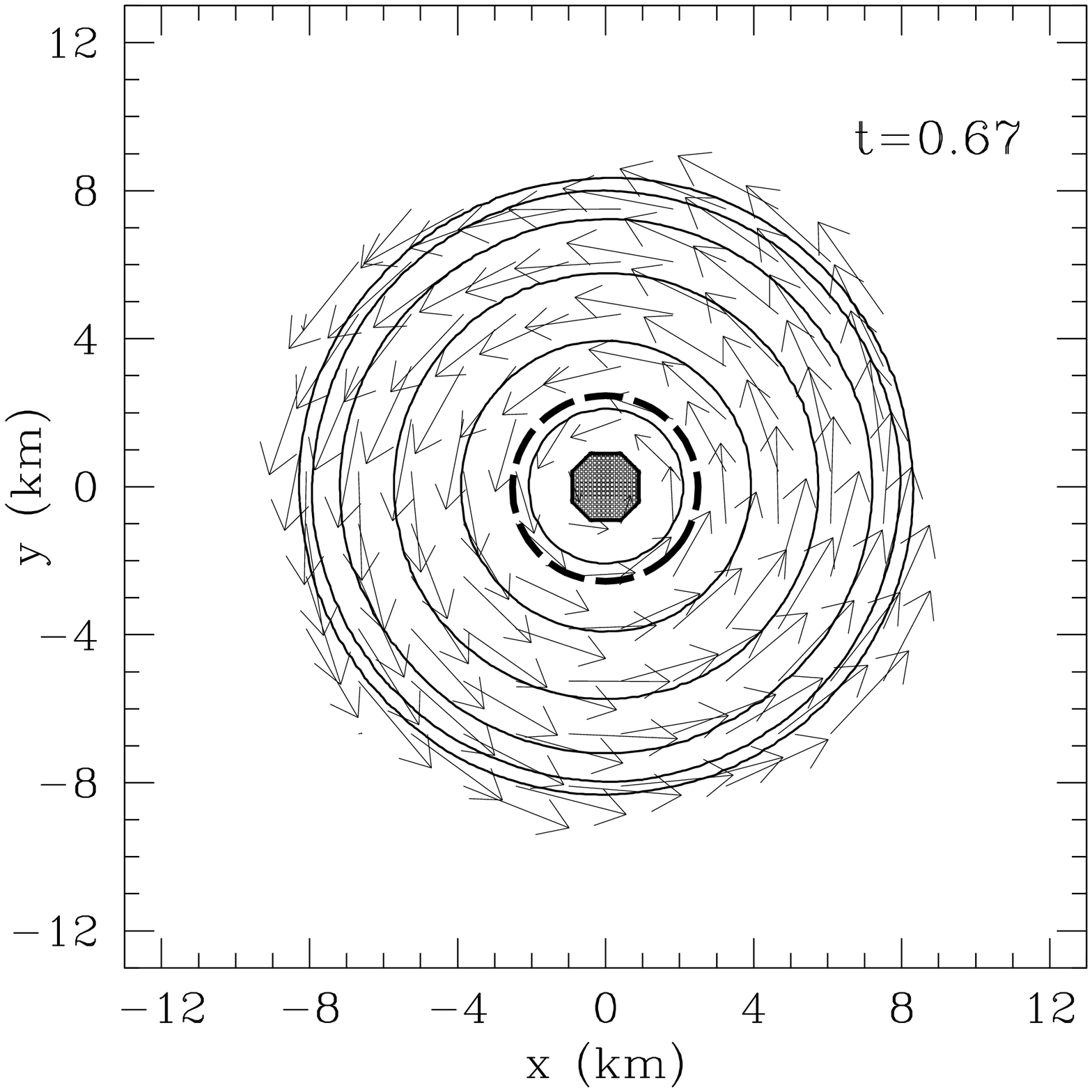} 
\hskip 1.0cm
\includegraphics[angle=0,width=6.2cm]{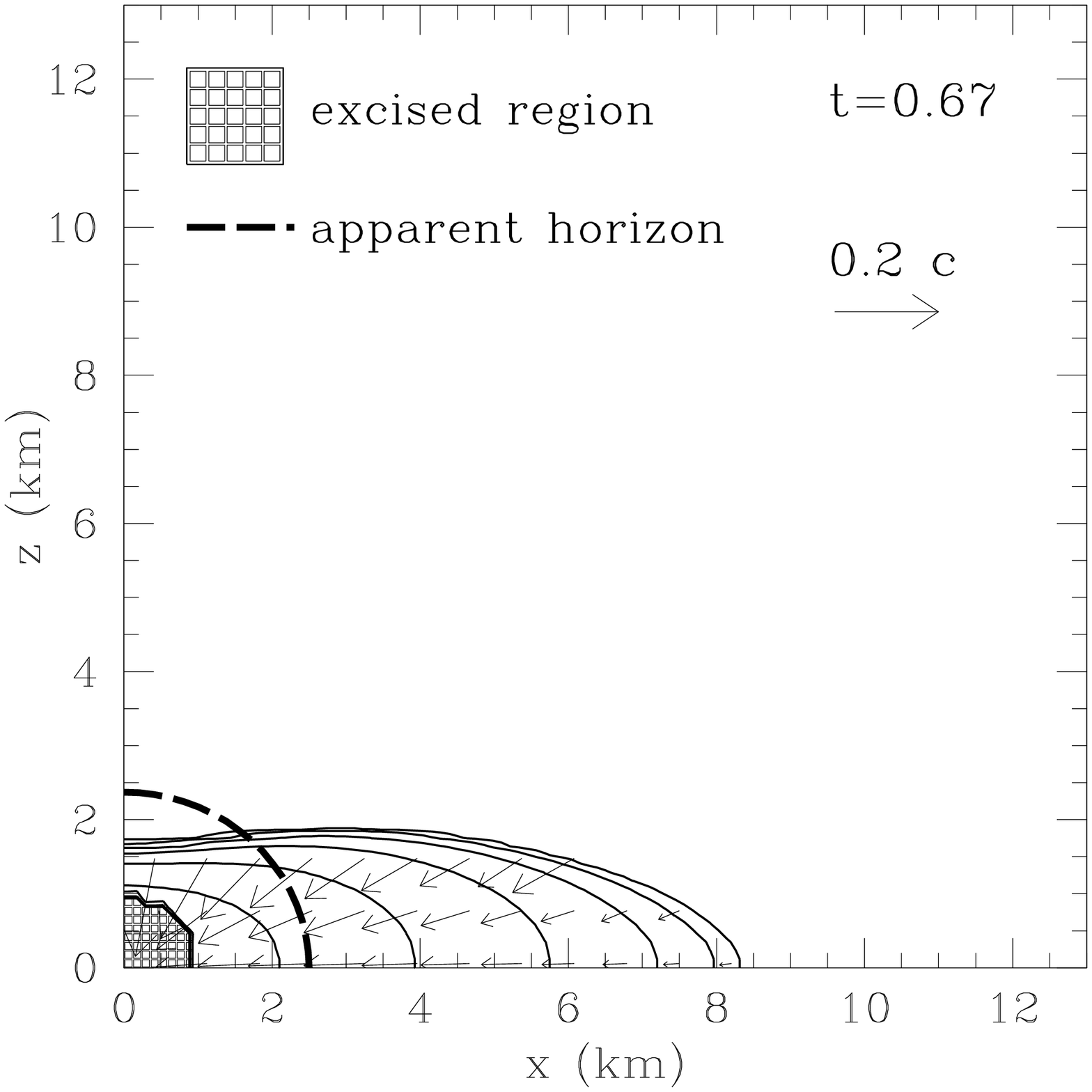} 
\caption{Collapse sequence for the rapidly rotating model D4. The conventions used in these panels
        are the same as in Fig.~\ref{figcollseq_D1}, which can be used for a comparison with the
        collapse of a slowly rotating model. Note that a region around the singularity that has formed
        is excised from the computational domain and is indicated as an area filled with
        squares. Also shown with a thick dashed line is the coordinate location of the apparent
        horizon.}
\label{figcollseq_D4} 
\end{figure*}

\begin{figure*}[htb]
\centering
\includegraphics[angle=0,width=7.5cm]{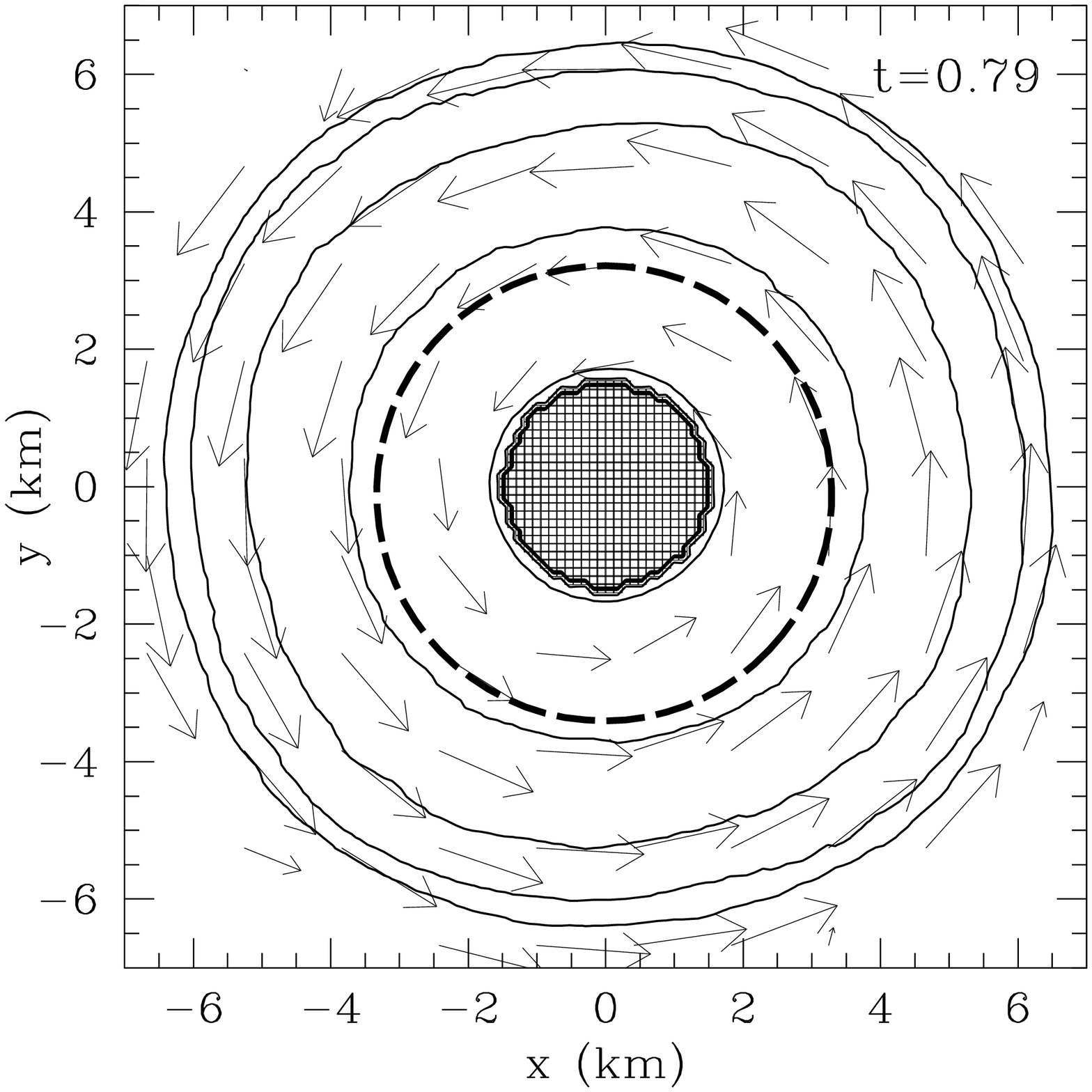} 
\hskip 1.0cm
\includegraphics[angle=0,width=7.5cm]{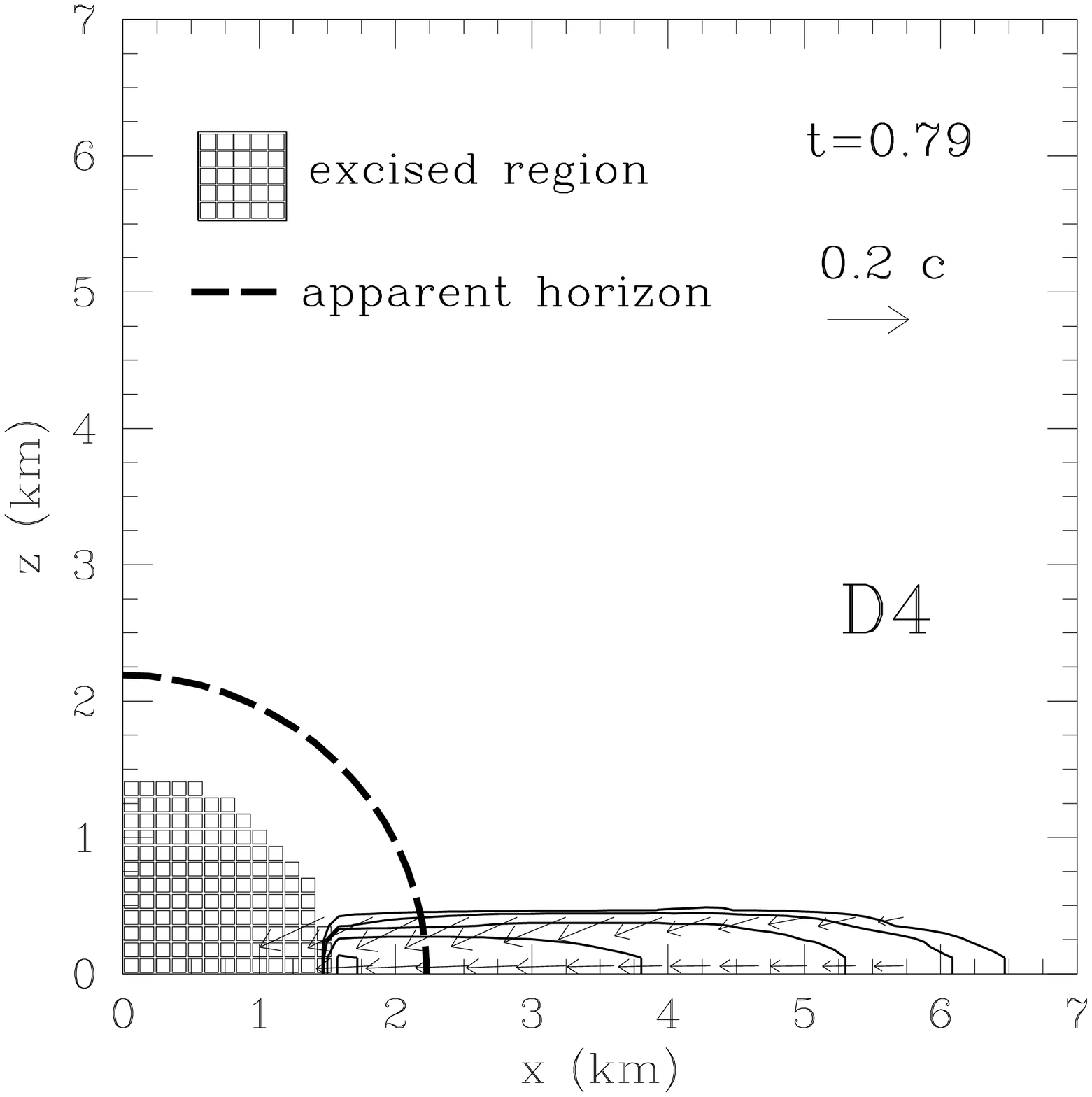} 
\caption{Magnified view of the final stages of the collapse of model D4. Note that a region around
        the singularity that has formed is excised from the computational domain and this is
        indicated as an area filled with squares. Also shown with a thick dashed line is the
        coordinate location of the apparent horizon. Note that because of the rapid rotation, this
        surface has significant departures from a two-sphere ({\it cf.}  Fig.~\ref{surfcs_spctm} for
        a clearer view).}
\label{figcollseq_D4z} 
\end{figure*}

        Overall, confirming what was already discussed by several authors
in the past, the gravitational collapse of the slowly rotating stellar
model D1 takes place in an almost spherical manner and we have found no
evidence of shock formation which could prevent the prompt collapse to a
black hole, nor appreciable deviations from axisymmetry ({\it cf.}\ left
panel of Fig.~\ref{figcollseq_D1z}). It is possible, although not likely,
that these qualitative features may be altered when a realistic EOS is
used, since in this case shocks may appear, whose heating could stall or
prevent the prompt collapse to a black hole.  However, as mentioned in
the Introduction, more dramatic changes are expected to appear if the
initial configurations are chosen to have larger initial angular momenta
and in particular when $J/M^2 \gtrsim 1$~\cite{Shibata03,Duez04}. A first
anticipation of the important corrections that centrifugal effects could
produce is presented in the following Section, where we examine the
dynamics of a rapidly rotating stellar model.

\subsection{Rapidly rotating stellar models}
\label{sec:d4}

        We next consider the dynamics of the matter during the collapse
of model D4 which, being rapidly rotating, is already rather flattened
initially ({\it i.e.}  $r_p/r_e =0.65$) and has the largest $J/M^2$
among the dynamically unstable models ({\it cf.}\ Fig.~\ref{figInitial}
and Table~\ref{tableInitial}).

        As for the slowly rotating star D1, we show in
Figs.~\ref{figcollseq_D4}--~\ref{figcollseq_D4z} some representative
snapshots of the evolution of this rapidly rotating model. The data has
been computed using the same resolution of $288^2\times 144$ zones and
the isocontour levels shown for the rest-mass density are the same used
in Fig.~\ref{figcollseq_D1}--\ref{figcollseq_D1z}. It is apparent from
the panels of Fig.~\ref{figcollseq_D4} referring to $t=0$, that model D4
is considerably more oblate than D1, as one would expect for a star
rotating at almost the mass-shedding limit.

\begin{figure}
\centering
\includegraphics[angle=0,width=8.5cm]{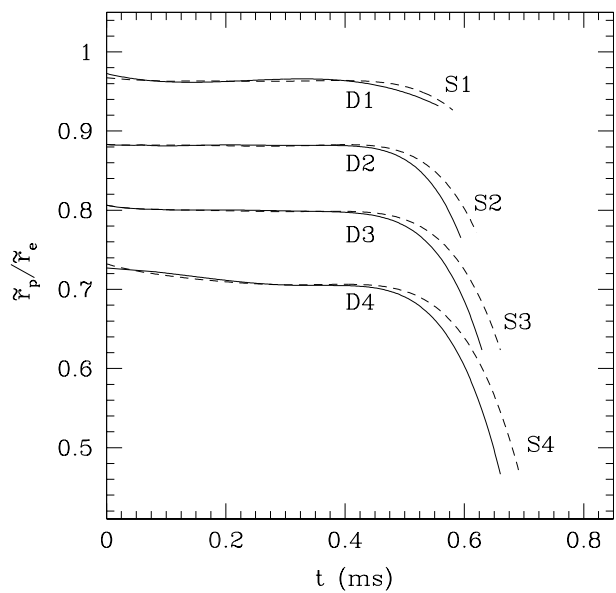} 
\caption{Ratio of the proper polar radius to the proper equatorial radius
        for all the initial models. Each curve ends at the time when, for
        each simulation, all the matter along the $z$-axis has fallen 
        below the apparent horizon.}
\label{figRadii_ratio} 
\end{figure}

        Also in this case, we believe that the sequence in Fig.~\ref{figcollseq_D4} is simple to
illustrate. Note, in particular, how the dynamics is very similar to the one discussed for model D1
up to a time $t\sim 0.49$ ms. However, as the collapse proceeds, significant differences between the
two models start to emerge and in the case of model D4 the large angular velocity of the progenitor
stellar model produces significant deviations from a spherical infall. Indeed, the parts of the star
around the rotation axis that are experiencing smaller centrifugal forces collapse more promptly
and, as a result, the configuration increases its oblateness.

        This is illustrated in Fig.~\ref{figRadii_ratio}, which shows the
time evolution of the ratio of the polar and equatorial proper radii for
all models in Table~\ref{tableInitial} (note that these ratios should not
be confused with those in Table~\ref{tableInitial} that refer, instead,
to coordinate radii). Each curve in Fig.~\ref{figRadii_ratio} extends
until all of the matter along the $z$-axis has fallen inside the apparent
horizon of the newly formed black hole. Clearly, in all cases the
oblateness increases as the collapse proceeds and this is much more
evident for those stellar models that are rapidly rotating initially. In
particular, for the most rapidly rotating models D4 and S4, the ratio
between polar and equatorial proper radii becomes as small as 0.45 at the
time when the matter on the rotation axis is below the apparent horizon.

        At about $t=0.64$ ms ({\it i.e.} at $t=0.649\ {\rm ms}=70.8\ M$ in the high-resolution
run), the collapse of model D4 produces an apparent horizon. Soon after this, the central regions of
the computational domain are excised, preventing the code from crashing and thus allowing for an
extended time evolution. The dynamics of the matter at this stage is shown in the lower panels of
Fig.~\ref{figcollseq_D4}, which refer to $t=0.67$ ms and where both the location of the apparent
horizon (thick dashed line) and of the effective excised region (area filled with squares) are
shown. By this time the star has flattened considerably, all of the matter near the rotation axis
has fallen inside the apparent horizon, but a disc of low-density matter remains near the equatorial
plane, orbiting at very high velocities $\gtrsim 0.2\ c$. The appearance of an effective barrier
preventing a purely radial infall of matter far from the rotational axis may be the consequence of
the larger initial angular momentum of the this collapsing matter and of the pressure wave
originating from the faster collapse along the ritational axis. Note, in fact, that the radial
velocity at the equator does not increase significantly at the stellar surface between $t\simeq
0.49$ and $t\simeq 0.67$ ms, but that it actually slightly decreases ({\it cf.}  the $(x,z)$ planes
in the mid and lower panels of Fig.~\ref{figcollseq_D4}). This is the opposite of what happens for
the radial velocity of the fluid elements in the stellar interior on the equatorial plane: it grows
also in this time interval. A more detailed discussion of this behaviour will be made in
Section~\ref{rtgs}.

        Note that the disc formed outside the apparent horizon is {\it
not} dynamically stable and, in fact, it rapidly accretes onto the newly
formed black hole. This is shown in Fig.~\ref{figcollseq_D4z}, which
offers a magnified view at a later time $t=0.79$ ms. At this stage the
disc is considerably flattened but also has large radial inward
velocities which induce it to be accreted rapidly onto the black hole. Note
that as the area of the apparent horizon increases, so does the excised
region, which is allowed to grow accordingly. This can be appreciated by
comparing the areas filled with squares in the lower panels of
Fig.~\ref{figcollseq_D4} (referring to $t=0.67$ ms) with the
corresponding ones in Fig.~\ref{figcollseq_D4z} (referring to $t=0.79$
ms).

        By a time $t=0.85$ ms, essentially all ({\it i.e.}\ more than
$99.9\%$) of the residual stellar matter has fallen within the trapped
surface of the apparent horizon and the black hole thus formed
approaches the Kerr solution (see
Section~\ref{sec:dynamics-horizons}). Note that a simple kinematic
explanation can be given for the instability of the disc formed during
this oblate collapse and comes from relating the position of the outer
edge of the disc when this first forms, with the location of the ISCO
of the newly formed Kerr black hole. Measuring accurately the mass and
spin of the black hole reveals, in fact, that the ISCO is located at
$x = 11.08$ km, which is always larger than the outer edge of the disc
({\it cf.}\ lower panels of Fig.~\ref{figcollseq_D4}).  Such behaviour
is not surprising since we are here dealing with initial models with a
moderate $J/M^2$, that collapse essentially on a dynamical timescale,
and for which pressure gradients cannot play an important role. As a
result, simple point-like particle motion in stationary spacetimes is
a sufficient approximation to the dynamics.

        Also for model D4, a more detailed discussion ({\it e.g.} of the
evolution of the distribution of angular velocity, or of the disc
rest mass) will be presented in Section~\ref{sec:discf_dr}. Here,
however, we can anticipate that when analysed more closely the rest mass
in the disc and outside the apparent horizon is effectively very small
and that the angular velocity shows appreciable departures from a uniform
profile.

\begin{figure}
\centering
\includegraphics[angle=0,width=8.5cm]{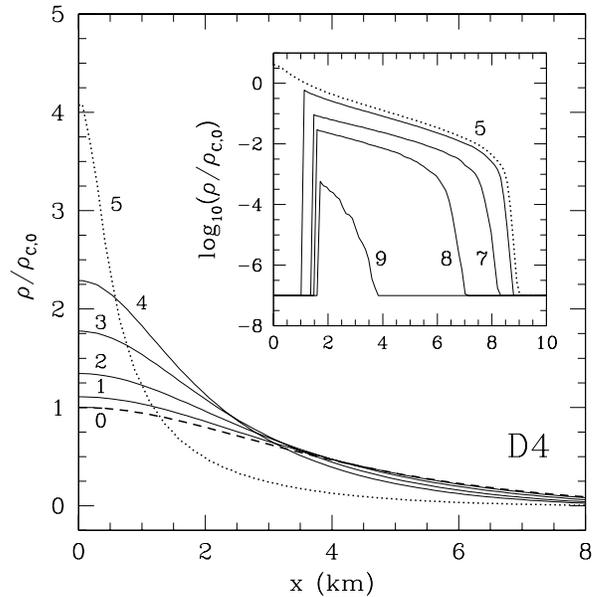} 
\caption{Rest-mass density of model D4 normalized to the initial value at
        the stellar centre. The profiles are measured along the $x$-axis
        on the equatorial plane and refer to different times (see main
        text for details). Line 5, shown as dotted, corresponds to the
        time when the apparent horizon is first found. The inset shows a
        magnified view of the final stages of the evolution using a
        logarithmic scale and also the location of the
        excised region as it grows in time.}
\label{figRho} 
\end{figure}

        A more quantitative description of the rest-mass density
evolution is presented in Fig.~\ref{figRho}, where different lines show
the profiles of the rest-mass density along the $x$-axis on the
equatorial plane. The values are normalized to the initial value at the
stellar centre, with different labels referring to different times and in
particular to $t=0.0$ (dashed line), $0.25, 0.40, 0.49, 0.54, 0.65, 0.67,
0.74, 0.79$ and $0.89$ ms, respectively. Line 5, furthermore, is shown as
dotted and refers to the time when the apparent horizon is first
formed. After this time, the excised region is cut from the computational
domain as shown in the inset of Fig.~\ref{figRho}, which illustrates the
final stages of the evolution. Note that as the matter falls into the
black hole, the apparent horizon increases its radius and thus the
location of the excised region moves outside. This is clearly shown in
the inset. Note also that the rest-mass density does not drop to zero
outside the stellar matter but is levelled off to the uniform value of the
atmosphere, whose rest-mass density is seven orders of magnitude smaller
than the initial central density. 

\begin{figure*}
\centering
\includegraphics[angle=0,width=8.0cm]{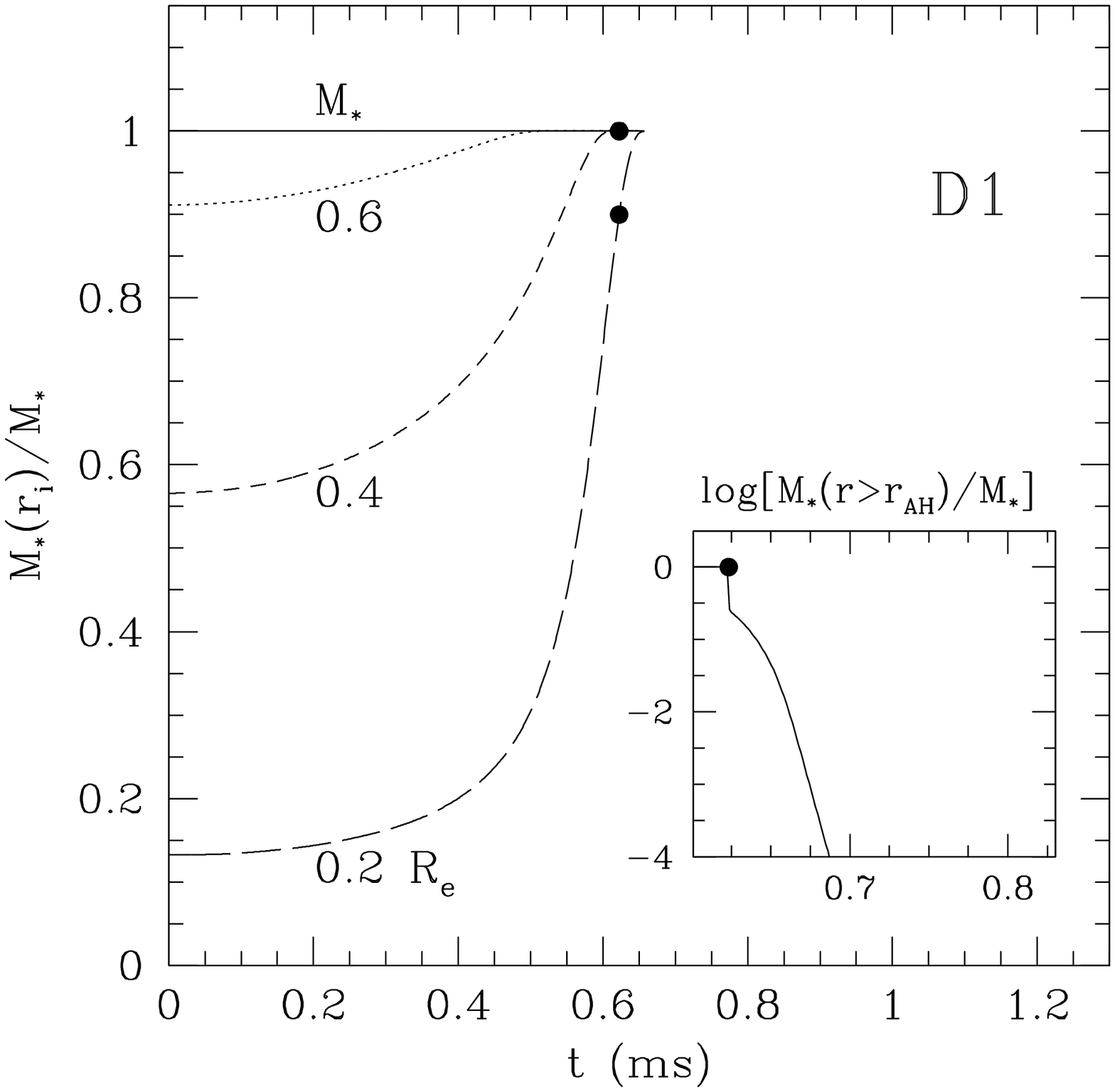}
\hskip 1.0 cm
\includegraphics[angle=0,width=8.0cm]{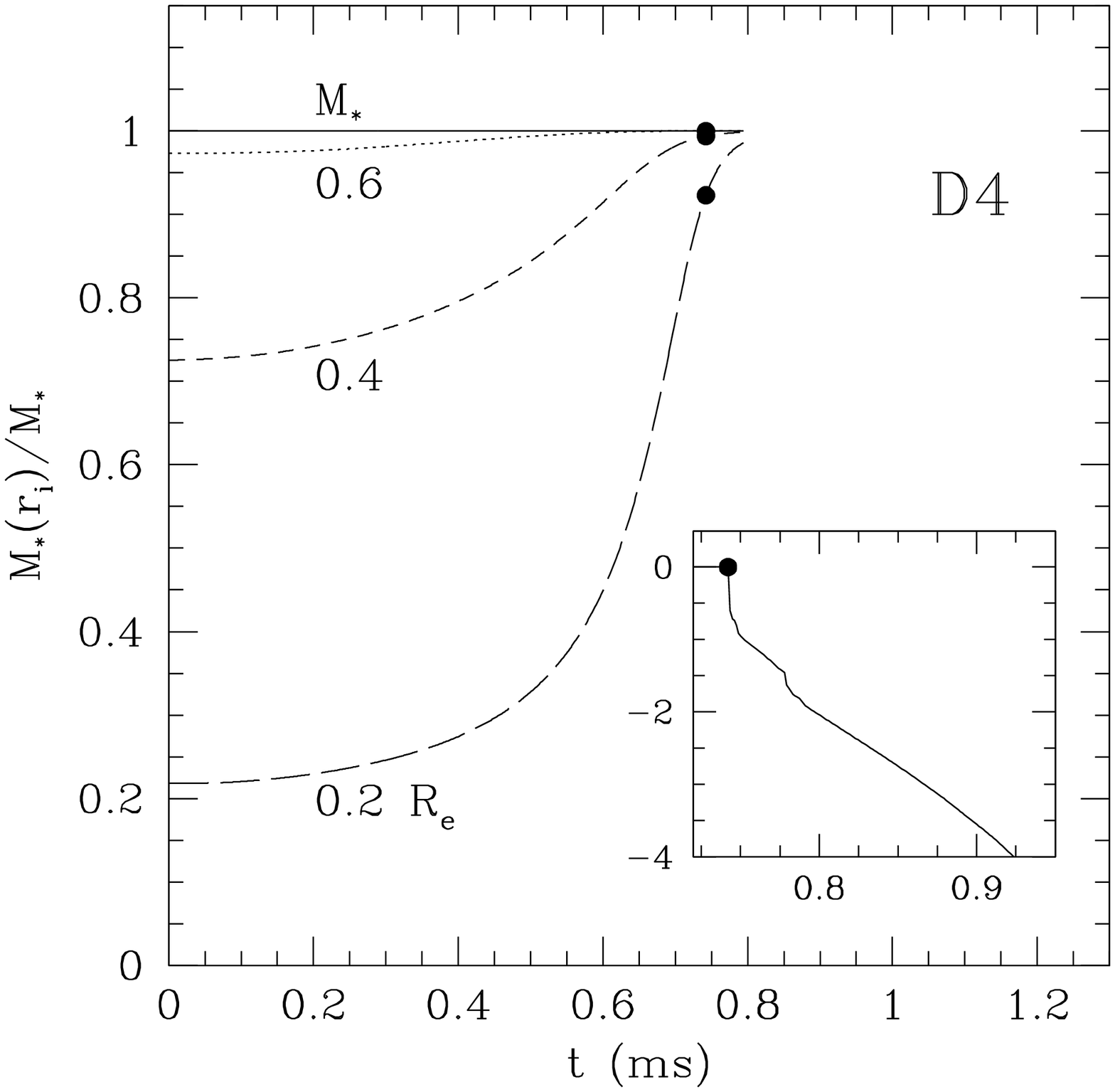}       
\caption{Evolution of the mass fraction versus time during the collapse
        of D1 (left panel) and D4 (right panel). The rest mass is
        measured within two-spheres of coordinate radii $r_i = 0.2, 0.4$
        and $0.6$ times the initial stellar equatorial circumferential
        radius $R_e$ ({\it cf.}\ Table~\ref{tableInitial}). Marked with
        filled dots on the different lines are the times at which the
        apparent horizon is first found (the data refers to a simulation
        with $96^2\times 48$ gridzones). The insets in both panels show,
        on a logarithmic scale, the evolution of the normalized rest mass
        outside the apparent horizon. Note that this is appreciably non-zero
        for a rather long time in case of model D4.}
\label{rmr}
\end{figure*}

        It should be remarked that such a tenuous atmosphere has no
dynamical impact and does not produce any increase of the mass of the
black hole that can be appreciated in our simulations. With such
rest-mass densities, in fact, it would take a time $\sim 10^4 M$ to
produce a net increase of $\sim 1\%$ in the black-hole mass. Clearly,
these systematic errors are well below the truncation errors, even at the
highest resolutions.

        The simulation ends at $t=0.91\ {\rm ms}\ \sim 99\ M$, when
the rest-mass density is everywhere at the atmosphere level and the
violations of the Hamiltonian constraint are large. By this time the
evolution has been carried for more than $28\%$ of the total time
using a singularity excising region. Also in this case, we do not find
evidence of shock formation nor of significant deviations from
axisymmetry.

        As mentioned in the Introduction, all simulations to-date agree
that no massive and stable discs form for initial models of neutron stars
that are uniformly rotating and when a polytropic EOS with $1 \le N \le
2$ is used. Our results corroborate this view and in turn imply that the
collapse of a rapidly rotating old and cold neutron star cannot lead to
the formation of the central engine believed to operate in a gamma-ray
burst, namely a rotating black hole surrounded by a
centrifugally-supported, self-gravitating torus. Relativistic simulations
with more appropriate initial data, accounting in particular for the
extended envelope of the massive progenitor star which is essential in the
so-called collapsar model of gamma-ray bursts~\cite{Woosley00}, will be
necessary to shed light on the mechanism responsible for such events.

        Convincing evidence has recently emerged~\cite{Duez04} that a
massive disc can be produced if the stellar models are initially rotating
differentially and with initial total angular momenta $J/M^2 \gtrsim 1$,
as it may happen for young and hot neutron stars. In this case, the
massive disc could emit intense gravitational radiation either through
its oscillations~\cite{Zanotti02} or as a result of the fragmentation
produced by non-axisymmetric instabilities~\cite{Duez04}.  We are
presently investigating this possibility and the results of our
investigation will be reported in a forthcoming paper.

\subsection{Disc formation and differential rotation}
\label{sec:discf_dr}

        We now discuss in more detail two interesting properties of the
matter dynamics in both slowly and rapidly rotating models: the evolution
of the rest mass outside the apparent horizon and the development of
differential rotation during the collapse.  

        In order to monitor the changes of the rest-mass distribution
during the collapse we define the rest mass within a two-sphere of
coordinate radius $r_i < R_e$ as (see, for instance, \cite{Shibata99c})
\begin{equation}
\label{mratio}
M_*(r_i) = \int_{r^{'}<r_i} \rho \alpha u^0 \sqrt{\gamma}
        \ {\rm d}^3{\bf x}^{'} \ , 
\end{equation}
where ${\rm d}^3{\bf x}^{'}$ is the 3D coordinate
volume element.  Shown in Fig.~\ref{rmr} is the evolution of the
rest masses measured within several representative two-spheres for
models D1 (left panel) and D4 (right panel), respectively. Different
lines refer to different coordinate radii for the two-spheres ({\it
i.e.} $r_i = 0.2, 0.4$ and $0.6\ R_e$, where $R_e$ is the initial
equatorial circumferential radius) and are normalized to the total
rest mass within the computational domain $M_*$, shown as a solid
line. Marked instead with filled dots are the values of $M_*(r_i)$ at
the times when the apparent horizon is first found; for simplicity,
the data shown in Fig.~\ref{rmr} refer to a simulation with
$96^2\times 48$ gridzones, but for this quantities higher resolutions
have just the effect of shifting the time at which the apparent
horizon is first found.

\begin{figure*}
\centering
\includegraphics[angle=0,width=7.5cm]{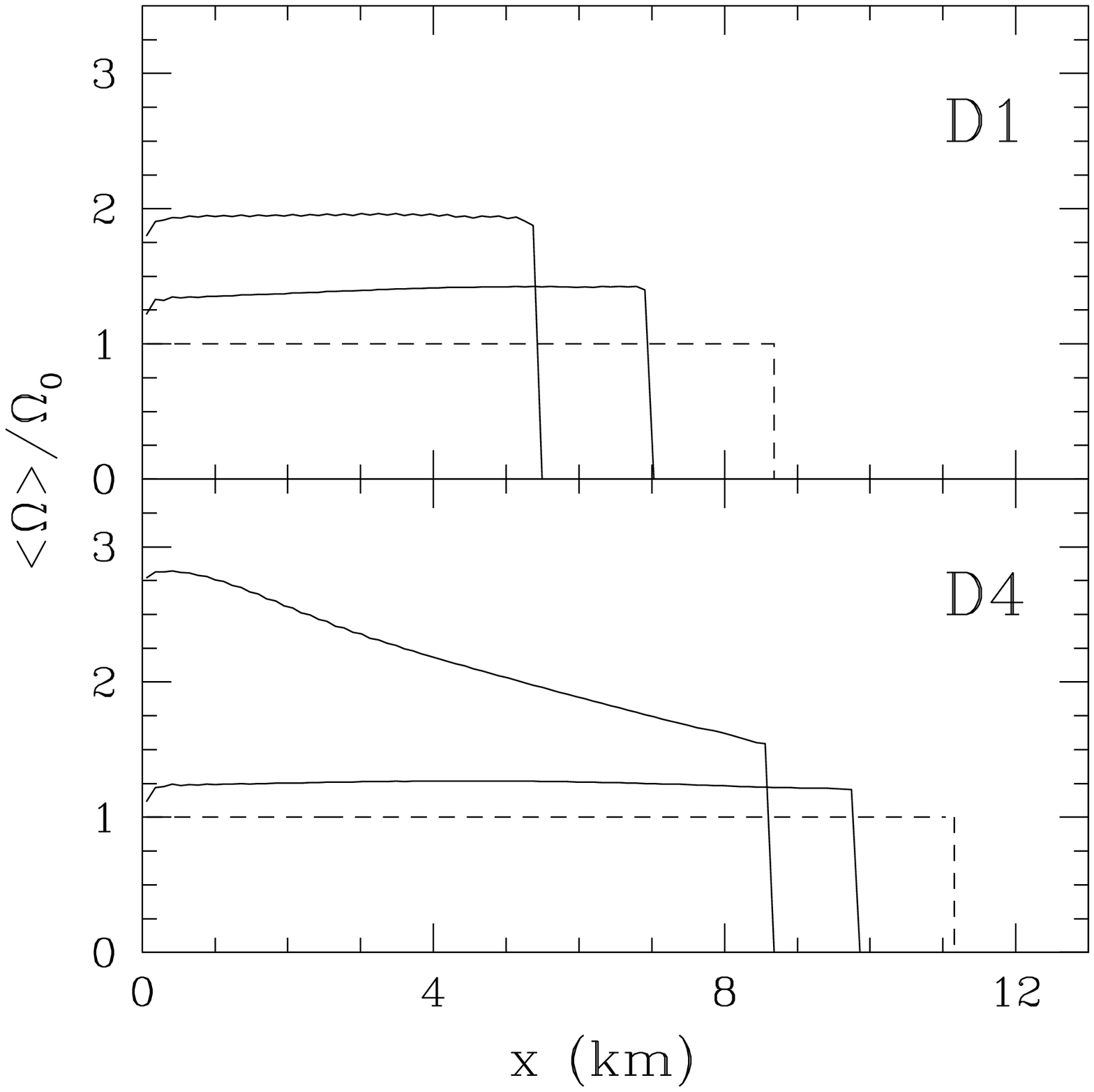} 
\hskip 1.5cm
\includegraphics[angle=0,width=7.5cm]{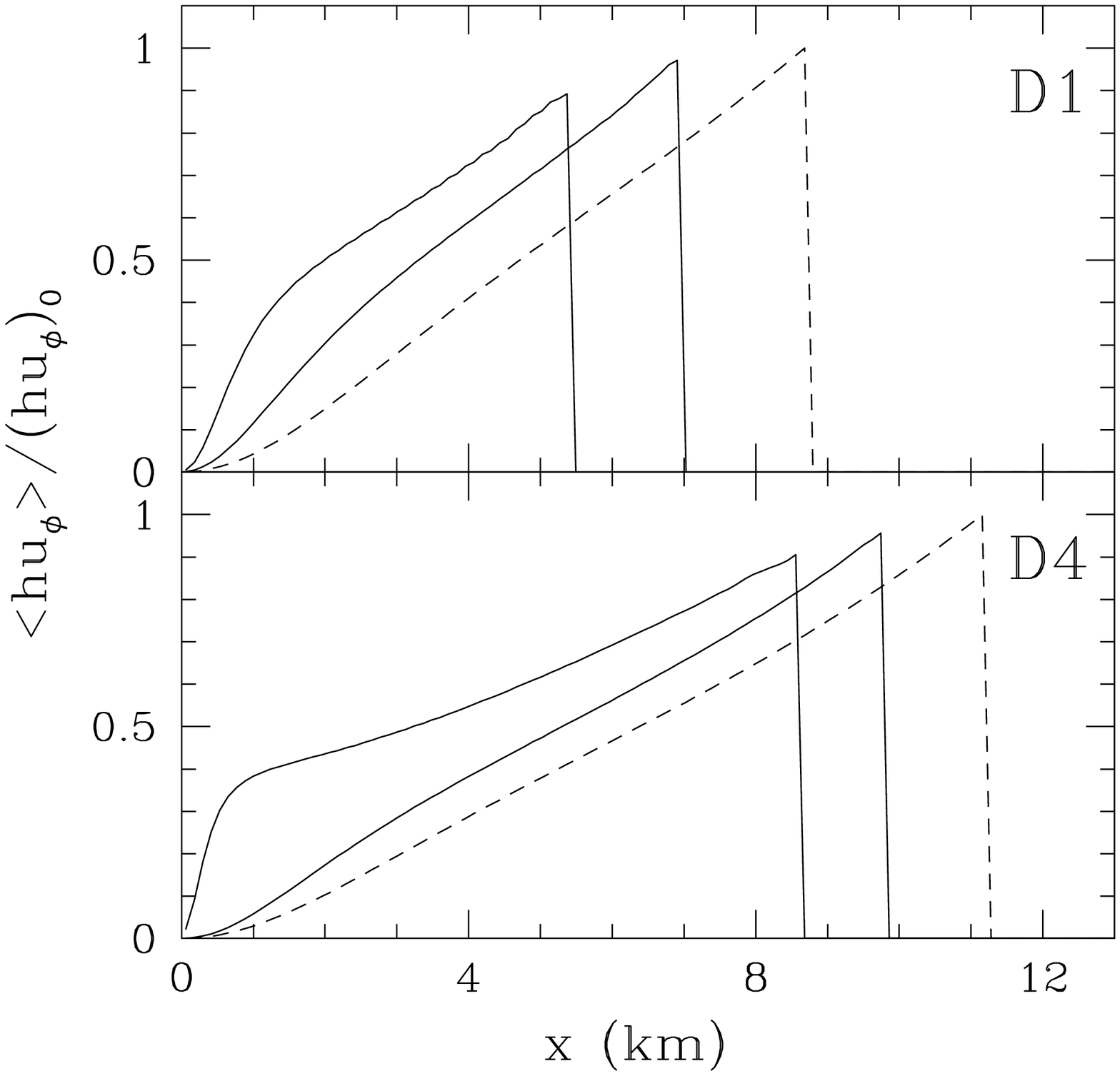} 
\caption{Evolution of the averaged angular velocity $\langle \Omega
        \rangle$ (left panel) and of the averaged angular momentum per
        unit mass $\langle h u_{\phi} \rangle$ (right panel). Both
        quantities are measured at the stellar equator, are normalized to
        the initial value at the stellar surface and refer to both
        models D1 (upper parts) and D4 (lower parts).}
\label{diff_rot} 
\end{figure*}

        As mentioned before, the excised region is not introduced
immediately after the apparent horizon has been found, but only when this
has grown to a sufficiently large size.  When this happens, the inner part of the
computational domain is removed and the integrals (\ref{mratio}) are no
longer meaningful. As a result, all of the curves in Fig.~\ref{rmr} are
truncated at the time when the excision region is first introduced, which
occurs at $t = 0.72$ ms and $t = 0.79$ ms for models D1 and D4,
respectively.

        A rapid comparison between the two panels of Fig.~\ref{rmr} is
sufficient to identify the differences in the rest-mass evolution in
slowly and rapidly rotating models. Firstly, the rest-mass
distribution is very different already initially, being more uniform
in D1 and more centrally concentrated in D4, as can be appreciated by
comparing $M_*$ at $r_i = 0.4 R_e$ and $0.6 R_e$. Secondly, the
rest-mass infall is much faster for the slowly rotating model D1,
while it is more progressive for model D4, as shown by the change in
the fractional mass ratio at $r_i=0.4 R_e$. Finally, the amount of
matter outside $r_i = 0.4 R_e$ at the time when the apparent horizon
is found, and which is very close to the amount of matter outside the
apparent horizon, is different in the two cases, being essentially
zero for model D1 and a few percent for model D4. A clearer view of
this is presented in the two insets of Fig.~\ref{rmr} which show, on a
logarithmic scale, the evolution of the normalized rest masses outside
the apparent horizons, {\it i.e.}\ $M_*(r>r_{_{\rm AH}})/M_*$, since
these first form and as they grow in time. It is interesting to note
the different behaviour in this case with a rapid decrease when the
rotation rate is small and a much slower one in the case of a rapidly
rotating progenitor (Note that the two insets cover the same timescale
although they refer to a different time interval.).

        Two additional comments are worth making. The first one is that
$M_*$ effectively includes also the rest mass in the atmosphere but this
is always $\lesssim 10^{-4}$ of the total rest mass. The second one is
that $M_*$ in Fig.~\ref{rmr} does not simply refer to the initial value
of the total rest mass but is effectively computed at each step and
appears constant in time because of the ability of the code to conserve
rest mass. A closer look at the solid curve in Fig.~\ref{rmr} reveals, in
fact, that $M_*$ varies over time to less than one part in $10^{4}$.

        An interesting question to ask at this stage is whether these
uniformly rotating models will develop any degree of differential
rotation as the collapse proceeds. Part of the interest in this comes
from the fact that neutron stars are thought to rotate differentially, at
least during the initial stages of their life. This is expected to hold
both when the neutron star is produced through a stellar core collapse,
in which case the differential rotation may be present already in the
stellar progenitor and is then amplified during collapse
\cite{Dimmelmeier02b}, and when the neutron star is the end-result of a
binary merger of neutron stars~\cite{Shibata02a}. However, as the neutron
star cools and grows older, dissipative viscous effects or the coupling
with non-turbulent magnetic fields are expected to bring the star into
uniform rotation (see \cite{Shapiro00,Shapiro03,Shapiro04} for a detailed
description of this process). It is therefore interesting to investigate
whether a degree of differential rotation will be produced also during
the final collapse of a uniformly rotating star to a Kerr black hole.  To
answer this question we have monitored both the averaged angular velocity
$\langle \Omega \rangle$, defined as
\begin{equation}
\label{omega}
\langle \Omega \rangle  \equiv \frac{1}{2}\left(
        \frac{u^\phi}{u^t}{\Bigg\vert_{x-{\rm axis}}} + 
        \frac{u^\phi}{u^t}{\Bigg\vert_{y-{\rm axis}}}  
        \right) \ ,
\end{equation}
and the corresponding averaged angular momentum per unit mass $\langle h
u_{\phi} \rangle$, which is a conserved quantity along the path lines of
fluid elements in an axisymmetric (but not necessarily stationary)
spacetime~\cite{Bardeen70}. Note that $u^{\phi}/u^t = \alpha v^{\phi} -
\beta^{\phi}$ and the average over the two different directions is here
used to compensate the small errors that are produced in the evaluation
of these quantities near the axes. 

        We note that our measure of the differential rotation will depend on the specific slicing
chosen. However, for the simulations reported here, the lengthscale of variation of the lapse
function at any given time is always larger than the stellar radius at that time, ensuring that the
events on the same timeslice are also close in proper time.  A useful measure of the differential
rotation that develops during collapse is the departure from unity of the ratio of the values of
$\Omega$ at the centre and at the surface of the star on the equatorial plane and it is instructive
to compare how this varies in the dynamics of the two models D1 and D4, which have been evolved
using the same slicing.

        The time evolution of $\langle \Omega \rangle$ and $\langle h
u_{\phi} \rangle$ is presented in the two panels Fig.~\ref{diff_rot},
whose lower parts refer to model D4 while the upper ones refer to model
D1. Both quantities are shown normalized to their initial value at the
stellar surface. Let us concentrate on the slowly rotating model
first. The different lines refer to three representative times which are
$t=0.0$ (shown as dashed), $t=0.45$ and $0.52$ ms,
respectively. Initially, the angular velocity is, by construction,
uniform throughout the star (left panel) and the corresponding specific
angular momentum grows linearly with the distance from the stellar centre (right
panel). As the collapse proceeds and the stellar size decreases, the
angular velocity is expected to increase while the angular momentum per
unit mass remains constant. This is indeed what happens for model D1,
whose specific angular momentum is conserved with an overall error at the
stellar surface which is always less than $10\%$ and which decreases with
resolution.  A similar behaviour is observed also much later in the
simulation, when the apparent horizon has been found and the singularity
has been excised. Overall, the angular velocity in the collapsing model
D1 grows like $\Omega(t) \propto r^{-n}_e$, where $n\simeq 1.5$ and
therefore less than it would do in the case of the collapse of a
Newtonian, uniform density star ({\it i.e.} $n=2$); which is a result of
relativistic and rotational effects (see \cite{Cumming00}).

        A comparison of the lower parts of the two panels in
Fig.~\ref{diff_rot} is sufficient to realize that the evolution of the
angular velocity is rather different for a rapidly rotating stellar
model. The different lines in this case refer to $t=0.0, 0.48$ and $0.65$
ms, respectively, and it is apparent that a non-negligible degree of
differential rotation develops as the collapse proceeds, with a
difference of a factor $\sim 2$ between the angular velocity of the inner
and outer parts of the collapsing matter as the apparent horizon first
appears. Clearly, this differential rotation is produced very rapidly and
will persist only for a very short time before the star is enclosed in a
trapped surface.

        It is difficult to establish, at this stage, whether the
differential rotation generated in this way could produce a phenomenology
observable in some astrophysical context and more detailed
investigations, in particular of the coupling of this differential
rotation with magnetic fields ~\cite{Rezzolla00,Spruit99}, are necessary.
Finally, it is worth remarking that while differential rotation develops
for model D4 but not for D1, the specific angular momentum is conserved
to the same accuracy in both models.


\section{Dynamics of the horizons}
\label{sec:dynamics-horizons}

        In order to investigate the formation of a black hole in our
simulations, we have used horizon finders, available through the {\tt
Cactus} framework, which compute both the {\it apparent} horizon and the
{\it event} horizon. The apparent horizon, which is defined as the
outermost closed surface on which all outgoing photons normal to the
surface have zero expansion, is calculated at every time step and its
location is used to set up the excised region inside the
horizon. Specific technical details about the handling of the excised
region for the fields are presented
in~\cite{Alcubierre00a,Alcubierre01a}, while a brief discussion of how
the hydrodynamical excision is performed in {\tt Whisky} was presented in
Section~\ref{sec:hydr-excis}.

        In contrast, the event horizon, which is an expanding null
surface composed of photons which will eventually find themselves
trapped, is computed {\it a posteriori}, once the simulation is finished,
by reconstructing the full spacetime from the 3D data each
simulation produces. In stationary black-hole systems, where no
mass-energy falls into the black hole, the apparent and event horizons
coincide, but generally (in dynamical spacetimes) the apparent horizon
lies inside the event horizon. We have here used the fast solver of
Thornburg~\cite{Thornburg2003:AH-finding} to locate the apparent horizon
at every time step, and the level-set finder of Diener~\cite{Diener03a}
to locate the event horizon after the simulation has been completed and
the data produced is post-processed.

        In all cases considered, we have found that the event horizon
rapidly grows to its asymptotic value after formation.  With a temporal
gap of $\sim 10 M$ after the formation of the event horizon, the apparent
horizon is found and then it rapidly approaches the event horizon, always
remaining within it. With the exception of the initial gap, the horizon
proper areas as extracted from the apparent and event horizon are very
close (see {\it e.g.,} Fig.~\ref{surfcs_spctm}).

\subsection{Measuring the Event-Horizon Mass}
\label{sec:mass_and_spin}

        We measure the mass of the newly formed black hole to estimate
the amount of energy that is emitted as gravitational radiation during
the collapse. In particular, we do a simple energy accounting, comparing
the mass of the black hole with the ADM mass of the spacetime computed by
the initial data solver on a compactified grid extending to spatial
infinity~\cite{Stergioulas95}. This value is slightly different (1\% in
the worst case) from the one which is instead computed on the finite
domain of our computational grid at the initial time and after the
constraints are solved. The difference between the two values can be used
to define an ``error-bar'' for our measure of the black-hole mass and
hence of the energy in gravitational waves ({\it cf.}
Fig.~\ref{figHor_mass}). Two notes are worth making about this error
before we go on to discuss how the black-hole mass is actually
measured. Firstly, the difference between the two masses represents the
truncation error produced by the finite size of the computational domain
and is conceptually distinct from the truncation error introduced by the
finite differencing. While the first is assumed to be constant in time,
the second in general grows with time (especially after the excision is
made) and is monitored through the calculation of the constraint
equations. Secondly, this error-bar sets a global lower limit on the
accuracy of our measure of asymptotic quantities and therefore on the
energy lost to gravitational waves during the collapse. No reliable
measure of this lost energy can be made below the error-bar even if the
constraint equations are solved to a larger precision (this will be
discussed in more detail in Section \ref{christodoulou}).

\begin{figure}
\centering 
\includegraphics[angle=0,width=8.5cm]{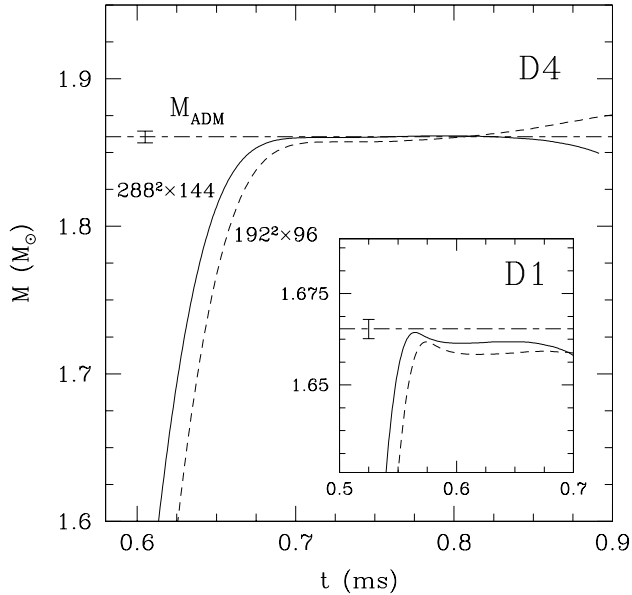}
\caption{Convergence of the measure of the black-hole mass as the
        resolution is increased. The curves refer to estimates using the
        event-horizon equatorial circumference [{\it i.e.} eq. (\ref{eq:mass
        from ce})] and have been obtained using $288^2\times 144$ and
        $192^2\times 96$ zones, respectively. Shown in the small inset
        are the results for model D1, while those for model D4 are in the
        main panel.}
\label{bh_mass_comp} 
\end{figure}

        The first and simplest method of approximating the black-hole
mass is to note that, for a Kerr (or Schwarzschild) black hole, the mass
can be found directly in terms of the event-horizon geometry as
\begin{equation}
\label{eq:mass from ce}
        M = \frac{C_{\textrm{eq}}}{4\pi} \ ,
\end{equation}
where $C_{\textrm{eq}}\equiv \int^{2\pi}_0 \sqrt{g_{\phi \phi}} d\phi$ is
the proper equatorial circumference. Provided there is a natural choice
of equatorial plane, it is expected that, as the black hole settles down
to Kerr, $C_{\textrm{eq}}$ will tend to the correct value. However, as
numerical errors build up at late times it may be impossible to reach
this asymptotic regime with sufficient accuracy.

        The estimate of $M$ coming from the use of (\ref{eq:mass from
ce}) is presented in Fig.~\ref{bh_mass_comp}, which shows the time
evolution of the event-horizon equatorial circumference. The two lines
refer to two different resolutions ($288^2\times 144$ and $192^2\times
96$ zones, respectively) and should be compared with the value of the ADM
mass $M_{\rm ADM}$ (indicated with a short-long dashed line), and with
the error-bars as deduced from the initial data. Shown in the small inset
are the results for model D1, while those for model D4 are in the main
panel. 

        Note that if a measure of the event horizon is not available,
eq.~(\ref{eq:mass from ce}) could be computed using the equatorial
circumference of the apparent horizon (this is what was done, for instance,
in~\cite{Duez04}). Doing so would yield results that are similar to those
shown in Fig.~\ref{bh_mass_comp}, although with a slightly larger
deviation from $M_{\rm ADM}$. This is because we have found the apparent
horizon to systematically underestimate the equatorial circumference. In
particular, in the high-resolution run for model D4, the differences
between the apparent and event-horizon equatorial circumferences
are~$\lesssim 2\ \%$.

\begin{figure}
\centering 
\includegraphics[angle=0,width=8.5cm]{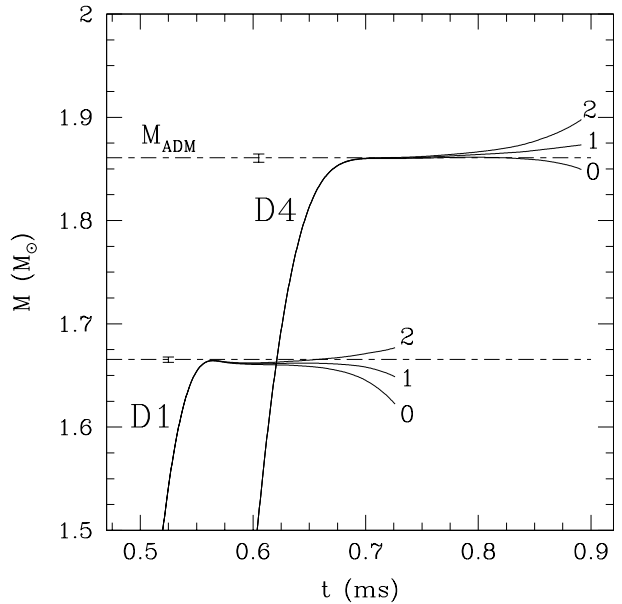} 
\caption{Evolution of the event-horizon mass $M = C_{eq}/4\pi$ for models
        D1 and D4. Different lines refer to the different initial guesses
        for the null surface and are numbered 0, 1 and 2. Note that they
        converge exponentially to the correct event-horizon surface for
        decreasing times. 
}
\label{eh_generators} 
\end{figure}

        Clearly, as the equatorial circumference grows, the agreement
with the expected ADM mass improves as it does with the use of higher
spatial resolution. However, equally evident is that the errors grow as
the collapse proceeds and this is due, in part, to the loss of strict
second-order convergence at later times, but also to the way the event
horizon is found. The level-set approach of~\cite{Diener03a}, in fact,
needs initial guesses for the null surface, which converge exponentially
to the correct event-horizon surface for decreasing times, hence
introduces a systematic error in the calculation of the event horizon at
late times. This is shown in Fig.~\ref{eh_generators}, which presents the
evolution of the event-horizon mass $M = C_{eq}/4\pi$ for models D1 and
D4. Different lines refer to the different initial guesses and are
numbered ``0'', ``1'' and ``2'', respectively (note that for the curves
shown in Fig.~\ref{bh_mass_comp} the initial guesses ``0'' and ``1'' were
used for cases D4 and D1, respectively).

\begin{figure*}
  \centering
  \includegraphics[angle=0,width=7.5cm]{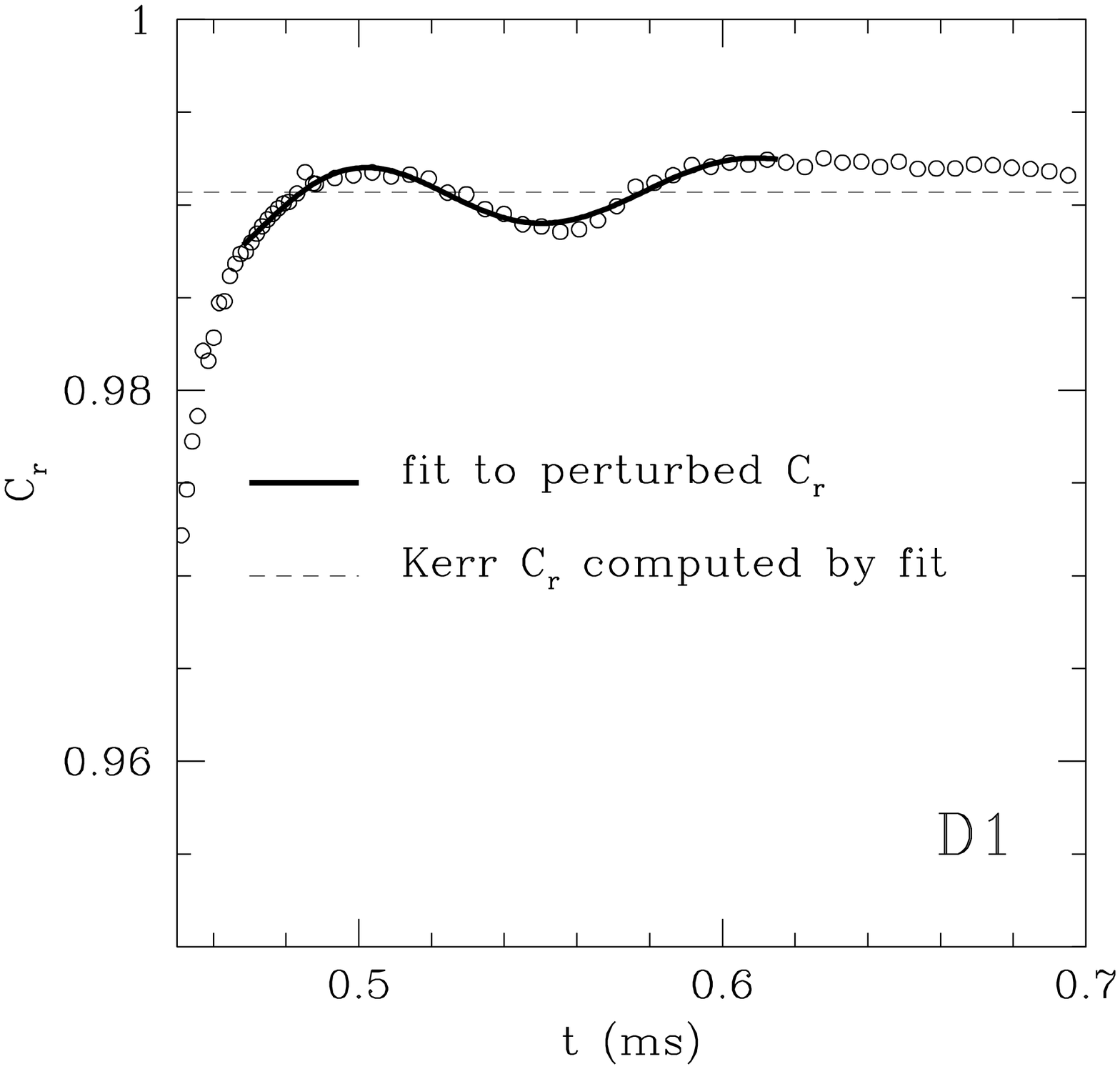} 
  \hskip 1.0 cm
  \includegraphics[angle=0,width=7.5cm]{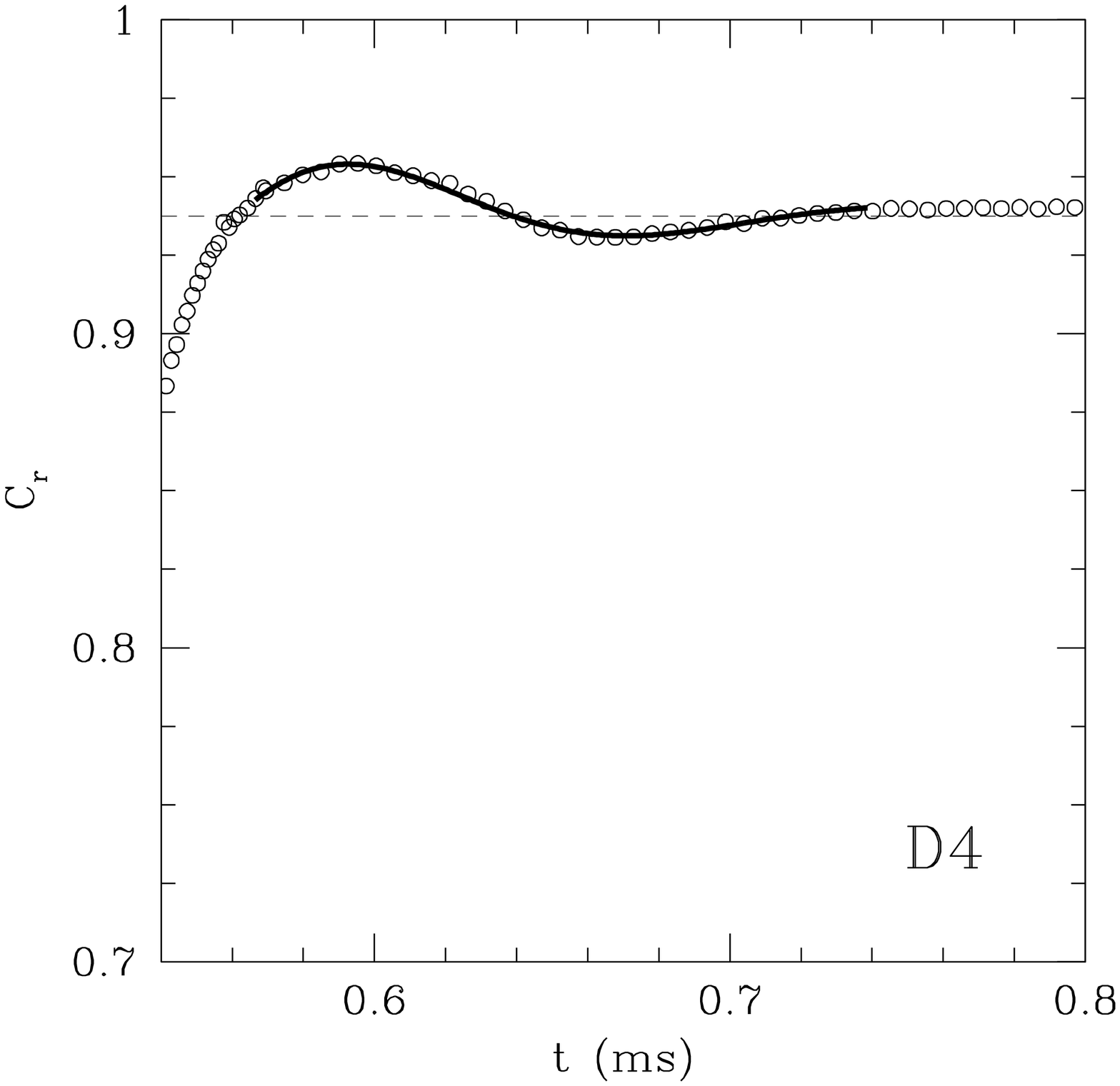} 
  \caption{Fitting the oblateness of the event horizon to QNMs of a Kerr
    black hole. The fit is shown with the solid line, while the open
    circles represent the computed values of $C_r$. The estimate for
    $C_r$ of a Kerr black hole having the fitted value of $a/M_{hor}$ is
    shown with a dashed line.}
  \label{fig:eh_modefitD1}
\end{figure*}

\subsection{Measuring the angular momentum of the black hole}

        A major difficulty in an accurate measurement of $M$ lies in the
calculation of its non-irreducible part, {\it i.e.} in the part that is
proportional to the black-hole angular momentum $J$. We now discuss a
number of different ways to calculate $J$ from the present simulations;
these measurements will then be used to obtain alternative estimates of
$M$ in Section~\ref{christodoulou}.

\subsubsection{Measuring $J$ from the Horizon Distortion}

        In a series of papers studying the dynamics of distorted
black-hole spacetimes, it was shown that the horizon geometry provides
a useful measure of the black-hole properties both in
vacuum~\cite{Anninos93a, Anninos94f, Anninos95c, Brandt94c} and when
these are accreting matter axisymmetrically~\cite{Brandt98}.
In particular, the idea is to look at the distortion of the horizon using
the ratio of polar and equatorial proper circumferences, $C_r \equiv
C_{\textrm{pol}} / C_{\textrm{eq}}$. For a perturbed Kerr black hole this
is expected to oscillate around the asymptotic Kerr value with the form
of a quasi-normal mode (QNM). By fitting to this mode we extract an
estimate of the angular momentum parameter $a/M_{hor}$ from the
relation~\cite{Brandt94b}
\begin{equation}
\label{eq:SmarrDistortion}
\frac{a}{M_{hor}} = \sqrt{1-\left( -1.55 + 2.55 C_r \right)^2} \ ,
\end{equation}
where we have indicated with $M_{hor}$ the black-hole mass as measured
from expression (\ref{eq:SmarrDistortion}), which coincides with $M$
only if the spacetime has become axisymmetric and stationary. The fit
through expression (\ref{eq:SmarrDistortion}) is expected to be accurate
to $\sim 2.5\%$~\cite{Brandt94b}.

        The fit itself depends on an initial guess for $a/M_{hor}$ and
we start from a Schwarzschild black hole and iterate till the desired
convergence is reached. This measure is not gauge invariant, although
eq.~\eqref{eq:SmarrDistortion} is independent of the spatial
coordinates up to the definition of the circumferential planes, but
works adequately with the gauges used here. The fit is best performed
shortly after black-hole formation as the oscillations are rapidly
damped. This minimizes numerical errors but in those cases where
matter continues to be accreted, it may lead to inaccurate estimates of
the angular momentum.

        Examples of the fitting procedure are shown in
Fig.~\ref{fig:eh_modefitD1}, in which the fit is shown as a solid
line, while the open circles represent the computed values of $C_r$;
these are slightly noisy as a result of the interpolation needed by the
level-set approach to find points on the horizon
two-surface~\cite{Diener03a}. The estimate for $C_r$ of a Kerr black hole
having the fitted value of $a/M_{hor}$ is shown as a dashed
line. Note that the values of $a/M_{hor}=0.21$ and $a/M_{hor}=0.54$ are
very close to the total $J/M^2$ of the initial stellar models, {\it i.e.}
0.2064 and 0.5433, 
as shown in Table~\ref{tab:eh_oblateness}. This demonstrates that, to
within numerical accuracy, the entire angular momentum of the
spacetime ends up in the black hole.

\begin{table}[htbp]
  \caption{Estimates of the black-hole angular momentum $J/M^2$ through the oblateness of the event
        horizon. The oscillations in the oblateness of the event horizon can be fitted to the normal
        modes of a Kerr black hole. Note that for each model the measured angular momentum is
        remarkably close to that of the initial spacetime $(J/M^2)_{\rm ADM}$. Also reported are the
        initial ADM mass, the value of the equatorial circumference as obtained through the fit
        $(C_r)_{\rm EH}$, and the corresponding value obtained through the estimated spin parameter
        $(C_r)_{\rm Kerr}$.}  \centering
  \begin{tabular}[c]{cccccc}
    \\
    \hline
    Model & $M_{\textrm{ADM}}$ & $(J/M)_{\textrm{ADM}}^2$ 
        & $(J / M^2)_{\rm EH}$& $(C_r)_{\rm EH}$ & $(C_r)_{\rm Kerr}$  \\ 
        \hline
    D1 & 1.6653 & 0.2064 & 0.21 & 0.99 & 0.9916 \\ 
    D2 & 1.7281 & 0.3625 & 0.36 & 0.97 & 0.9734 \\ 
    D3 & 1.7966 & 0.4685 & 0.47 & 0.95 & 0.9544 \\ 
    D4 & 1.8606 & 0.5433 & 0.54 & 0.94 & 0.9372 \\ 
    \hline
  \end{tabular}
  \label{tab:eh_oblateness}
\end{table}

\begin{figure*}
\centering
\includegraphics[angle=0,width=7.5cm]{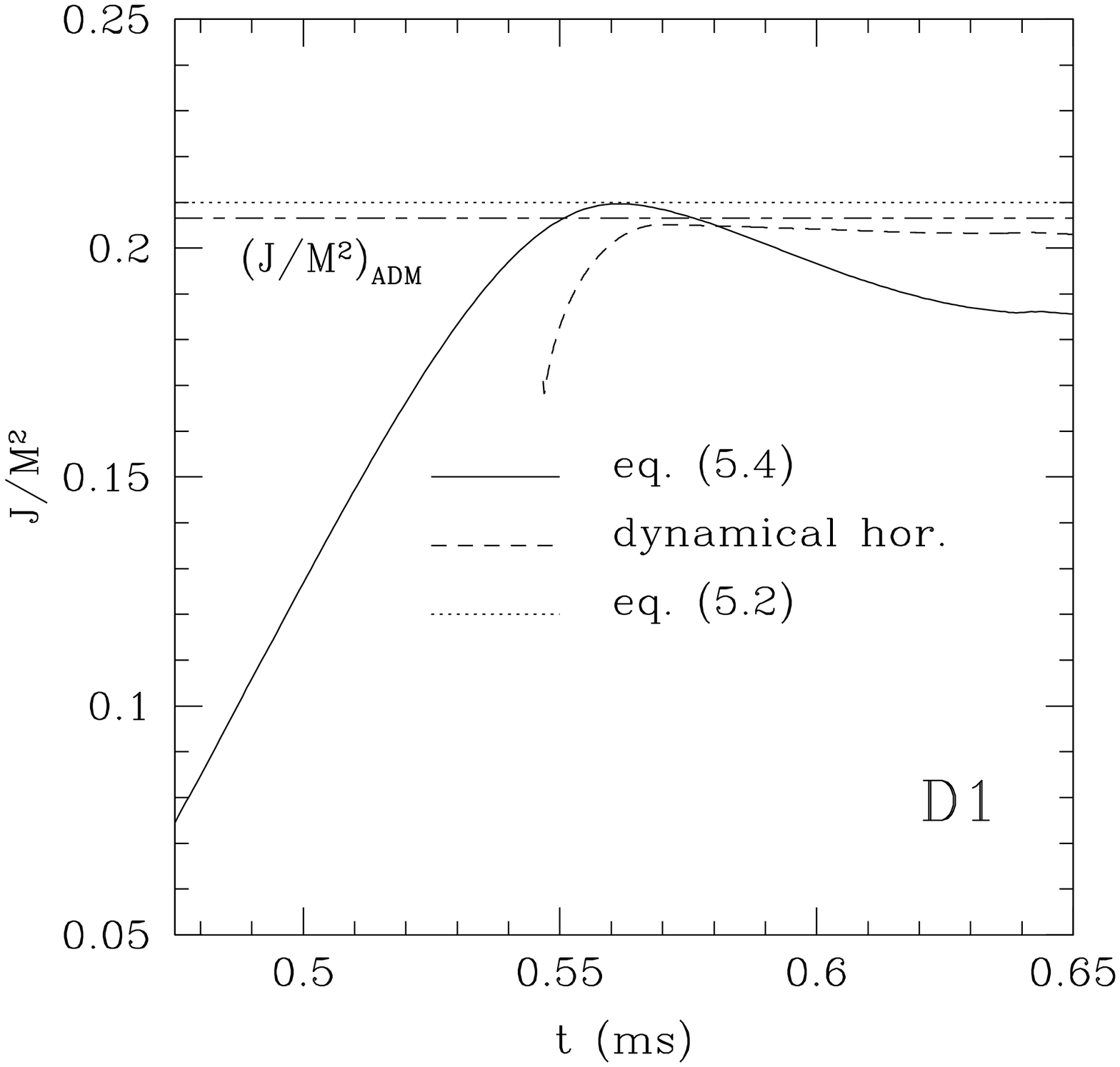} 
\hskip 1.0 cm
\includegraphics[angle=0,width=7.5cm]{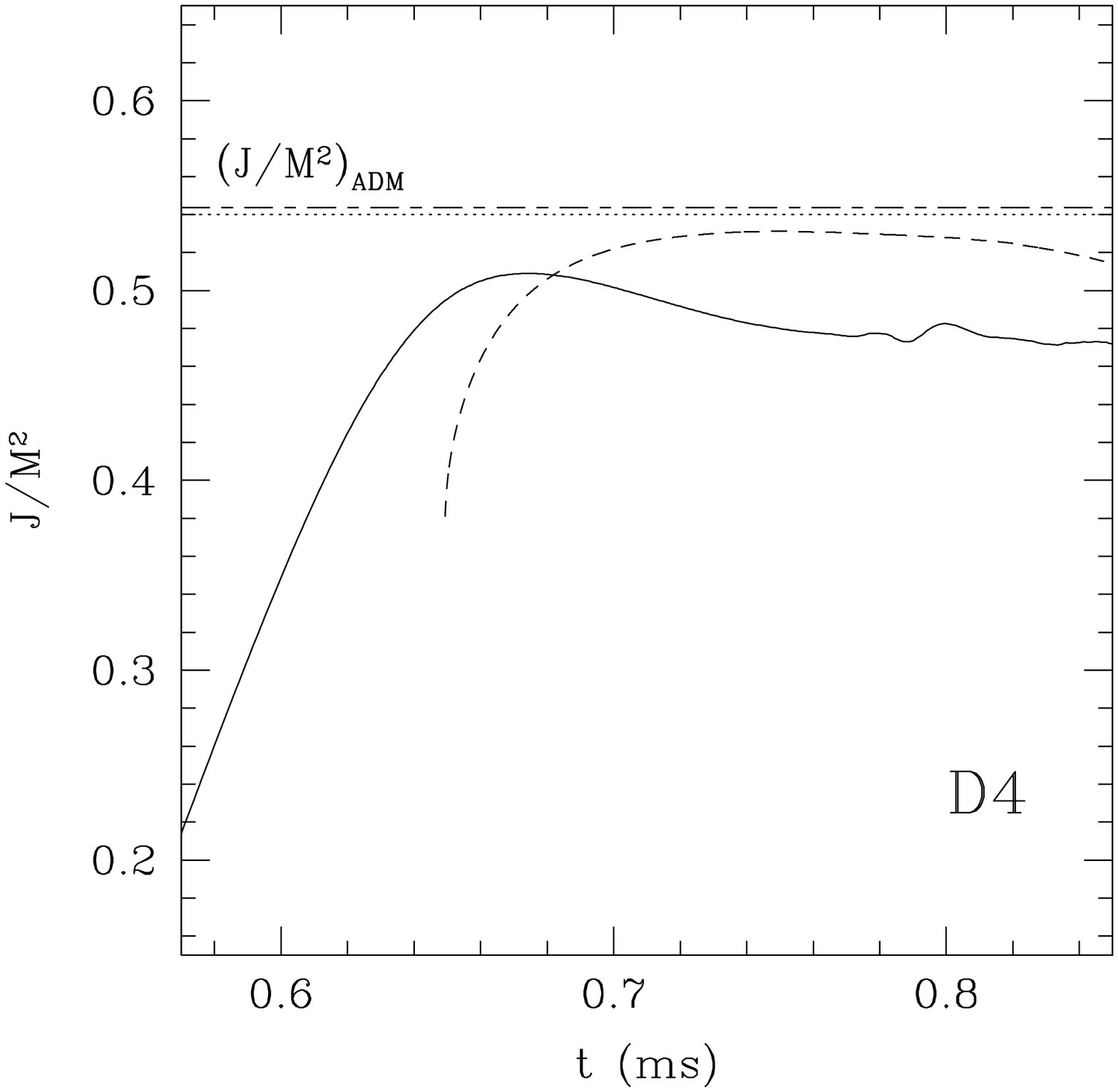} 
\caption{Comparison of the different measures of the angular momentum for
        the cases D1 (left panel) and D4 (right panel). The estimate
        using the fit to the circumference ratio (see left panel of
        Fig.~\ref{fig:eh_modefitD1}) is also shown. The dynamical-horizon
        spin measurement is considerably more accurate at late times as
        the event-horizon surfaces will diverge exponentially at this
        point. Shown with the horizontal short-long dashed lines are the
        values of $(J/M^2)_{\rm ADM}$ in the two cases as measured from
        the initial data (see main text for details).}
\label{fig:ehspin_D4} 
\end{figure*}

        Using expression~\eqref{eq:SmarrDistortion} to
estimate the angular momentum $J$ introduces an error, if the black hole
has not yet settled to a Kerr solution. Having this in mind, however, it
is possible to estimate the angular momentum as 
\begin{equation}
J=\left(\frac{a}{M_{hor}}\right) M_{hor} M \simeq 
  \left(\frac{a}{M_{hor}}\right) M^2\ .
\label{Jdist2}
\end{equation}

\subsubsection{Measuring $J$ with the dynamical-horizon framework}

        A second method of approximating $J$ and hence measuring $M$
is to use the {\em isolated} and {\it dynamical}-horizon frameworks of
Ashtekar and collaborators~\cite{Ashtekar99a,Ashtekar00a,Ashtekar01a,
Ashtekar-etal-2002-dynamical-horizons,Dreyer-etal-2002-isolated-horizons}.
This assumes the existence of an axisymmetric Killing vector field
intrinsic to a marginally trapped surface such as an apparent horizon.
This gives an unambiguous definition of the spin of the black hole and
hence of its total mass. If there is an energy flux
across the horizon, the isolated-horizon framework needs to be
extended to the {\em dynamical}-horizon
formalism~\cite{Ashtekar-etal-2002-dynamical-horizons,Ashtekar03a}.

In practice, our approach to the dynamical-horizon framework has been
through the use of a code by Schnetter which implements the algorithm
of~\cite{Dreyer-etal-2002-isolated-horizons} to calculate the horizon
quantities.  The advantage of the dynamical-horizon framework is that
it gives a measure of mass and angular momentum which is accurately
computed locally, without a global reconstruction of the spacetime.
One possible disadvantage is that the horizon itself is required to be
(close to) axisymmetric; so in case it deviates largely
from axial symmetry, no information can be found. However, because
arbitrarily large distortions are allowed as long as they are
axisymmetric, we have not encountered problems in applying the
dynamical-horizon framework to the horizons found in our simulations.

\subsubsection{Measuring $J$ from the Angular Velocity of the Event Horizon}

        A third method for computing $J$ only applies if an event
horizon is found and if its angular velocity has been measured. In a Kerr
background, in fact, the generators of the event horizon rotate with a
constant angular velocity $\omega\equiv - g_{t\phi}/g_{\phi\phi}=
\sqrt{g_{tt}/g_{\phi\phi}}$. In this case the generator velocity
can be directly related to the angular momentum parameter as
\begin{equation}
\label{eq:j and a over m from omega}
\frac{a}{M} = \frac{J}{M^2} = \left[ \frac{A \omega^2}{\pi} \left(1 -
      \frac{A \omega^2}{4 \pi} \right)\right]^{1/2} \ .
\end{equation}

As with the previous approximations, expression (\ref{eq:j and a over m
from omega}) is strictly valid only for a Kerr black hole and will
therefore contain a systematic error which, however, decays rapidly as
the black-hole perturbations are damped. On the other hand, the event
horizon generator velocities have the considerable advantage that
everything is measured instantaneously and the values of $\omega$ are
valid whether or not the background is an isolated Kerr black hole.  The
disadvantage, though, is that, as mentioned above, the numerical event
horizon surfaces become systematically less accurate at late times ({\it
cf.}  Fig.~\ref{eh_generators}).

\subsubsection{Comparison of angular momentum measurements}

        A detailed comparison of the three different methods for
measuring the angular momentum of the black hole is shown in
Fig.~\ref{fig:ehspin_D4}. The measurement of angular momentum using the
angular velocity of the generators is shown as solid lines. Both for
slowly (left panel) and rapidly (right panel) rotating stellar models,
the event horizon has zero area (and thus zero angular momentum) when it
is first formed.  However, as the rotating matter collapses, the event
horizon-area and angular momentum grow, the black hole is spun up and, to
numerical accuracy, the total angular momentum of the spacetime is
contained within the black hole ({\it cf.}\
Fig.~\ref{fig:eh_modefitD1}). At late times, the estimate using the
generator velocities of the event horizon drifts away, probably due to a
combination of gauge effects and the systematic errors in the trial
guesses for the null surfaces.

        In the case of the slowly rotating model D1, in particular, the
estimate from the dynamical-horizon finder is perfectly stable ({\it
cf.}\ dashed line in the left panel of Fig.~\ref{fig:ehspin_D4}),
indicating that an approximately stationary Kerr black hole has been
formed by the time the simulation is terminated. In the case of the
rapidly rotating model D4, however, this is no longer the case as matter
continues to be accreted also at later times, when the errors have also
increased considerably. As a result, the measure of the spin through the
dynamical-horizon finder is less accurate and does not seem to have
stabilized by the time the simulation ends ({\it cf.}\ dashed line in the
right panel of Fig.~\ref{fig:ehspin_D4}). This may indicate that the
final black hole has not settled down to a Kerr black hole on the
timescales considered here.

\subsection{black-hole mass from the Christodoulou formula}
\label{christodoulou}

        It was shown by Christodoulou that, in the axisymmetric and
stationary spacetime of a Kerr black hole, the square of the black-hole
mass $M$ is 
given by
\cite{Christodoulou70}
\begin{equation}
M^2 = M^2_{\rm irr} + \frac{4 \pi J^2}{A} =
        \frac{A}{16\pi}+\frac{4 \pi J^2}{A}\ ,
\label{MAH}
\end{equation}
where $M_{\rm irr}$ is the irreducible mass, $A$ is the event-horizon proper
area, and $J$ is the black-hole angular momentum. As the black hole
approaches a stationary state at late times, the apparent and event
horizons will tend to coincide and in that case the mass of the black
hole is very well approximated by the above formula. 

        We have applied the above formula, using the various methods for
measuring the angular momentum $J$. In particular, using the method for
obtaining $J$ from the distortion of the event horizon, through
eq. (\ref{Jdist2}), the black-hole mass is given by
\begin{equation}
\label{eq:m total from a over m quadratic}
M^2 = \frac{A}{8\pi} \left( \frac{M_{hor}}{a} \right)^2 
        \left[1 - \sqrt{ 1 - \left( \frac{a}{M_{hor}} \right)^2}\right] \ .
\end{equation}
If, on the other hand, $J$ is found from the angular velocity $\omega$ of
the event horizon, then it is possible to use (\ref{eq:j and a over m
from omega}) in (\ref{MAH}) and obtain
\begin{equation}
\label{eq:m total from omega}
  M^2 = \frac{A}{16\pi - 4 A \omega^2}\ .
\end{equation}
In the framework of dynamical horizons, expression
(\ref{MAH}) holds for {\it any} axisymmetric isolated or dynamical
horizon, independently of whether it is stationary.

\begin{figure*}
\centering
\includegraphics[angle=0,width=7.5cm]{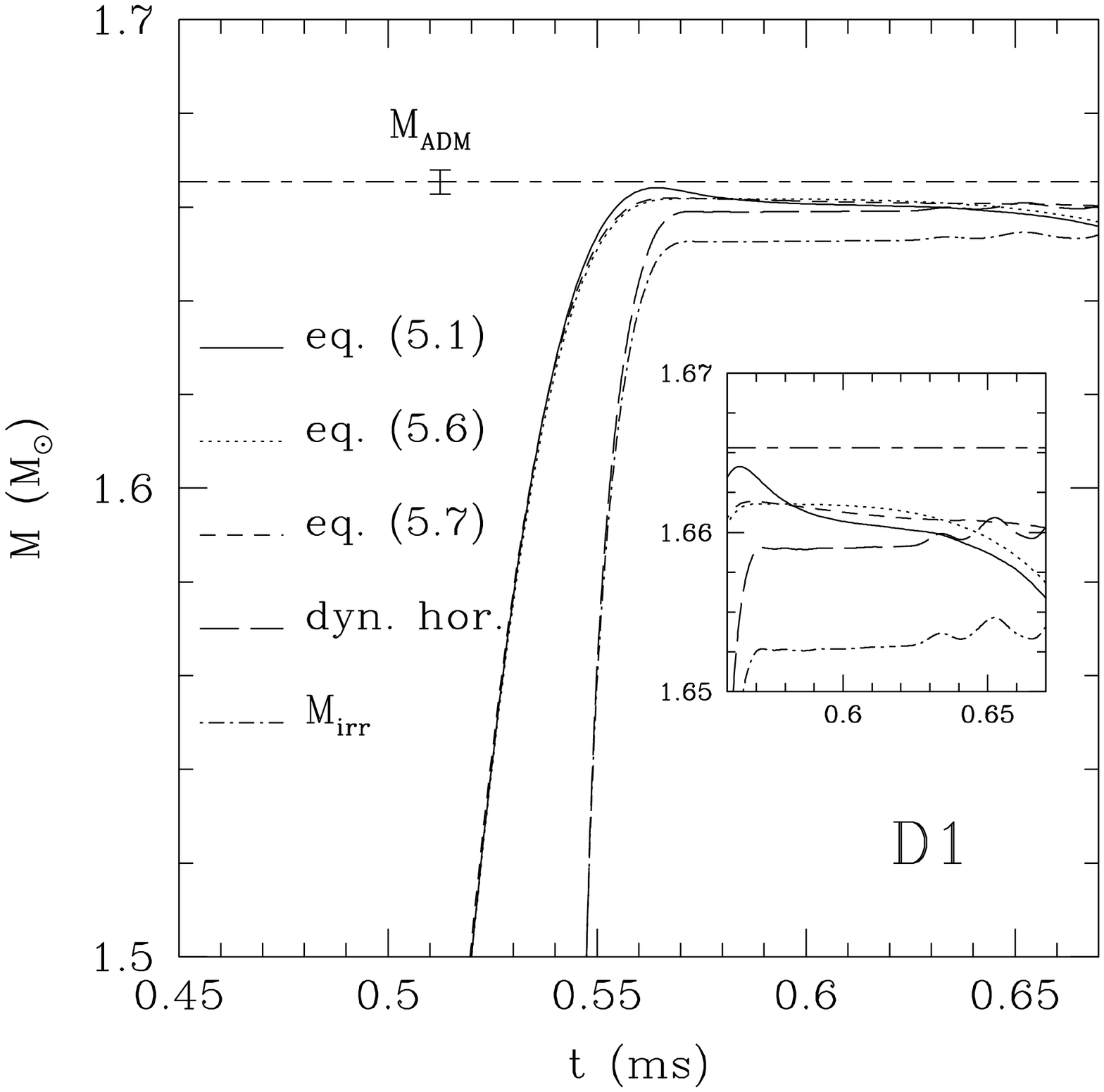} 
\hskip 1.0 cm
\includegraphics[angle=0,width=7.5cm]{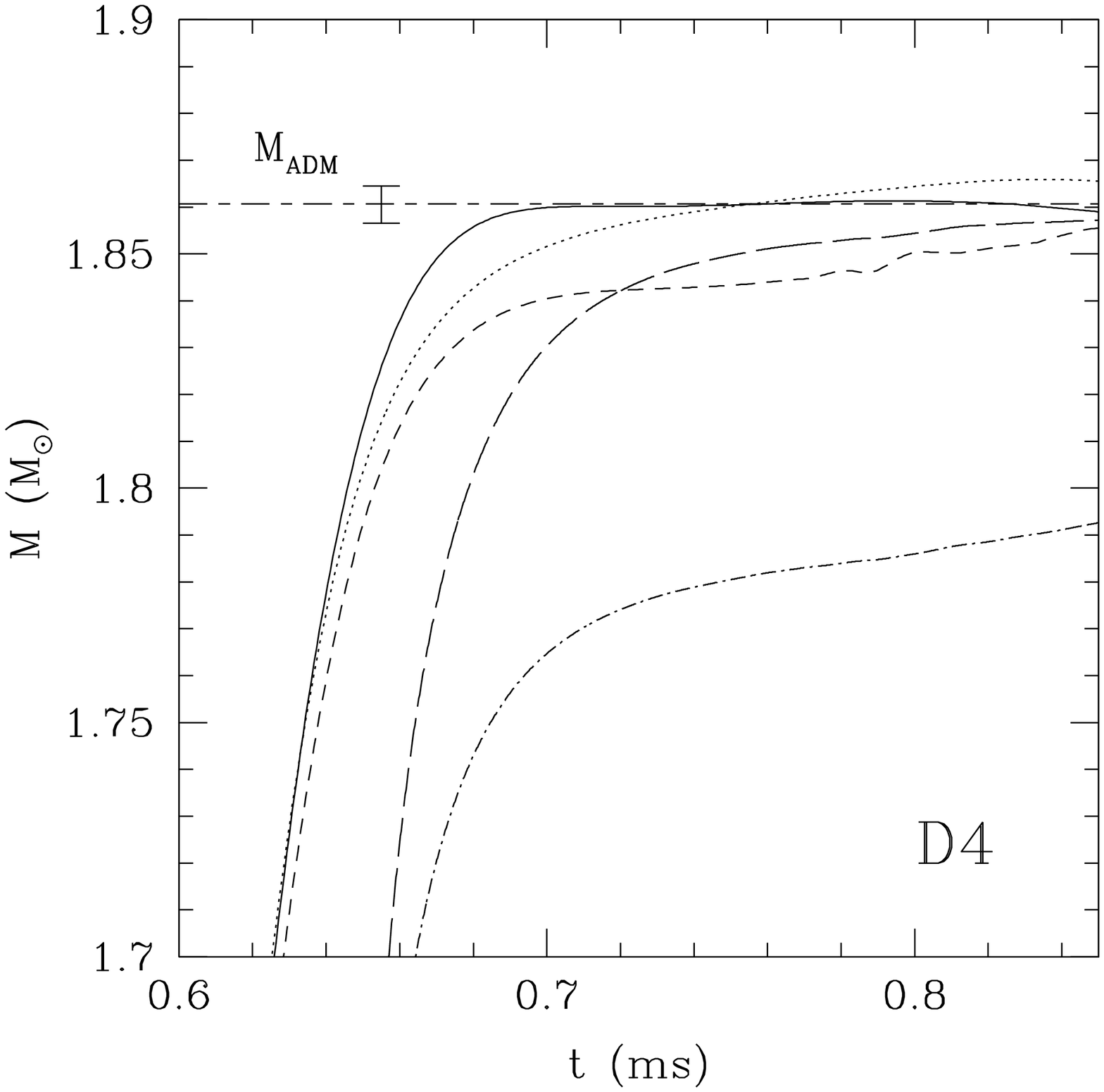} 
\caption{Comparison of the five different approaches used in the measure
        of the total black-hole mass for the collapse of models D1 and
        D4. Different lines refer to the different methods discussed in
        the main text. The left panel (model D1), shows that
        the different methods are overall comparable when the rotation is slow, but that
        differences are already present (this is as shown in the inset).
        The right panel (model D4) shows that the different measures can
        be considerably different when the rotation is large. 
}
\label{figHor_mass} 
\end{figure*}

        Figure~\ref{figHor_mass} collects the four different ways of
measuring the black-hole mass for the collapse of models D1 and D4. The
different lines refer to the different approaches we have outlined
above. In addition, we display the irreducible mass $M_{\rm irr}$. The left
panel of Fig.~\ref{figHor_mass}, in particular, shows the results of the
different measures for the slowly rotating model D1. Because in this case
all of the matter rapidly collapses into the black hole, the different
estimates of the total mass agree very well. However, already in this
slowly rotating case the irreducible mass of the apparent horizon is
noticeably lower. The left panel also shows that while the different
methods provide comparable estimates, only the one corresponding to
eq.~(\ref{eq:mass from ce}) ({\it i.e.} the solid line) falls for some
time within the error-bar provided by the initial estimate of $M_{\rm{ADM}}$
(this is particularly evident in the inset). Because when this happens
the norms of the Hamiltonian constraint have not yet started to grow
exponentially and the largest value of the constraint violation is about
an order of magnitude smaller ({\it i.e.} the L$_1$ norm of the
Hamiltonian constraint is $\sim 4.9 \times 10^{-4}$ at $t=0.56$ ms) we
can use the error-bar in $M_{\rm ADM}$ to place an upper bound of $0.5\%\
M_{\textrm{ADM}}$ to the energy lost through the emission of
gravitational radiation in this case. Clearly, the true bound is
certainly considerably lower and we expect that with accuracies
comparable to the ones of 2D simulations, our estimates of the efficiency
of gravitational radiation emission could converge to the values of Stark
and Piran~\cite{Stark85}.

        The right panel of Fig.~\ref{figHor_mass}, on the other hand,
shows the results of the different mass measures for the rapidly
rotating model D4. In this case, the contribution from the spin energy
is considerably larger and noticeable differences appear among the
different approaches. Since all seem to have systematic errors, this
makes it less trivial to establish which method is to prefer. On one
hand, those methods using information from the event-horizon
equatorial circumference or that fit the perturbations of the event
horizon [{\it i.e.}\ eqs.~(\ref{eq:mass from ce}) and (\ref{eq:m total
from a over m quadratic})] seem to provide accurate estimates at
earlier times but suffer of the overall inaccuracy at later stages,
when the initial guesses for the null surface are distinct. It is
indeed at these early times that these measurements are within the
error-bar provided by the initial estimate of $M_{\rm{ADM}}$. On the
other hand, those methods that measure the angular velocity of the
null generators [{\it i.e.}\ eq. (\ref{eq:m total from omega})] or
that use the dynamical-horizon framework, produce reasonably accurate
estimates, that converge with resolution, that monotonically grow in
time and that are within the error-bar of the initial estimate of
$M_{\rm{ADM}}$. Furthermore, in the case of the dynamical-horizon
framework, this is not only physically expected, given that a small
but non-zero fraction of the matter continues to be accreted nearly until
the end of the simulation, but it is also guaranteed analytically.

        Because of these differences in the measures of $M$ and
because the black hole does not have time to settle down to a constant
total mass, the upper bound on the energy emission is more
conservative than in case D1. In particular, taking again as a
reference the time when the estimate relative to eq.~(\ref{eq:mass
from ce}) is within the error-bar ({\it i.e.} at $t=0.70$ ms) and the
largest value of the constraint violation is about an order of
magnitude smaller ({\it i.e.}  the L$_1$ norm of the Hamiltonian
constraint is $\sim 1.2 \times 10^{-3}$) and is not yet growing
exponentially, we place an upper bound of $1\%\ M_{\textrm{ADM}}$ on
the energy lost through gravitational radiation.  Once again, we
expect the true value to be considerably smaller.

        One obvious and expected result is that the irreducible mass in
the collapse of model D4 (the dot-dashed line in the right panel of
Fig.~\ref{figHor_mass}) deviates by a large amount from the actual black
hole mass, since it does not include the rotational energy of the black
hole.

        Finally, we will make a comment on the different methods used for
measuring the mass and spin of a black hole in a numerical
simulation. Although the direct comparison of many different methods
employed here have provided valuable information on the dynamics of the
system, we have found the dynamical-horizon framework to be simple to
implement, accurate and not particularly affected by the errors from
which equivalent approaches seem to suffer, as shown in our
Figs.~\ref{figHor_mass} and \ref{fig:ehspin_D4}. As a result, we
recommend its use as a standard tool in numerical relativity simulations.

\subsection{Reconstructing the global spacetime}
\label{rtgs}

        All of the results presented and discussed in the previous
Sections 
describe only 
a small portion of the spacetime which has been solved during the
collapse. In addition to this, it is interesting and instructive to
collect all of these pieces of information into a {\it global}
description of the spacetime and look for those features which mark the
difference between the collapse of slowly and rapidly rotating stellar
models. As we discuss below, these features emerge in a very transparent
way within a global view of the spacetime.

\begin{figure*}
\centering
\includegraphics[angle=0,width=7.5cm]{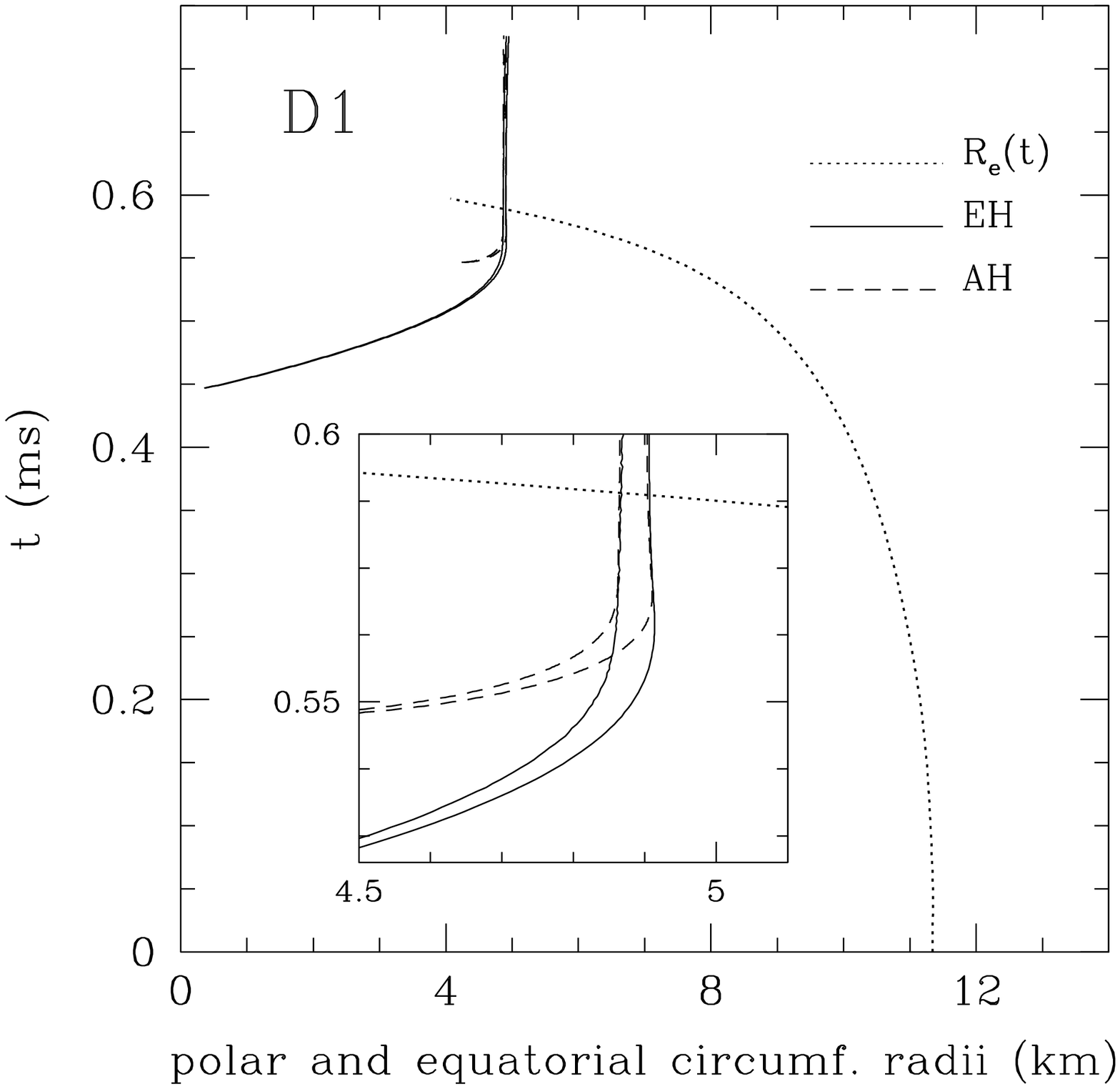}       
\hskip 1.0 cm
\includegraphics[angle=0,width=7.5cm]{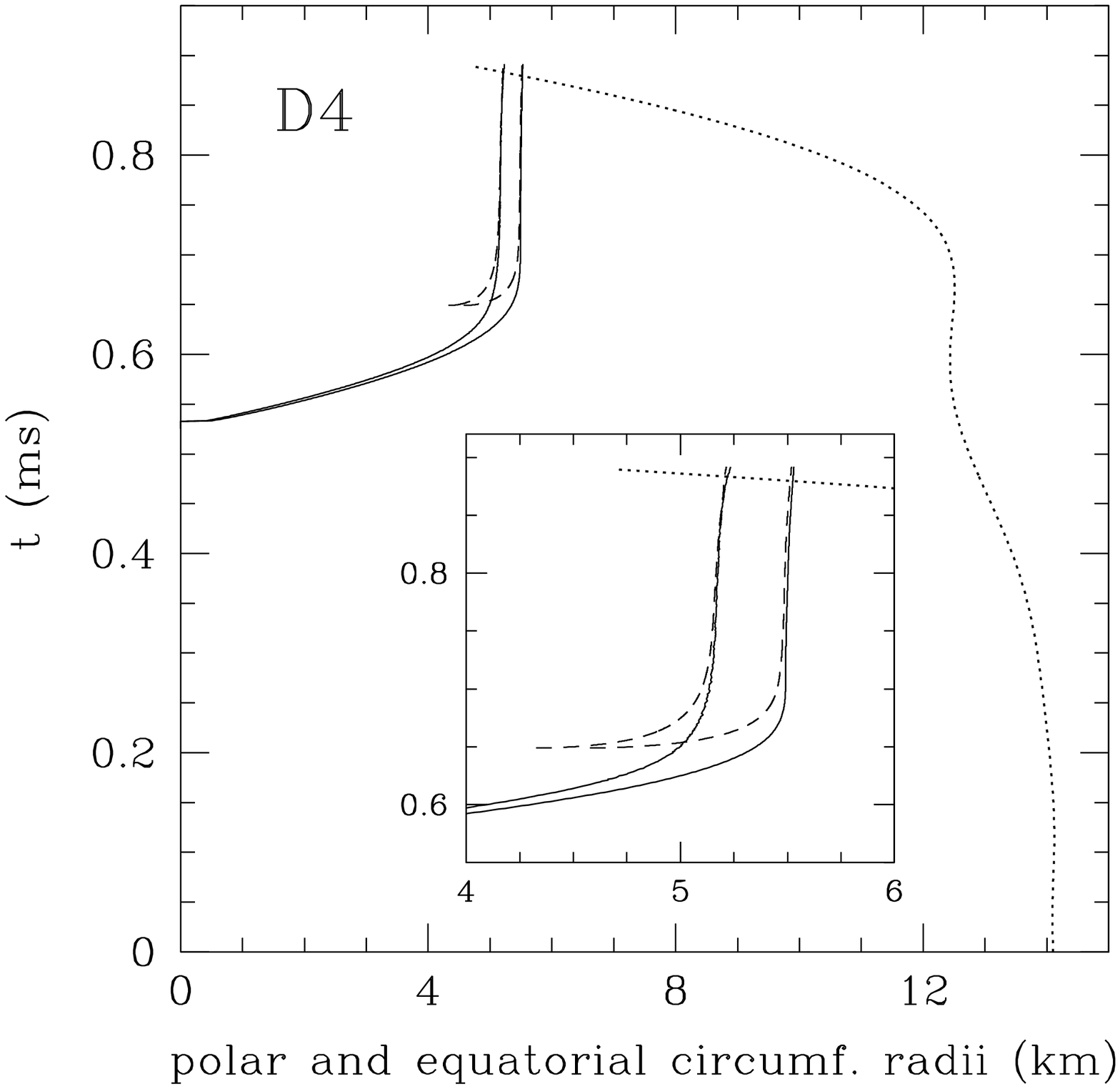}       
\caption{Evolution of the most relevant surfaces during the collapse for
        the cases D1 and D4. Solid, dashed and dotted lines represent the
        worldlines of the circumferential radii of the event horizon, of
        the apparent horizon and of the stellar surface,
        respectively. Note that for the horizons we plot both the
        equatorial and the polar circumferential radii, while only the
        equatorial circumferential radius is shown for the stellar
        surface. Shown in the insets are the magnified views of the
        worldlines during the final stages of the collapse. }
\label{surfcs_spctm}
\end{figure*}

        To construct this view, we use the worldlines of the most
representative surfaces during the collapse, namely those of the
equatorial stellar surface, of the apparent horizon and of the event
horizon. For all of them we need to use properly defined quantities and,
in particular, circumferential radii. The results of this spacetime
reconstruction are shown in Fig.~\ref{surfcs_spctm}, whose left and right
panels refer to the collapse of models D1 and D4, respectively. The
different lines indicate the worldlines of the circumferential radius of
the stellar surface (dotted line), as well as of the apparent horizon
(dashed line) and of the event horizon (solid line). Note that for the
horizons we show both the equatorial and the polar circumferential radii,
with the latter being always smaller than the former. For the stellar
surface, on the other hand, we show the equatorial circumferential radius
only. This is because the calculation of the stellar polar
circumferential radius requires a line integral along the stellar surface
on a given polar slice. Along this contour one must use a line element
which is suitably fitted to the stellar surface and diagonalized (see
\cite{Brandt94c} for a detailed discussion). In the case of model D4,
however, this is difficult to compute at late times, when the disc is
formed and the line integral becomes inaccurate.

        Note that in both panels of Fig.~\ref{surfcs_spctm} the event
horizon grows from an essentially zero size to its asymptotic
value. In contrast, the apparent horizon grows from an initially non-zero
size and, as it should, is always contained within the event
horizon. At late times, the worldlines merge to the precision
at which we can compute them. A rapid look at the two panels of
Fig.~\ref{surfcs_spctm} is sufficient to appreciate the different
properties in the dynamics of the collapse of slowly and rapidly rotating
models.

        Firstly, in the case of model D1, the differences between the
equatorial and polar circumferential radii of the two trapped surfaces
are very small and emerge only in the inset that offers a magnified view
of the worldlines during the final stages of the collapse. This is not
the case for model D4, for which the differences are much more evident
and can be appreciated also in the main panel. Of course, this is what
one expects given that the ratio of these two quantities 
depends on 
$a/M$ and is $\sim 1$ for a slowly rotating black hole ({\it cf.}
Table~\ref{tab:eh_oblateness}).

        Secondly, the worldlines of the stellar equatorial
circumferential radius are very different in the two cases. In the slowly
rotating model D1, in particular, the star collapses smoothly and the
worldline always has negative slope, thus reaching progressively smaller
radii as the evolution proceeds ({\it cf.} left panel of
Fig.~\ref{surfcs_spctm}). By time $t\simeq 0.59$ ms, the stellar
equatorial circumferential radius has shrunk below the corresponding
value of the event horizon. In the case of the rapidly rotating model
D4, on the other hand, this is no longer true and after an initial phase
which is similar to the one described for D1, the worldline does not
reach smaller radii. Rather, the stellar surface slows its inward motion
and, at around $t \sim 0.6$ ms, the stellar equatorial circumferential radius
does not vary appreciably. Indeed, the right panel of
Fig.~\ref{surfcs_spctm} shows that at this stage the stellar surface
moves to slightly larger radii. This behaviour marks the phase in which a
flattened configuration has been produced and the material at the outer
edge of the disc experiences a stall ({\it cf.} the middle
and lower panels of Fig.~\ref{figcollseq_D4}). As the collapse proceeds,
however, also this material will not be able to sustain its orbital
motion and, after $t \sim 0.7$ ms, the worldline moves to smaller radii
again. By a time $t\simeq 0.9$ ms, the stellar equatorial circumferential
radius has shrunk below the corresponding value of the event horizon.


\section{Conclusion}
\label{sec:conclusion}

        Although 3D numerical relativity has been a very active research
area for several years now, there are still a number of technical issues
to be addressed and physical problems to be investigated in
detail. Separate progress has been made so far in obtaining long-term
stable evolutions of vacuum spacetimes and of spacetimes with
matter. Both of them have posed significant numerical problems because of
the existence of horizons containing physical singularities, in one case,
and the development of non-linear hydrodynamical phenomena such as shocks,
in the other. In black-hole vacuum spacetimes, these problems have
successfully been dealt with by using better suited formulations of the
Einstein equations and by employing excision techniques for the regions
of the spacetime containing the singularity. In spacetimes containing
matter, on the other hand, sophisticated numerical techniques (such as
the HRSC methods) have been employed to accurately track the dynamics of
the shocks.

        Here, we have combined these two different approaches by
implementing excision techniques within a {\it forming} horizon, thus
following the dynamics of the matter as it accretes onto the developing
black hole. We have shown that doing so allows the numerical evolution to
proceed uninhibited from fully regular initial conditions of matter in
equilibrium and devoid of trapped surfaces, up to a vacuum spacetime
featuring an event horizon enclosing an excised physical
singularity. This new important ability in numerical relativity
evolutions will help in a more detailed investigation of complex
astrophysical systems, such as the coalescence of neutron star binaries,
considered as a prime candidate for the detection of gravitational waves,
and of the collapse of stellar cores, considered as the progenitors of
gamma-ray bursts.

        Our hydrodynamical excision technique is implemented within a new
3D general-relativistic numerical evolution code that combines
state-of-the-art numerical methods for the spacetime evolution ({\it
i.e.} the NOK formulation of the Einstein equations with Gamma-driver
shift conditions) with an accurate hydrodynamical evolution employing
several high-order HRSC methods. The evolution of the spacetime and of
the hydrodynamics is coupled transparently through the method of lines,
which allows for the straightforward implementation of various different
time-integrators.

        As a first astrophysical problem for this novel setup, we have here focused on the collapse
of rapidly rotating relativistic stars to Kerr black holes. The stars are assumed to be in uniform
rotation and dynamically unstable to axisymmetric perturbations. While the collapse of slowly
rotating initial models proceeds with the matter remaining nearly uniformly rotating, the dynamics
is shown to be very different in the case of initial models rotating near the mass-shedding limit,
for which strong differential rotation develops.
 Although the stars become highly flattened during collapse, attaining a disc-like shape, the
collapse cannot be halted because the specific angular momentum is not sufficient for a stable disc
to form. Instead, the matter in the disc spirals towards the black hole and angular momentum is
transferred inward to produce a spinning black hole.

        Several different approaches have been employed to compute the mass and angular momentum of
the newly formed Kerr black hole. Besides more traditional methods involving the measure of the
geometrical properties of the apparent {\it and} event horizons, we have fitted the oscillations of
the perturbed Kerr black hole to specific quasi-normal modes obtained by linear perturbation theory.
In addition, we have also considered the recently proposed {\it isolated} and {\it dynamical}
horizon frameworks, finding it to be simple to implement and yielding estimates which are accurate
and more robust than those of other methods. This variety of approaches has allowed for the
determination of both the mass and angular momentum of the black hole with an accuracy unprecedented
for a 3D simulation. These measures, in turn, have allowed us to set upper limits on the energy and
angular momentum that could be lost during the collapse in the form of gravitational radiation.

        Work using mesh-refinement techniques is already in
progress to extract more precise information and waveforms for the
gravitational radiation recorded at large distances from the
collapsing stars. This aspect of the gravitational collapse has not
yet been considered in full 3D simulations and will be reported in a
forthcoming paper~\cite{Baiotti04}. Finally, all of the techniques
discussed here will also be applied to the study of the collapse of
differentially rotating stars, governed by more realistic and
non-isentropic EOSs. The expectation is that initial data with $J/M^2
\gtrsim 1 $ can be constructed in this case, whose collapse could lead
to the formation of a massive disc orbiting around the newly formed
Kerr black hole~\cite{Shibata03,Shibata03b,Duez04}.


\appendix

\section{Numerical methods}
\label{sec:numerical-methods}

        In this Appendix we focus on the numerical methods that the {\tt
Whisky} code incorporates for the solution of the general relativistic
hydrodynamics equations. The corresponding methods for the spacetime
equations are those implemented in the {\tt Cactus} code and they have
been reported elsewhere and the interested reader is addressed
to~\cite{Alcubierre99d,Alcubierre02a} for more details.

        As mentioned in the main text, our code uses high-resolution
shock-capturing methods based on reconstruction evolution
methods. In this approach, all variables ${\bf q}$ are represented on the
numerical grid by cell-integral averages. The function is then {\it
reconstructed} within each cell, usually by piecewise polynomials in a
way that preserves conservation of the variables ${\bf q}$. This gives
two values at each cell boundary which are then used to solve
(approximately) the Riemann problem, giving the flux through the cell
boundary. A Method of Lines approach is then used to update in
time. We will here give brief descriptions of each method, but further
details can be found in~\cite{Baiotti03a}.

\subsection{Time update: the method of lines}
\label{sec:method-lines}

        The reconstruction methods guarantee that a prescribed order of
accuracy is retained in space. However, the need to retain a high-order
accuracy also in time can complicate considerably the evolution from a
time-level to the following one. As a way to handle this efficiently, we
have chosen to follow a MoL approach~\cite{Laney98,Toro99}.  Here, the
continuum equations are considered to be discretized in space only. The
resulting system of ordinary differential equations (ODEs) can then be
solved numerically with any stable solver. This method minimizes the
coupling between the spacetime and hydrodynamics solvers and allows for a
transparent implementation of different evolution schemes.

        MoL itself does not have a precise truncation error but, rather, it acquires the truncation
order of the time-integrator employed.  Several integrators are available in our implementation of
MoL, including the second-order Iterative Crank Nicholson (ICN) solver and Runge-Kutta (RK) solvers
of first to fourth-order accuracy. The second and third-order RK solvers are known to be TVD whilst
the fourth-order is known to not be TVD~\cite{Shu88,Gottlieb98}. As the coupling between the
spacetime and the hydrodynamics is only second-order accurate, we typically use the ICN solver.

The calculation of the right hand side to feed to the ODE splits
into the following parts:

\begin{enumerate}
  
\item Calculation of the source terms ${\mathbf s}({\mathbf
q}(x^{(1)}_{j_1},x^{(2)}_{j_2},x^{(3)}_{j_3}))$ at all the grid points.
 
\item For each direction $x^{(i)}$:

\begin{itemize}

  \item Reconstruction of the data ${\mathbf q}$ to both sides of a cell
  boundary. In this way, two values ${\mathbf q}_{_L}$ and ${\mathbf
  q}_{_R}$ of ${\mathbf q}_{j_i+1/2}$ are determined at the cell
  boundary; ${\mathbf q}_{_L}$ is obtained from cell $j_i$ (left cell)
  and ${\mathbf q}_{_R}$ from cell $j_i+1$ (right cell)
  (see Appendix~\ref{sec:reconstr-meth} for more details).
    
  \item Solution at cell boundary of the approximate Riemann problem
  having the values ${\mathbf q}_{_{L,R}}$ as initial data
  (see Appendix~\ref{sec:riem-probl-solv} for more details).
 
  \item Calculation of the inter-cell flux ${\mathbf
  f}^{(x^{(i)})}({\mathbf q}_{j_i+1/2})$, that is, of the flux across the
  boundary between a cell ({\it e.g.,} the $j_i$-th) and its closest
  neighbour ({\it e.g.,} the $(j_i+1)$-th).

\end{itemize}

\item Recovery of the primitive variables and computation of the
stress-energy tensor for use in the Einstein equations.

\end{enumerate}

        As a result of steps 1.--3., the core of the {\tt Whisky} code is
effectively represented by two routines. One that reconstructs the
function ${\mathbf q}$ at the boundaries of a computational cell and
another one that calculates the inter-cell flux ${\mathbf f}$ at this
cell boundary.

\subsection{Reconstruction methods}
\label{sec:reconstr-meth}

        For the reconstruction procedure we have implemented several different approaches, including
slope-limited TVD methods, the Piecewise Parabolic Method ~\cite{Colella84} and the Essentially
Non-Oscillatory method~\cite{Harten87}.

        The TVD method uses limiters to avoid oscillations at shocks: we typically use the Van Leer
monotized centred method, although a variety of others (minmod, Superbee) have also been
implemented. This method is simple and computationally the least expensive to implement, but is at
most second-order accurate, dropping to first-order at local extrema.

        The PPM method of Colella and Woodward~\cite{Colella84} is a composite reconstruction method
that has special treatments for shocks, where the reconstruction is
modified to retain monotonicity, and contact surfaces, where the reconstruction is modified to
sharpen the jump. PPM contains a number of tunable parameters, but we always use those suggested
by~\cite{Colella84}. PPM is third-order accurate at most.

        The ENO methods have a large number of variants, with the common
theme that the ``least oscillatory'' stencil amongst all possible
stencils of a given order is chosen. In practice we use it in its
simplest form: direct piecewise polynomial reconstruction of the
variables, as described in~\cite{Shu99}. ENO has no tunable parameters
besides the order of accuracy, which may be arbitrary.

        All these methods are stable in the presence of shocks. By
default we use PPM as this seems to be the best balance between accuracy
and computational efficiency, as shown, for example, in~\cite{Font99},
and is our best choice for all the test evolutions we present in
Appendix~\ref{sec:numerical-tests}.

\subsection{Approximate Riemann problem solvers}
\label{sec:riem-probl-solv}

        Once a reconstruction procedure has provided data on either side
of each cell boundary, this is then used to specify the initial states of
the semi-infinite piecewise constant Riemann problems. Since the exact
solution of the Riemann problem~\cite{Pons00} is still too costly to use,
even when recast in an efficient form~\cite{Rezzolla01,Rezzolla03}, we
have here implemented three different approximate Riemann solvers.

        The first and simplest method implemented is the HLLE
method~\cite{Harten83}.  This approximates the solution by only two waves
with the intermediate state given by the conservation of the
mass-flux. This method is very efficient but diffusive. The second method
is the Roe solver~\cite{Roe81}. This solves a linearized problem at each
boundary, approximating every wave by either a shock or a contact
discontinuity. This method is less efficient but very accurate. However,
it may have problems near sonic points. The third method is the Marquina
solver~\cite{Donat96,Donat98}. This is similar to the Roe solver, except
that at possible sonic points a Lax-Friedrichs flux (analogous to the
HLLE method) is used, ensuring that the solution does not contain
rarefaction shocks. Note that we use the modified method
of~\cite{Aloy99b} instead of the original method.

      Both the Roe and Marquina solvers require the computation of the
eigenvalues and eigenvectors (from both the right and left cell) of
the linearized Jacobian matrices ${\mathbf A}_{_L}$ and ${\mathbf
A}_{_R}$ given by \mbox{${\mathbf f}_{_L} = {\mathbf A}_{_L}{\mathbf
q}_{_L}$} and \mbox{${\mathbf f}_{_R} = {\mathbf A}_{_R}{\mathbf
q}_{_R}$}. We use an implemention of the analytic expression for the
left eigenvectors~\cite{Ibanez01}, thus avoiding the computationally
expensive inversion of the three $5\times 5$ matrices of the right
eigenvectors, associated to each spatial direction. We also use a
compact version of the flux formula (a variant on the methods
described in~\cite{Aloy99a}) to increase speed and accuracy.  These
improvements bring a $\sim 40\%$ reduction of the computational time
spent in the solution of the hydrodynamics equations and a $\sim
5-10\%$ reduction in evolutions involving also the time integration of
the Einstein equations. The small overall gain in efficiency is due to
the fact that only around half of the computational time is spent
computing the update of the hydrodynamic variables, with the other
half spent in the update of the spacetime field variables. Of the time
spent computing the hydrodynamic update terms around one third is
spent solving the Riemann problem. These improvements were made by
Joachim Frieben at the Universidad de Valencia.

\subsection{Treatment of the atmosphere}
\label{sec:atm-treatment}

        At least mathematically, the region outside our initial stellar
models is assumed to be perfect vacuum. Independently of whether this
represents a physically realistic description of a compact star, the
vacuum represents a singular limit of the equations~\eqref{eq:consform1}
and must be treated artificially. We have here followed a standard
approach in computational fluid-dynamics and added a tenuous
``atmosphere'' filling the computational domain outside the star. This
approach, which was implemented also in~\cite{Font02c,Shibata02a}, was
instead not used in the simulations presented in~\cite{Duez:2002bn},
where new strategies were suggested. Unfortunately, these corrections
would not have been effective here because of the conservation form of
our hydrodynamical equations~\eqref{eq:consform1}, which we have used to
guarantee the correct evolution of shocks.

        We have treated the atmosphere as a perfect fluid governed by the
same polytropic EOS used for the bulk matter, but having a zero
coordinate velocity. Furthermore, the rest-mass density is several
(usually 7) orders of magnitude smaller than the initial central
density. Note that the atmosphere used for the calculation of the initial
data and the one evolved during the simulations need not be the
same. Indeed, for the initial stellar models used for the collapse
calculations we have typically set the atmosphere to be two orders of
magnitude smaller than the evolved one to minimise spurious matter
accretion onto the black hole. In the pulsation tests presented in
Appendix~\ref{sec:numerical-tests}, on the other hand, the initial and
evolved atmospheres are the same.

        The evolution of the hydrodynamical equations in gridzones where
the atmosphere is present is the same as the one used in the bulk of the
flow. Furthermore, when the rest mass in a gridzone falls below the
threshold set by the atmosphere, that gridzone is simply not updated in
time.

        As mentioned in Section~\ref{sec:d4}, the use of a tenuous
atmosphere has no dynamical impact and does not produce any increase of
the mass of the black hole that can be appreciated in our
simulations. With the rest-mass densities used here for the atmosphere,
in fact, and using the mass accretion rates measured once the apparent
surface is first found, we have estimated that a net increase of $\sim
1\%$ in the black-hole mass would require an integration time of $\sim
10^4 M$. These systematic errors are well below our truncation errors,
even at the highest resolutions we can afford.

\section{Numerical tests}
\label{sec:numerical-tests}

        Several tests have been performed to assess the stability and
accuracy of our code, some of which have also provided new interesting
physical results. Here, we briefly discuss some of the most
representative of these tests and results, referring the interested
reader to~\cite{Baiotti03a} for more details.

\begin{figure}
\centering
\includegraphics[angle=0,width=8.5cm]{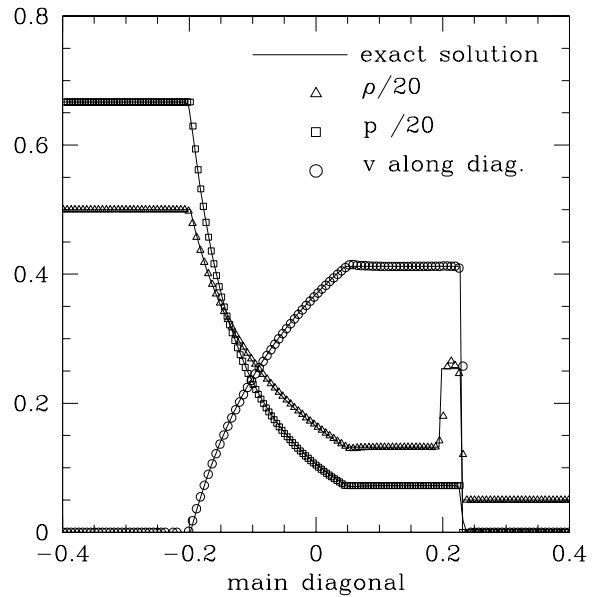}
\caption{Solution of a Riemann problem set on the main diagonal of the
cubic grid. The figure shows the comparison of the hydrodynamical
variables evolved by {\tt Whisky}, indicated with symbols, with the
exact solution. The numerical simulation was obtained with the van
Leer reconstruction method and the Roe solver, on a grid with $140^3$ points.}
\label{figShock} 
\end{figure}

        First of all, we consider a standard shock-tube test, setting as
initial data a global Riemann problem, in particular one in which the
initial discontinuity is orthogonal to the main diagonal of our cubic
grid. More precisely the initial data consists of a left ($L$) and right
($R$) states with values given respectively by
\begin{alignat}{5}
\rho_R &=& 1;\;\; p_R &= 1.666 \times 10^{-6};\;\; &v_R &= 0\ ,
\nonumber\\
\rho_L &=& 10;\;\; p_L &= 13.333; \;\;&v_L &= 0\ .
\nonumber 
\end{alignat}
In Fig.~\ref{figShock} we show the solution at a given time together with
the exact solution. The excellent agreement of the two sets of curves is
particularly remarkable if one bears in mind that the evolution is fully
3D and not simply along a coordinate axis.

\begin{figure}
\centering
\includegraphics[angle=0,width=8.5cm]{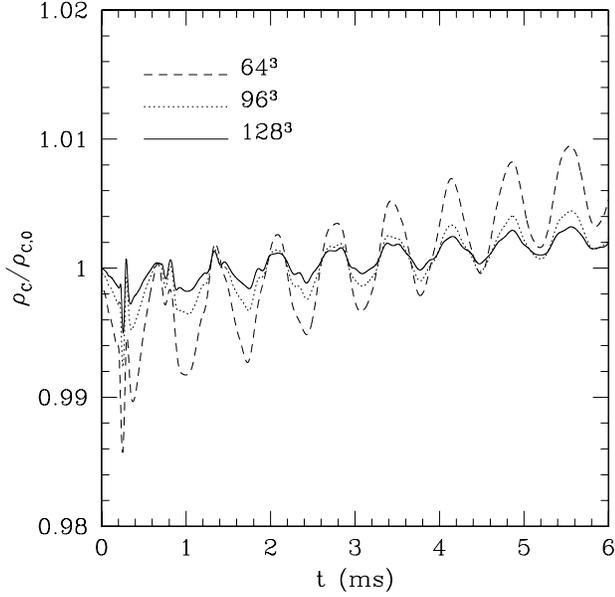}
\caption{Central mass-density, normalized to the initial value, in a
stable TOV star (\mbox{$ M = 1.4\, M_\odot $} and polytropic exponent
$ \Gamma = 2 $) evolution at different resolutions. PPM and Marquina
were used for all runs.}  \label{figTOVdyn rho}
\end{figure}

        Next, we consider the evolution of a stable relativistic
polytropic spherical (TOV) star. As this is a static solution, no
evolution is expected. Yet as shown in Fig.~\ref{figTOVdyn rho}, both a
small periodic oscillation and a small secular increase of the central
density of the star are detected during the numerical evolution of the
equations. Both effects have, however, a single explanation. Since the
initial data contains also a small truncation error, this is responsible
for triggering radial oscillations which appear as periodic variations in
the central density. As the resolution is increased, the truncation error
is reduced and so is the amplitude of the oscillation. The secular
growth, on the other hand, is a purely numerical problem, probably
related to the violation of the constraint equations. As for the
oscillation amplitude, also the secular growth converges to zero with
increasing resolution. Note that the convergence rate is not exactly
second-order but slightly smaller, because the reconstruction schemes are
only first-order accurate at local extrema ({\it i.e.}  the centre and
the surface of the star) thus increasing the overall truncation
error~\cite{alcubierre00b}.

        In order to further investigate the accuracy of our
implementation of the hydrodynamics equations, we have suppressed the
spacetime evolution and solved just the hydrodynamics equations in the
fixed spacetime of the initial TOV solution. This approximation is
referred to as the ``Cowling approximation'' and is widely used in
perturbative studies of oscillating stars. In this case, in addition to
the confirmation of the convergence rate already checked in fully evolved
runs, we have also compared the frequency spectrum of the numerically
induced oscillation with the results obtained by an independent 2D
code~\cite{Font99} and with perturbative analyses. 

\begin{figure}
\centering
\includegraphics[angle=0,width=8.5cm]{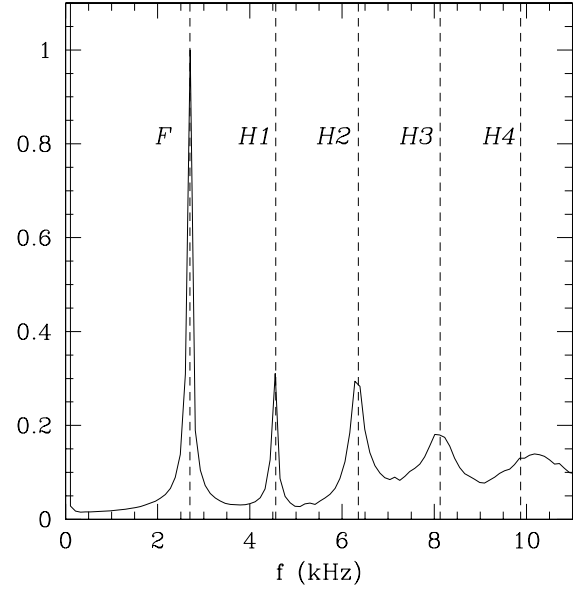}       
\caption{Power spectrum of the central mass-density evolution of an $
M = 1.4\, M_\odot $, $ \Gamma = 2 $ stable TOV star performed with $
128^3 $ grid points. The units of the vertical axis are arbitrary.}
\label{figFreq}
\end{figure}

        In Fig.~\ref{figFreq} we show a comparison between the two codes
reporting the power spectrum of the central density oscillations computed
with the {\tt Whisky} code and the corresponding frequencies as obtained
with perturbative techniques and with the 2D code. Clearly the agreement
is very good with an error below 1\% in the fundamental frequency. The
fact that the frequencies computed with the code coincide with the
physical eigenfrequencies calculated through perturbative analysis allows
us to study with our code the physical properties of linear normal-modes
of oscillation even if such oscillations are generated numerically.

\begin{figure}
\centering
\includegraphics[angle=0,width=8.5cm]{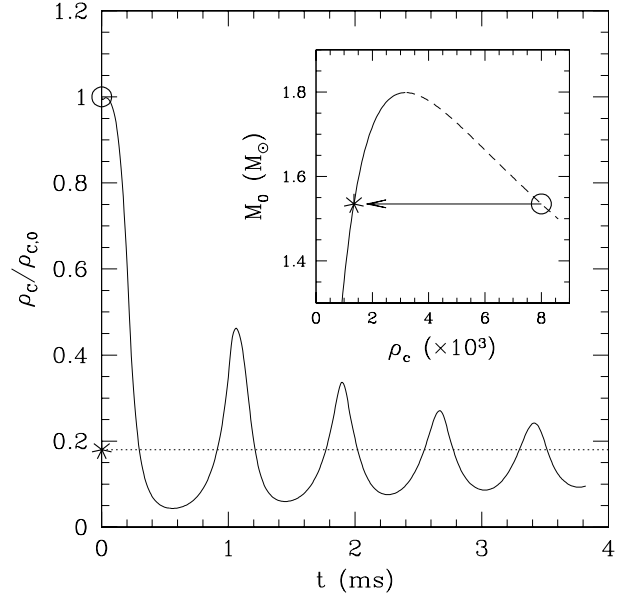}
\caption{Normalized central mass-density evolution of an $ M = 1.4\,
M_\odot $, $ \Gamma = 2 $ unstable TOV star performed with $ 96^3 $ grid
points.}
\label{figMigration}
\end{figure}

        The last test performed in the linear regime consists of the
evolution of stationary solutions of rapidly rotating stars, with angular
velocity up to 95\% of the allowed mass-shedding limit for uniformly
rotating stars. In this case, a number of small improvements on the
boundary and gauge conditions have allowed us to extend considerably the
timescale of our evolutions of stable rapidly rotating stars, which can
now be evolved for about 10 ms, a timescale which is 3 times larger than
the one previously reported in~\cite{Font02c}. In analogy with the
non-rotating case, the truncation error triggers quasi-radial oscillations
in the star. Such pulsations converge to zero with increasing
resolution. Determining the frequency spectrum of fully relativistic and
rapidly rotating stars is an important achievement, allowing the
investigation of a parameter space which is astrophysically relevant but
too difficult to treat with current perturbative techniques. More details
on this investigation will be presented in a separate paper.

        Finally, we have considered tests of the non-linear dynamics of
isolated spherical relativistic stars. To this purpose we have
constructed TOV solutions that are placed on the unstable branch of
the equilibrium configurations (see inset of
Fig.~\ref{figMigration}). The truncation error in the initial data for
a TOV solution is sufficient to move the model to a different
configuration and in {\tt Whisky} this leads to a rapid migration
toward a stable configuration of equal rest mass but smaller central
density. Such a violent expansion produces large amplitude radial
oscillations in the star that are either at constant amplitude, if the
polytropic EOS (\ref{poly}) is used, or are damped through shock
heating, if the ideal fluid EOS (\ref{id fluid}) is used and the
equation for $\tau$ is evolved in time. A summary of this dynamics is
presented in Fig.~\ref{figMigration}, which shows the time series of
the normalized central density for a TOV solution. Note that the
asymptotic central density tends to a value corresponding to a
rest mass slightly smaller than the initial one (straight dotted
line). This is the energy loss due to the internal dissipation.


\acknowledgments 

        It is a pleasure to thank Joachim Frieben, {Jos\'e
        M$^{\underline{\mbox{a}}}$. Ib\'a\~{n}ez} and Andrea Nerozzi
        who have participated to the development and testing of the
        code. We are also very grateful to Peter Diener, Jonathan
        Thornburg, Erik Schnetter, Badri Krishnan and Frank Herrmann
        for their help in the study of the trapped surfaces, to Abhay
        Ashtekar for clarifying some aspects of the dynamical-horizon
        framework, and to John Miller and K\=oji Ury\=u for carefully
        reading the manuscript.  We are also grateful to the {\tt
        Cactus}-code team, for their efforts in producing an efficient
        infrastructure for numerical relativity.  Financial support
        for this work has been provided by the MIUR and EU Network
        Programme (Research Training Network contract
        HPRN-CT-2000-00137).  J.A.F.  acknowledges financial support
        from the Spanish Ministerio de Ciencia y Tecnolog\'{\i}a
        (grant AYA 2001-3490-C02-01).  The computations were performed
        on the {\it Albert100} cluster at the University of Parma and
        the {\em Peyote} cluster at the Albert Einstein Institute.


\bibliographystyle{apsrev}

\bibliography{references}

\begin{thebibliography}{122}
\expandafter\ifx\csname natexlab\endcsname\relax\def\natexlab#1{#1}\fi
\expandafter\ifx\csname bibnamefont\endcsname\relax
  \def\bibnamefont#1{#1}\fi
\expandafter\ifx\csname bibfnamefont\endcsname\relax
  \def\bibfnamefont#1{#1}\fi
\expandafter\ifx\csname citenamefont\endcsname\relax
  \def\citenamefont#1{#1}\fi
\expandafter\ifx\csname url\endcsname\relax
  \def\url#1{\texttt{#1}}\fi
\expandafter\ifx\csname urlprefix\endcsname\relax\def\urlprefix{URL }\fi
\providecommand{\bibinfo}[2]{#2}
\providecommand{\eprint}[2][]{\url{#2}}

\bibitem[{\citenamefont{Cook et~al.}(1998)\citenamefont{Cook, Huq, Klasky,
  Scheel, Abrahams, Anderson, Anninos, Baumgarte, Bishop, Brandt
  et~al.}}]{Cook97a}
\bibinfo{author}{\bibfnamefont{G.~B.} \bibnamefont{Cook}},
  \bibinfo{author}{\bibfnamefont{M.~F.} \bibnamefont{Huq}},
  \bibinfo{author}{\bibfnamefont{S.~A.} \bibnamefont{Klasky}},
  \bibinfo{author}{\bibfnamefont{M.~A.} \bibnamefont{Scheel}},
  \bibinfo{author}{\bibfnamefont{A.~M.} \bibnamefont{Abrahams}},
  \bibinfo{author}{\bibfnamefont{A.}~\bibnamefont{Anderson}},
  \bibinfo{author}{\bibfnamefont{P.}~\bibnamefont{Anninos}},
  \bibinfo{author}{\bibfnamefont{T.~W.} \bibnamefont{Baumgarte}},
  \bibinfo{author}{\bibfnamefont{N.}~\bibnamefont{Bishop}},
  \bibinfo{author}{\bibfnamefont{S.~R.} \bibnamefont{Brandt}},
  \bibnamefont{et~al.}, \bibinfo{journal}{Phys. Rev. Lett}
  \textbf{\bibinfo{volume}{80}}, \bibinfo{pages}{2512} (\bibinfo{year}{1998}).

\bibitem[{\citenamefont{Abrahams et~al.}(1998)\citenamefont{Abrahams, Rezzolla,
  Rupright, Anderson, Anninos, Baumgarte, Bishop, Brandt, Browne, Camarda
  et~al.}}]{Abrahams97a}
\bibinfo{author}{\bibfnamefont{A.~M.} \bibnamefont{Abrahams}},
  \bibinfo{author}{\bibfnamefont{L.}~\bibnamefont{Rezzolla}},
  \bibinfo{author}{\bibfnamefont{M.~E.} \bibnamefont{Rupright}},
  \bibinfo{author}{\bibfnamefont{A.}~\bibnamefont{Anderson}},
  \bibinfo{author}{\bibfnamefont{P.}~\bibnamefont{Anninos}},
  \bibinfo{author}{\bibfnamefont{T.~W.} \bibnamefont{Baumgarte}},
  \bibinfo{author}{\bibfnamefont{N.~T.} \bibnamefont{Bishop}},
  \bibinfo{author}{\bibfnamefont{S.~R.} \bibnamefont{Brandt}},
  \bibinfo{author}{\bibfnamefont{J.~C.} \bibnamefont{Browne}},
  \bibinfo{author}{\bibfnamefont{K.}~\bibnamefont{Camarda}},
  \bibnamefont{et~al.}, \bibinfo{journal}{Phys. Rev. Lett.}
  \textbf{\bibinfo{volume}{80}}, \bibinfo{pages}{1812} (\bibinfo{year}{1998}).

\bibitem[{\citenamefont{G\'omez et~al.}(1998)\citenamefont{G\'omez, Lehner,
  Marsa, Winicour, Abrahams, Anderson, Anninos, Baumgarte, Bishop, Brandt
  et~al.}}]{Gomez98a}
\bibinfo{author}{\bibfnamefont{R.}~\bibnamefont{G\'omez}},
  \bibinfo{author}{\bibfnamefont{L.}~\bibnamefont{Lehner}},
  \bibinfo{author}{\bibfnamefont{R.}~\bibnamefont{Marsa}},
  \bibinfo{author}{\bibfnamefont{J.}~\bibnamefont{Winicour}},
  \bibinfo{author}{\bibfnamefont{A.~M.} \bibnamefont{Abrahams}},
  \bibinfo{author}{\bibfnamefont{A.}~\bibnamefont{Anderson}},
  \bibinfo{author}{\bibfnamefont{P.}~\bibnamefont{Anninos}},
  \bibinfo{author}{\bibfnamefont{T.~W.} \bibnamefont{Baumgarte}},
  \bibinfo{author}{\bibfnamefont{N.~T.} \bibnamefont{Bishop}},
  \bibinfo{author}{\bibfnamefont{S.~R.} \bibnamefont{Brandt}},
  \bibnamefont{et~al.}, \bibinfo{journal}{Phys. Rev. Lett.}
  \textbf{\bibinfo{volume}{80}}, \bibinfo{pages}{3915} (\bibinfo{year}{1998}).

\bibitem[{\citenamefont{Anderson and Matzner}(2004)}]{AM:03}
\bibinfo{author}{\bibfnamefont{M.}~\bibnamefont{Anderson}} \bibnamefont{and}
  \bibinfo{author}{\bibfnamefont{R.~A.} \bibnamefont{Matzner}},
  \bibinfo{journal}{submitted to Phys. Rev. D}  (\bibinfo{year}{2004}),
  \eprint{gr-qc/0307055}.

\bibitem[{\citenamefont{Schnetter}(2003)}]{Schnetter03c}
\bibinfo{author}{\bibfnamefont{E.}~\bibnamefont{Schnetter}}, Ph.D. thesis,
  \bibinfo{school}{Universit{\"a}t T{\"u}bingen},
  \bibinfo{address}{T{\"u}bingen, Germany} (\bibinfo{year}{2003}),
  \bibinfo{note}{{URN}: urn:nbn:de:bsz:21-opus-8191}, \eprint{{URL}:
  http://w210.ub.uni-tuebingen.de/dbt/volltexte/2003/819/}.

\bibitem[{\citenamefont{Bonazzola et~al.}(2004)\citenamefont{Bonazzola,
  Gourgoulhon, Grandcl{\'e}ment, and Novak}}]{Bonazzola03a}
\bibinfo{author}{\bibfnamefont{S.}~\bibnamefont{Bonazzola}},
  \bibinfo{author}{\bibfnamefont{E.}~\bibnamefont{Gourgoulhon}},
  \bibinfo{author}{\bibfnamefont{P.}~\bibnamefont{Grandcl{\'e}ment}},
  \bibnamefont{and} \bibinfo{author}{\bibfnamefont{J.}~\bibnamefont{Novak}},
  \bibinfo{journal}{submitted to Phys. Rev. D}  (\bibinfo{year}{2004}),
  \bibinfo{note}{gr-qc/0307082}.

\bibitem[{\citenamefont{Lehner}(2001)}]{Lehner01a}
\bibinfo{author}{\bibfnamefont{L.}~\bibnamefont{Lehner}},
  \bibinfo{journal}{Class. Quantum Grav.} \textbf{\bibinfo{volume}{18}},
  \bibinfo{pages}{R25} (\bibinfo{year}{2001}).

\bibitem[{\citenamefont{Nakamura et~al.}(1987)\citenamefont{Nakamura, Oohara,
  and Kojima}}]{Nakamura87}
\bibinfo{author}{\bibfnamefont{T.}~\bibnamefont{Nakamura}},
  \bibinfo{author}{\bibfnamefont{K.}~\bibnamefont{Oohara}}, \bibnamefont{and}
  \bibinfo{author}{\bibfnamefont{Y.}~\bibnamefont{Kojima}},
  \bibinfo{journal}{Prog. Theor. Phys. Suppl.} \textbf{\bibinfo{volume}{90}},
  \bibinfo{pages}{1} (\bibinfo{year}{1987}).

\bibitem[{\citenamefont{Shibata and Nakamura}(1995)}]{Shibata95}
\bibinfo{author}{\bibfnamefont{M.}~\bibnamefont{Shibata}} \bibnamefont{and}
  \bibinfo{author}{\bibfnamefont{T.}~\bibnamefont{Nakamura}},
  \bibinfo{journal}{Phys. Rev. D} \textbf{\bibinfo{volume}{52}},
  \bibinfo{pages}{5428} (\bibinfo{year}{1995}).

\bibitem[{\citenamefont{Baumgarte and Shapiro}(1999)}]{Baumgarte99}
\bibinfo{author}{\bibfnamefont{T.~W.} \bibnamefont{Baumgarte}}
  \bibnamefont{and} \bibinfo{author}{\bibfnamefont{S.~L.}
  \bibnamefont{Shapiro}}, \bibinfo{journal}{Physical Review D}
  \textbf{\bibinfo{volume}{59}}, \bibinfo{pages}{024007}
  (\bibinfo{year}{1999}).

\bibitem[{\citenamefont{Shibata et~al.}(2000)\citenamefont{Shibata, Baumgarte,
  and Shapiro}}]{Shibata99e}
\bibinfo{author}{\bibfnamefont{M.}~\bibnamefont{Shibata}},
  \bibinfo{author}{\bibfnamefont{T.~W.} \bibnamefont{Baumgarte}},
  \bibnamefont{and} \bibinfo{author}{\bibfnamefont{S.~L.}
  \bibnamefont{Shapiro}}, \bibinfo{journal}{Phys. Rev. D}
  \textbf{\bibinfo{volume}{61}}, \bibinfo{pages}{044012}
  (\bibinfo{year}{2000}).

\bibitem[{\citenamefont{Shibata}(2003{\natexlab{a}})}]{Shibata03}
\bibinfo{author}{\bibfnamefont{M.}~\bibnamefont{Shibata}},
  \bibinfo{journal}{Phys. Rev. D} \textbf{\bibinfo{volume}{67}},
  \bibinfo{pages}{024033} (\bibinfo{year}{2003}{\natexlab{a}}).

\bibitem[{\citenamefont{Alcubierre
  et~al.}(2000{\natexlab{a}})\citenamefont{Alcubierre, Br\"{u}gmann,
  Dramlitsch, Font, Papadopoulos, Seidel, Stergioulas, and
  Takahashi}}]{Alcubierre99d}
\bibinfo{author}{\bibfnamefont{M.}~\bibnamefont{Alcubierre}},
  \bibinfo{author}{\bibfnamefont{B.}~\bibnamefont{Br\"{u}gmann}},
  \bibinfo{author}{\bibfnamefont{T.}~\bibnamefont{Dramlitsch}},
  \bibinfo{author}{\bibfnamefont{J.}~\bibnamefont{Font}},
  \bibinfo{author}{\bibfnamefont{P.}~\bibnamefont{Papadopoulos}},
  \bibinfo{author}{\bibfnamefont{E.}~\bibnamefont{Seidel}},
  \bibinfo{author}{\bibfnamefont{N.}~\bibnamefont{Stergioulas}},
  \bibnamefont{and}
  \bibinfo{author}{\bibfnamefont{R.}~\bibnamefont{Takahashi}},
  \bibinfo{journal}{Phys. Rev. D} \textbf{\bibinfo{volume}{62}},
  \bibinfo{pages}{044034} (\bibinfo{year}{2000}{\natexlab{a}}).

\bibitem[{\citenamefont{Font et~al.}(2002)\citenamefont{Font, Goodale, Iyer,
  Miller, Rezzolla, Seidel, Stergioulas, Suen, and Tobias}}]{Font02c}
\bibinfo{author}{\bibfnamefont{J.~A.} \bibnamefont{Font}},
  \bibinfo{author}{\bibfnamefont{T.}~\bibnamefont{Goodale}},
  \bibinfo{author}{\bibfnamefont{S.}~\bibnamefont{Iyer}},
  \bibinfo{author}{\bibfnamefont{M.}~\bibnamefont{Miller}},
  \bibinfo{author}{\bibfnamefont{L.}~\bibnamefont{Rezzolla}},
  \bibinfo{author}{\bibfnamefont{E.}~\bibnamefont{Seidel}},
  \bibinfo{author}{\bibfnamefont{N.}~\bibnamefont{Stergioulas}},
  \bibinfo{author}{\bibfnamefont{W.~M.} \bibnamefont{Suen}}, \bibnamefont{and}
  \bibinfo{author}{\bibfnamefont{M.}~\bibnamefont{Tobias}},
  \bibinfo{journal}{Phys. Rev. D} \textbf{\bibinfo{volume}{65}},
  \bibinfo{pages}{084024} (\bibinfo{year}{2002}).

\bibitem[{\citenamefont{Yoneda and Shinkai}(2002)}]{Yoneda02a}
\bibinfo{author}{\bibfnamefont{G.}~\bibnamefont{Yoneda}} \bibnamefont{and}
  \bibinfo{author}{\bibfnamefont{H.}~\bibnamefont{Shinkai}},
  \bibinfo{journal}{Phys. Rev. D} \textbf{\bibinfo{volume}{66}},
  \bibinfo{pages}{124003} (\bibinfo{year}{2002}).

\bibitem[{\citenamefont{Kawamura et~al.}(2003)\citenamefont{Kawamura, Oohara,
  and Nakamura}}]{Kawamura03}
\bibinfo{author}{\bibfnamefont{M.}~\bibnamefont{Kawamura}},
  \bibinfo{author}{\bibfnamefont{K.}~\bibnamefont{Oohara}}, \bibnamefont{and}
  \bibinfo{author}{\bibfnamefont{T.}~\bibnamefont{Nakamura}},
  \bibinfo{journal}{submitted to Prog. Theor. Phys.}  (\bibinfo{year}{2003}),
  \eprint{astro-ph/0306481}.

\bibitem[{\citenamefont{Seidel and Suen}(1992)}]{Seidel92a}
\bibinfo{author}{\bibfnamefont{E.}~\bibnamefont{Seidel}} \bibnamefont{and}
  \bibinfo{author}{\bibfnamefont{W.-M.} \bibnamefont{Suen}},
  \bibinfo{journal}{Phys. Rev. Lett.} \textbf{\bibinfo{volume}{69}},
  \bibinfo{pages}{1845} (\bibinfo{year}{1992}).

\bibitem[{\citenamefont{Brandt et~al.}(2000{\natexlab{a}})\citenamefont{Brandt,
  Correll, G\'omez, Huq, Laguna, Lehner, Marronetti, Matzner, Neilsen, Pullin
  et~al.}}]{Brandt00}
\bibinfo{author}{\bibfnamefont{S.}~\bibnamefont{Brandt}},
  \bibinfo{author}{\bibfnamefont{R.}~\bibnamefont{Correll}},
  \bibinfo{author}{\bibfnamefont{R.}~\bibnamefont{G\'omez}},
  \bibinfo{author}{\bibfnamefont{M.~F.} \bibnamefont{Huq}},
  \bibinfo{author}{\bibfnamefont{P.}~\bibnamefont{Laguna}},
  \bibinfo{author}{\bibfnamefont{L.}~\bibnamefont{Lehner}},
  \bibinfo{author}{\bibfnamefont{P.}~\bibnamefont{Marronetti}},
  \bibinfo{author}{\bibfnamefont{R.~A.} \bibnamefont{Matzner}},
  \bibinfo{author}{\bibfnamefont{D.}~\bibnamefont{Neilsen}},
  \bibinfo{author}{\bibfnamefont{J.}~\bibnamefont{Pullin}},
  \bibnamefont{et~al.}, \bibinfo{journal}{Phys. Rev. Lett.}
  \textbf{\bibinfo{volume}{85}}, \bibinfo{pages}{5496}
  (\bibinfo{year}{2000}{\natexlab{a}}).

\bibitem[{\citenamefont{Alcubierre and Br\"ugmann}(2001)}]{Alcubierre00a}
\bibinfo{author}{\bibfnamefont{M.}~\bibnamefont{Alcubierre}} \bibnamefont{and}
  \bibinfo{author}{\bibfnamefont{B.}~\bibnamefont{Br\"ugmann}},
  \bibinfo{journal}{Phys. Rev. D} \textbf{\bibinfo{volume}{63}},
  \bibinfo{pages}{104006} (\bibinfo{year}{2001}).

\bibitem[{\citenamefont{Kidder et~al.}(2000{\natexlab{a}})\citenamefont{Kidder,
  Scheel, {T}eukolsky, Carlson, and Cook}}]{Kidder00a}
\bibinfo{author}{\bibfnamefont{L.~E.} \bibnamefont{Kidder}},
  \bibinfo{author}{\bibfnamefont{M.~A.} \bibnamefont{Scheel}},
  \bibinfo{author}{\bibfnamefont{S.~A.} \bibnamefont{{T}eukolsky}},
  \bibinfo{author}{\bibfnamefont{E.~D.} \bibnamefont{Carlson}},
  \bibnamefont{and} \bibinfo{author}{\bibfnamefont{G.~B.} \bibnamefont{Cook}},
  \bibinfo{journal}{Phys. Rev. D} \textbf{\bibinfo{volume}{62}},
  \bibinfo{pages}{084032} (\bibinfo{year}{2000}{\natexlab{a}}).

\bibitem[{\citenamefont{Kidder et~al.}(2001)\citenamefont{Kidder, Scheel, and
  {T}eukolsky}}]{Kidder01a}
\bibinfo{author}{\bibfnamefont{L.~E.} \bibnamefont{Kidder}},
  \bibinfo{author}{\bibfnamefont{M.~A.} \bibnamefont{Scheel}},
  \bibnamefont{and} \bibinfo{author}{\bibfnamefont{S.~A.}
  \bibnamefont{{T}eukolsky}}, \bibinfo{journal}{Phys. Rev. D}
  \textbf{\bibinfo{volume}{64}}, \bibinfo{pages}{064017}
  (\bibinfo{year}{2001}).

\bibitem[{\citenamefont{Alcubierre
  et~al.}(2001{\natexlab{a}})\citenamefont{Alcubierre, Br\"ugmann, Pollney,
  Seidel, and Takahashi}}]{Alcubierre01a}
\bibinfo{author}{\bibfnamefont{M.}~\bibnamefont{Alcubierre}},
  \bibinfo{author}{\bibfnamefont{B.}~\bibnamefont{Br\"ugmann}},
  \bibinfo{author}{\bibfnamefont{D.}~\bibnamefont{Pollney}},
  \bibinfo{author}{\bibfnamefont{E.}~\bibnamefont{Seidel}}, \bibnamefont{and}
  \bibinfo{author}{\bibfnamefont{R.}~\bibnamefont{Takahashi}},
  \bibinfo{journal}{Phys. Rev. D} \textbf{\bibinfo{volume}{64}},
  \bibinfo{pages}{61501 (R)} (\bibinfo{year}{2001}{\natexlab{a}}).

\bibitem[{\citenamefont{Yo et~al.}(2002)\citenamefont{Yo, Baumgarte, and
  Shapiro}}]{Yo02a}
\bibinfo{author}{\bibfnamefont{H.-J.} \bibnamefont{Yo}},
  \bibinfo{author}{\bibfnamefont{T.}~\bibnamefont{Baumgarte}},
  \bibnamefont{and} \bibinfo{author}{\bibfnamefont{S.}~\bibnamefont{Shapiro}},
  \bibinfo{journal}{Phys. Rev. D} \textbf{\bibinfo{volume}{66}},
  \bibinfo{pages}{084026} (\bibinfo{year}{2002}).

\bibitem[{\citenamefont{Laguna and Shoemaker}(2002)}]{Laguna02}
\bibinfo{author}{\bibfnamefont{P.}~\bibnamefont{Laguna}} \bibnamefont{and}
  \bibinfo{author}{\bibfnamefont{D.}~\bibnamefont{Shoemaker}},
  \bibinfo{journal}{Class. Quantum Grav.} \textbf{\bibinfo{volume}{19}},
  \bibinfo{pages}{3679} (\bibinfo{year}{2002}).

\bibitem[{\citenamefont{Choptuik et~al.}(2003)\citenamefont{Choptuik,
  Hirschmann, Liebling, and Pretorius}}]{Choptuik:2003ac}
\bibinfo{author}{\bibfnamefont{M.~W.} \bibnamefont{Choptuik}},
  \bibinfo{author}{\bibfnamefont{E.~W.} \bibnamefont{Hirschmann}},
  \bibinfo{author}{\bibfnamefont{S.~L.} \bibnamefont{Liebling}},
  \bibnamefont{and}
  \bibinfo{author}{\bibfnamefont{F.}~\bibnamefont{Pretorius}},
  \bibinfo{journal}{Phys. Rev.} \textbf{\bibinfo{volume}{D68}},
  \bibinfo{pages}{044007} (\bibinfo{year}{2003}).

\bibitem[{\citenamefont{Sperhake et~al.}(2003)\citenamefont{Sperhake, Smith,
  Kelly, Laguna, and Shoemaker}}]{Sperhake:2003fc}
\bibinfo{author}{\bibfnamefont{U.}~\bibnamefont{Sperhake}},
  \bibinfo{author}{\bibfnamefont{K.~L.} \bibnamefont{Smith}},
  \bibinfo{author}{\bibfnamefont{B.}~\bibnamefont{Kelly}},
  \bibinfo{author}{\bibfnamefont{P.}~\bibnamefont{Laguna}}, \bibnamefont{and}
  \bibinfo{author}{\bibfnamefont{D.}~\bibnamefont{Shoemaker}}
  (\bibinfo{year}{2003}), \eprint{gr-qc/0307015}.

\bibitem[{\citenamefont{May and White}(1967)}]{MayWhite67}
\bibinfo{author}{\bibfnamefont{M.~M.} \bibnamefont{May}} \bibnamefont{and}
  \bibinfo{author}{\bibfnamefont{R.~H.} \bibnamefont{White}}, in
  \emph{\bibinfo{booktitle}{Methods in Computational Physics}}, edited by
  \bibinfo{editor}{\bibfnamefont{B.}~\bibnamefont{Alder}},
  \bibinfo{editor}{\bibfnamefont{F.}~\bibnamefont{S.}}, \bibnamefont{and}
  \bibinfo{editor}{\bibfnamefont{R.}~\bibnamefont{M.}}
  (\bibinfo{publisher}{Academic Press}, \bibinfo{address}{London},
  \bibinfo{year}{1967}), vol.~\bibinfo{volume}{7}, p. \bibinfo{pages}{129}.

\bibitem[{\citenamefont{Duez et~al.}(2004)\citenamefont{Duez, Shapiro, and
  Yo}}]{Duez04}
\bibinfo{author}{\bibfnamefont{M.~D.} \bibnamefont{Duez}},
  \bibinfo{author}{\bibfnamefont{S.~L.} \bibnamefont{Shapiro}},
  \bibnamefont{and} \bibinfo{author}{\bibfnamefont{H.-J.} \bibnamefont{Yo}},
  \bibinfo{journal}{Phys. Rev. D} \textbf{\bibinfo{volume}{69}},
  \bibinfo{pages}{104016} (\bibinfo{year}{2004}).

\bibitem[{\citenamefont{Nakamura and Sasaki}(1981)}]{Nakamura81b}
\bibinfo{author}{\bibfnamefont{T.}~\bibnamefont{Nakamura}} \bibnamefont{and}
  \bibinfo{author}{\bibfnamefont{M.}~\bibnamefont{Sasaki}},
  \bibinfo{journal}{Phys. Lett.} \textbf{\bibinfo{volume}{106 B}},
  \bibinfo{pages}{69} (\bibinfo{year}{1981}).

\bibitem[{\citenamefont{Nakamura}(1983)}]{Nakamura83}
\bibinfo{author}{\bibfnamefont{T.}~\bibnamefont{Nakamura}},
  \bibinfo{journal}{Prog. Theor. Phys.} \textbf{\bibinfo{volume}{70}},
  \bibinfo{pages}{1144} (\bibinfo{year}{1983}).

\bibitem[{\citenamefont{Bardeen and Piran}(1983)}]{Bardeen83}
\bibinfo{author}{\bibfnamefont{J.~M.} \bibnamefont{Bardeen}} \bibnamefont{and}
  \bibinfo{author}{\bibfnamefont{T.}~\bibnamefont{Piran}},
  \bibinfo{journal}{Phys. Reports} \textbf{\bibinfo{volume}{96}},
  \bibinfo{pages}{205} (\bibinfo{year}{1983}).

\bibitem[{\citenamefont{Stark and Piran}(1985)}]{Stark85}
\bibinfo{author}{\bibfnamefont{R.~F.} \bibnamefont{Stark}} \bibnamefont{and}
  \bibinfo{author}{\bibfnamefont{T.}~\bibnamefont{Piran}},
  \bibinfo{journal}{Phys. Rev. Lett.} \textbf{\bibinfo{volume}{55}},
  \bibinfo{pages}{891} (\bibinfo{year}{1985}).

\bibitem[{\citenamefont{Stark and Piran}(1986)}]{stark86}
\bibinfo{author}{\bibfnamefont{R.~F.} \bibnamefont{Stark}} \bibnamefont{and}
  \bibinfo{author}{\bibfnamefont{T.}~\bibnamefont{Piran}}, in
  \emph{\bibinfo{booktitle}{Proceedings of the Fourth Marcell Grossman Meeting
  on General Relativity, Rome, 1985}}, edited by
  \bibinfo{editor}{\bibfnamefont{R.}~\bibnamefont{Ruffini}}
  (\bibinfo{publisher}{Elsevier Science Publisher}, \bibinfo{year}{1986}), p.
  \bibinfo{pages}{327}.

\bibitem[{\citenamefont{Stark and Piran}(1987)}]{stark87}
\bibinfo{author}{\bibfnamefont{R.~F.} \bibnamefont{Stark}} \bibnamefont{and}
  \bibinfo{author}{\bibfnamefont{T.}~\bibnamefont{Piran}},
  \bibinfo{journal}{Comp. Phys. Rep.} \textbf{\bibinfo{volume}{5}},
  \bibinfo{pages}{221} (\bibinfo{year}{1987}).

\bibitem[{\citenamefont{Alcubierre
  et~al.}(2001{\natexlab{b}})\citenamefont{Alcubierre, Benger, Br\"ugmann,
  Lanfermann, Nerger, Seidel, and Takahashi}}]{alcubierre00b}
\bibinfo{author}{\bibfnamefont{M.}~\bibnamefont{Alcubierre}},
  \bibinfo{author}{\bibfnamefont{W.}~\bibnamefont{Benger}},
  \bibinfo{author}{\bibfnamefont{B.}~\bibnamefont{Br\"ugmann}},
  \bibinfo{author}{\bibfnamefont{G.}~\bibnamefont{Lanfermann}},
  \bibinfo{author}{\bibfnamefont{L.}~\bibnamefont{Nerger}},
  \bibinfo{author}{\bibfnamefont{E.}~\bibnamefont{Seidel}}, \bibnamefont{and}
  \bibinfo{author}{\bibfnamefont{R.}~\bibnamefont{Takahashi}},
  \bibinfo{journal}{Phys. Rev. Lett.} \textbf{\bibinfo{volume}{87}},
  \bibinfo{pages}{271103} (\bibinfo{year}{2001}{\natexlab{b}}).

\bibitem[{\citenamefont{Frauendiener}(2002)}]{Frauendiener02}
\bibinfo{author}{\bibfnamefont{J.}~\bibnamefont{Frauendiener}},
  \bibinfo{journal}{Phys. Rev. D} \textbf{\bibinfo{volume}{66}},
  \bibinfo{pages}{104027} (\bibinfo{year}{2002}).

\bibitem[{\citenamefont{Shibata}(2000)}]{Shibata00a}
\bibinfo{author}{\bibfnamefont{M.}~\bibnamefont{Shibata}},
  \bibinfo{journal}{Prog. Theor. Phys.} \textbf{\bibinfo{volume}{104}},
  \bibinfo{pages}{325} (\bibinfo{year}{2000}).

\bibitem[{\citenamefont{Shibata}(2003{\natexlab{b}})}]{Shibata03b}
\bibinfo{author}{\bibfnamefont{M.}~\bibnamefont{Shibata}},
  \bibinfo{journal}{Astrophys. J.} \textbf{\bibinfo{volume}{595}},
  \bibinfo{pages}{992} (\bibinfo{year}{2003}{\natexlab{b}}).

\bibitem[{\citenamefont{Baiotti et~al.}(2003)\citenamefont{Baiotti, Hawke,
  Montero, and Rezzolla}}]{Baiotti03a}
\bibinfo{author}{\bibfnamefont{L.}~\bibnamefont{Baiotti}},
  \bibinfo{author}{\bibfnamefont{I.}~\bibnamefont{Hawke}},
  \bibinfo{author}{\bibfnamefont{P.}~\bibnamefont{Montero}}, \bibnamefont{and}
  \bibinfo{author}{\bibfnamefont{L.}~\bibnamefont{Rezzolla}}, in
  \emph{\bibinfo{booktitle}{Computational Astrophysics in Italy: Methods and
  Tools}}, edited by
  \bibinfo{editor}{\bibfnamefont{R.}~\bibnamefont{Capuzzo-Dolcetta}}
  (\bibinfo{publisher}{Mem. Soc. Astron. It. Suppl.},
  \bibinfo{address}{Trieste}, \bibinfo{year}{2003}), vol.~\bibinfo{volume}{1},
  p. \bibinfo{pages}{327}.

\bibitem[{\citenamefont{Font}(2003)}]{Font03}
\bibinfo{author}{\bibfnamefont{J.~A.} \bibnamefont{Font}},
  \bibinfo{journal}{Living Rev. Relativity} \textbf{\bibinfo{volume}{6}},
  \bibinfo{pages}{4} (\bibinfo{year}{2003}),
  \bibinfo{note}{http://relativity.livingreviews.org/Articles/lrr-2003-4}.

\bibitem[{eun()}]{eunetworkweb}
\bibinfo{note}{European RTN on Sources of Gravitational Waves, \hbox{\tt
  www.eu-network.org}}.

\bibitem[{\citenamefont{Rupright et~al.}(1998)\citenamefont{Rupright, Abrahams,
  and Rezzolla}}]{Rupright98}
\bibinfo{author}{\bibfnamefont{M.~E.} \bibnamefont{Rupright}},
  \bibinfo{author}{\bibfnamefont{A.~M.} \bibnamefont{Abrahams}},
  \bibnamefont{and} \bibinfo{author}{\bibfnamefont{L.}~\bibnamefont{Rezzolla}},
  \bibinfo{journal}{Phys. Rev. D} \textbf{\bibinfo{volume}{58}},
  \bibinfo{pages}{044005} (\bibinfo{year}{1998}).

\bibitem[{\citenamefont{Rezzolla et~al.}(1999)\citenamefont{Rezzolla, Abrahams,
  Matzner, Rupright, and Shapiro}}]{Rezzolla99a}
\bibinfo{author}{\bibfnamefont{L.}~\bibnamefont{Rezzolla}},
  \bibinfo{author}{\bibfnamefont{A.~M.} \bibnamefont{Abrahams}},
  \bibinfo{author}{\bibfnamefont{R.~A.} \bibnamefont{Matzner}},
  \bibinfo{author}{\bibfnamefont{M.}~\bibnamefont{Rupright}}, \bibnamefont{and}
  \bibinfo{author}{\bibfnamefont{S.}~\bibnamefont{Shapiro}},
  \bibinfo{journal}{Phys. Rev. D} \textbf{\bibinfo{volume}{59}},
  \bibinfo{pages}{064001} (\bibinfo{year}{1999}).

\bibitem[{\citenamefont{Schnetter et~al.}(2004)\citenamefont{Schnetter, Hawley,
  and Hawke}}]{Schnetter-etal-03b}
\bibinfo{author}{\bibfnamefont{E.}~\bibnamefont{Schnetter}},
  \bibinfo{author}{\bibfnamefont{S.~H.} \bibnamefont{Hawley}},
  \bibnamefont{and} \bibinfo{author}{\bibfnamefont{I.}~\bibnamefont{Hawke}},
  \bibinfo{journal}{Class. Quantum Grav.} \textbf{\bibinfo{volume}{21}},
  \bibinfo{pages}{1465} (\bibinfo{year}{2004}).

\bibitem[{\citenamefont{Baiotti et~al.}(2005)\citenamefont{Baiotti, Hawke,
  Rezzolla, and Schnetter}}]{Baiotti04}
\bibinfo{author}{\bibfnamefont{L.}~\bibnamefont{Baiotti}},
  \bibinfo{author}{\bibfnamefont{I.}~\bibnamefont{Hawke}},
  \bibinfo{author}{\bibfnamefont{L.}~\bibnamefont{Rezzolla}}, \bibnamefont{and}
  \bibinfo{author}{\bibfnamefont{E.}~\bibnamefont{Schnetter}},
  \bibinfo{journal}{{\it work in progress}}  (\bibinfo{year}{2005}).

\bibitem[{\citenamefont{{\tt Cactus Computational Toolkit}}()}]{Cactusweb}
\bibinfo{author}{\bibnamefont{{\tt Cactus Computational Toolkit}}},
  \emph{\bibinfo{title}{{\tt www.cactuscode.org}}}.

\bibitem[{\citenamefont{Font et~al.}(2000{\natexlab{a}})\citenamefont{Font,
  Miller, Suen, and Tobias}}]{Font98b}
\bibinfo{author}{\bibfnamefont{J.~A.} \bibnamefont{Font}},
  \bibinfo{author}{\bibfnamefont{M.}~\bibnamefont{Miller}},
  \bibinfo{author}{\bibfnamefont{W.~M.} \bibnamefont{Suen}}, \bibnamefont{and}
  \bibinfo{author}{\bibfnamefont{M.}~\bibnamefont{Tobias}},
  \bibinfo{journal}{Phys. Rev. D} \textbf{\bibinfo{volume}{61}},
  \bibinfo{pages}{044011} (\bibinfo{year}{2000}{\natexlab{a}}).

\bibitem[{\citenamefont{Collela and Woodward}(1984)}]{Colella84}
\bibinfo{author}{\bibfnamefont{P.}~\bibnamefont{Collela}} \bibnamefont{and}
  \bibinfo{author}{\bibfnamefont{P.~R.} \bibnamefont{Woodward}},
  \bibinfo{journal}{J. Comput. Phys.} \textbf{\bibinfo{volume}{54}},
  \bibinfo{pages}{174} (\bibinfo{year}{1984}).

\bibitem[{\citenamefont{Harten et~al.}(1987)\citenamefont{Harten, Engquist,
  Osher, and Chakrabarty}}]{Harten87}
\bibinfo{author}{\bibfnamefont{A.}~\bibnamefont{Harten}},
  \bibinfo{author}{\bibfnamefont{B.}~\bibnamefont{Engquist}},
  \bibinfo{author}{\bibfnamefont{S.}~\bibnamefont{Osher}}, \bibnamefont{and}
  \bibinfo{author}{\bibfnamefont{S.~R.} \bibnamefont{Chakrabarty}},
  \bibinfo{journal}{J. Comput. Phys.} \textbf{\bibinfo{volume}{71}},
  \bibinfo{pages}{2311} (\bibinfo{year}{1987}).

\bibitem[{\citenamefont{Harten et~al.}(1983)\citenamefont{Harten, Lax, and van
  Leer}}]{Harten83}
\bibinfo{author}{\bibfnamefont{A.}~\bibnamefont{Harten}},
  \bibinfo{author}{\bibfnamefont{P.~D.} \bibnamefont{Lax}}, \bibnamefont{and}
  \bibinfo{author}{\bibfnamefont{B.}~\bibnamefont{van Leer}},
  \bibinfo{journal}{SIAM Rev.} \textbf{\bibinfo{volume}{25}},
  \bibinfo{pages}{35} (\bibinfo{year}{1983}).

\bibitem[{\citenamefont{Aloy et~al.}(1999{\natexlab{a}})\citenamefont{Aloy,
  Ib{\'a}nez, Mart\'\i, and M\"uller}}]{Aloy99b}
\bibinfo{author}{\bibfnamefont{M.~A.} \bibnamefont{Aloy}},
  \bibinfo{author}{\bibfnamefont{J.~M.} \bibnamefont{Ib{\'a}nez}},
  \bibinfo{author}{\bibfnamefont{J.~M.} \bibnamefont{Mart\'\i}},
  \bibnamefont{and} \bibinfo{author}{\bibfnamefont{E.}~\bibnamefont{M\"uller}},
  \bibinfo{journal}{Astroph. J. Supp.} \textbf{\bibinfo{volume}{122}},
  \bibinfo{pages}{151} (\bibinfo{year}{1999}{\natexlab{a}}).

\bibitem[{\citenamefont{Ib{\'a}nez et~al.}(2001)\citenamefont{Ib{\'a}nez, Aloy,
  Font, Mart\'{\i}, Miralles, and Pons}}]{Ibanez01}
\bibinfo{author}{\bibfnamefont{J.}~\bibnamefont{Ib{\'a}nez}},
  \bibinfo{author}{\bibfnamefont{M.}~\bibnamefont{Aloy}},
  \bibinfo{author}{\bibfnamefont{J.}~\bibnamefont{Font}},
  \bibinfo{author}{\bibfnamefont{J.}~\bibnamefont{Mart\'{\i}}},
  \bibinfo{author}{\bibfnamefont{J.}~\bibnamefont{Miralles}}, \bibnamefont{and}
  \bibinfo{author}{\bibfnamefont{J.}~\bibnamefont{Pons}}, in
  \emph{\bibinfo{booktitle}{Godunov methods: theory and applications}}, edited
  by \bibinfo{editor}{\bibfnamefont{E.}~\bibnamefont{Toro}}
  (\bibinfo{publisher}{Kluwer Academic/Plenum Publishers},
  \bibinfo{year}{2001}).

\bibitem[{\citenamefont{Aloy et~al.}(1999{\natexlab{b}})\citenamefont{Aloy,
  Pons, and Ib{\'a}nez}}]{Aloy99a}
\bibinfo{author}{\bibfnamefont{M.~A.} \bibnamefont{Aloy}},
  \bibinfo{author}{\bibfnamefont{J.~A.} \bibnamefont{Pons}}, \bibnamefont{and}
  \bibinfo{author}{\bibfnamefont{J.~M.} \bibnamefont{Ib{\'a}nez}},
  \bibinfo{journal}{Comput. Phys. Commun.} \textbf{\bibinfo{volume}{120}},
  \bibinfo{pages}{115} (\bibinfo{year}{1999}{\natexlab{b}}).

\bibitem[{\citenamefont{Arnowitt et~al.}(1962)\citenamefont{Arnowitt, Deser,
  and Misner}}]{Arnowitt62}
\bibinfo{author}{\bibfnamefont{R.}~\bibnamefont{Arnowitt}},
  \bibinfo{author}{\bibfnamefont{S.}~\bibnamefont{Deser}}, \bibnamefont{and}
  \bibinfo{author}{\bibfnamefont{C.~W.} \bibnamefont{Misner}}, in
  \emph{\bibinfo{booktitle}{Gravitation: An Introduction to Current Research}},
  edited by \bibinfo{editor}{\bibfnamefont{L.}~\bibnamefont{Witten}}
  (\bibinfo{publisher}{John Wiley}, \bibinfo{address}{New York},
  \bibinfo{year}{1962}), pp. \bibinfo{pages}{227--265}.

\bibitem[{\citenamefont{Richtmyer and Morton}(1967)}]{Richtmyer67}
\bibinfo{author}{\bibfnamefont{R.~D.} \bibnamefont{Richtmyer}}
  \bibnamefont{and} \bibinfo{author}{\bibfnamefont{K.}~\bibnamefont{Morton}},
  \emph{\bibinfo{title}{Difference Methods for Initial Value Problems}}
  (\bibinfo{publisher}{Interscience Publishers}, \bibinfo{address}{New York},
  \bibinfo{year}{1967}).

\bibitem[{\citenamefont{York}(1979)}]{York79}
\bibinfo{author}{\bibfnamefont{J.~W.} \bibnamefont{York}, \bibfnamefont{Jr.}},
  in \emph{\bibinfo{booktitle}{Sources of Gravitational Radiation}}, edited by
  \bibinfo{editor}{\bibfnamefont{L.~L.} \bibnamefont{Smarr}}
  (\bibinfo{publisher}{Cambridge University Press},
  \bibinfo{address}{Cambridge, UK}, \bibinfo{year}{1979}),
  p.~\bibinfo{pages}{83}.

\bibitem[{\citenamefont{Alcubierre et~al.}(2003)\citenamefont{Alcubierre,
  Br\"ugmann, Diener, Koppitz, Pollney, Seidel, and Takahashi}}]{Alcubierre02a}
\bibinfo{author}{\bibfnamefont{M.}~\bibnamefont{Alcubierre}},
  \bibinfo{author}{\bibfnamefont{B.}~\bibnamefont{Br\"ugmann}},
  \bibinfo{author}{\bibfnamefont{P.}~\bibnamefont{Diener}},
  \bibinfo{author}{\bibfnamefont{M.}~\bibnamefont{Koppitz}},
  \bibinfo{author}{\bibfnamefont{D.}~\bibnamefont{Pollney}},
  \bibinfo{author}{\bibfnamefont{E.}~\bibnamefont{Seidel}}, \bibnamefont{and}
  \bibinfo{author}{\bibfnamefont{R.}~\bibnamefont{Takahashi}},
  \bibinfo{journal}{Phys. Rev. D} \textbf{\bibinfo{volume}{67}},
  \bibinfo{pages}{084023} (\bibinfo{year}{2003}).

\bibitem[{\citenamefont{Sarbach et~al.}(2002)\citenamefont{Sarbach, Calabrese,
  Pullin, and Tiglio}}]{Sarbach02a}
\bibinfo{author}{\bibfnamefont{O.}~\bibnamefont{Sarbach}},
  \bibinfo{author}{\bibfnamefont{G.}~\bibnamefont{Calabrese}},
  \bibinfo{author}{\bibfnamefont{J.}~\bibnamefont{Pullin}}, \bibnamefont{and}
  \bibinfo{author}{\bibfnamefont{M.}~\bibnamefont{Tiglio}},
  \bibinfo{journal}{Phys. Rev. D} \textbf{\bibinfo{volume}{66}},
  \bibinfo{pages}{064002} (\bibinfo{year}{2002}).

\bibitem[{\citenamefont{Bona et~al.}(2003)\citenamefont{Bona, Ledvinka,
  Palenzuela, and Zacek}}]{Bona:2003qn}
\bibinfo{author}{\bibfnamefont{C.}~\bibnamefont{Bona}},
  \bibinfo{author}{\bibfnamefont{T.}~\bibnamefont{Ledvinka}},
  \bibinfo{author}{\bibfnamefont{C.}~\bibnamefont{Palenzuela}},
  \bibnamefont{and} \bibinfo{author}{\bibfnamefont{M.}~\bibnamefont{Zacek}},
  \bibinfo{journal}{Phys. Rev.} \textbf{\bibinfo{volume}{D69}},
  \bibinfo{pages}{064036} (\bibinfo{year}{2003}).

\bibitem[{\citenamefont{Nagy et~al.}(2004)\citenamefont{Nagy, Ortiz, and
  Reula}}]{Nagy04}
\bibinfo{author}{\bibfnamefont{G.}~\bibnamefont{Nagy}},
  \bibinfo{author}{\bibfnamefont{O.~E.} \bibnamefont{Ortiz}}, \bibnamefont{and}
  \bibinfo{author}{\bibfnamefont{O.~A.} \bibnamefont{Reula}}
  (\bibinfo{year}{2004}), \bibinfo{note}{gr-qc/0402123}.

\bibitem[{\citenamefont{Alcubierre
  et~al.}(2000{\natexlab{b}})\citenamefont{Alcubierre, Allen, Br\"{u}gmann,
  Seidel, and Suen}}]{Alcubierre99e}
\bibinfo{author}{\bibfnamefont{M.}~\bibnamefont{Alcubierre}},
  \bibinfo{author}{\bibfnamefont{G.}~\bibnamefont{Allen}},
  \bibinfo{author}{\bibfnamefont{B.}~\bibnamefont{Br\"{u}gmann}},
  \bibinfo{author}{\bibfnamefont{E.}~\bibnamefont{Seidel}}, \bibnamefont{and}
  \bibinfo{author}{\bibfnamefont{W.}~\bibnamefont{Suen}},
  \bibinfo{journal}{Phys. Rev. D} \textbf{\bibinfo{volume}{62}},
  \bibinfo{pages}{124011} (\bibinfo{year}{2000}{\natexlab{b}}).

\bibitem[{\citenamefont{Balakrishna et~al.}(1996)\citenamefont{Balakrishna,
  Daues, Seidel, Suen, Tobias, and Wang}}]{Balakrishna96a}
\bibinfo{author}{\bibfnamefont{J.}~\bibnamefont{Balakrishna}},
  \bibinfo{author}{\bibfnamefont{G.}~\bibnamefont{Daues}},
  \bibinfo{author}{\bibfnamefont{E.}~\bibnamefont{Seidel}},
  \bibinfo{author}{\bibfnamefont{W.-M.} \bibnamefont{Suen}},
  \bibinfo{author}{\bibfnamefont{M.}~\bibnamefont{Tobias}}, \bibnamefont{and}
  \bibinfo{author}{\bibfnamefont{E.}~\bibnamefont{Wang}},
  \bibinfo{journal}{Class. Quantum Grav.} \textbf{\bibinfo{volume}{13}},
  \bibinfo{pages}{L135} (\bibinfo{year}{1996}).

\bibitem[{\citenamefont{Bona et~al.}(1995)\citenamefont{Bona, Mass{\'o},
  Seidel, and Stela}}]{Bona94b}
\bibinfo{author}{\bibfnamefont{C.}~\bibnamefont{Bona}},
  \bibinfo{author}{\bibfnamefont{J.}~\bibnamefont{Mass{\'o}}},
  \bibinfo{author}{\bibfnamefont{E.}~\bibnamefont{Seidel}}, \bibnamefont{and}
  \bibinfo{author}{\bibfnamefont{J.}~\bibnamefont{Stela}},
  \bibinfo{journal}{Phys. Rev. Lett.} \textbf{\bibinfo{volume}{75}},
  \bibinfo{pages}{600} (\bibinfo{year}{1995}).

\bibitem[{\citenamefont{Alcubierre}(1997)}]{Alcubierre97a}
\bibinfo{author}{\bibfnamefont{M.}~\bibnamefont{Alcubierre}},
  \bibinfo{journal}{Phys. Rev. D} \textbf{\bibinfo{volume}{55}},
  \bibinfo{pages}{5981} (\bibinfo{year}{1997}).

\bibitem[{\citenamefont{Alcubierre and Mass{\'o}}(1998)}]{Alcubierre97b}
\bibinfo{author}{\bibfnamefont{M.}~\bibnamefont{Alcubierre}} \bibnamefont{and}
  \bibinfo{author}{\bibfnamefont{J.}~\bibnamefont{Mass{\'o}}},
  \bibinfo{journal}{Phys. Rev. D} \textbf{\bibinfo{volume}{57}},
  \bibinfo{pages}{4511} (\bibinfo{year}{1998}).

\bibitem[{\citenamefont{Smarr and York}(1978)}]{Smarr78b}
\bibinfo{author}{\bibfnamefont{L.}~\bibnamefont{Smarr}} \bibnamefont{and}
  \bibinfo{author}{\bibfnamefont{J.}~\bibnamefont{York}},
  \bibinfo{journal}{Phys. Rev. D} \textbf{\bibinfo{volume}{17}},
  \bibinfo{pages}{2529} (\bibinfo{year}{1978}).

\bibitem[{\citenamefont{Mart\'{\i} et~al.}(1991)\citenamefont{Mart\'{\i},
  Ib{\'a}nez, and Miralles}}]{Marti91}
\bibinfo{author}{\bibfnamefont{J.~M.} \bibnamefont{Mart\'{\i}}},
  \bibinfo{author}{\bibfnamefont{J.~M.} \bibnamefont{Ib{\'a}nez}},
  \bibnamefont{and} \bibinfo{author}{\bibfnamefont{J.~A.}
  \bibnamefont{Miralles}}, \bibinfo{journal}{Phys. Rev. D}
  \textbf{\bibinfo{volume}{43}}, \bibinfo{pages}{3794} (\bibinfo{year}{1991}).

\bibitem[{\citenamefont{Banyuls et~al.}(1997)\citenamefont{Banyuls, Font,
  Ib{\'a}nez, Mart\'{\i}, and Miralles}}]{Banyuls97}
\bibinfo{author}{\bibfnamefont{F.}~\bibnamefont{Banyuls}},
  \bibinfo{author}{\bibfnamefont{J.~A.} \bibnamefont{Font}},
  \bibinfo{author}{\bibfnamefont{J.~M.} \bibnamefont{Ib{\'a}nez}},
  \bibinfo{author}{\bibfnamefont{J.~M.} \bibnamefont{Mart\'{\i}}},
  \bibnamefont{and} \bibinfo{author}{\bibfnamefont{J.~A.}
  \bibnamefont{Miralles}}, \bibinfo{journal}{Astrophys. J.}
  \textbf{\bibinfo{volume}{476}}, \bibinfo{pages}{221} (\bibinfo{year}{1997}).

\bibitem[{\citenamefont{Zwerger}(1995)}]{Zwerger95}
\bibinfo{author}{\bibfnamefont{T.}~\bibnamefont{Zwerger}}, Ph.D. thesis,
  \bibinfo{school}{Technische Universit\"at M\"unchen},
  \bibinfo{address}{M\"unchen, Germany} (\bibinfo{year}{1995}).

\bibitem[{\citenamefont{Zwerger and M{\"u}ller}(1997)}]{Zwerger97}
\bibinfo{author}{\bibfnamefont{T.}~\bibnamefont{Zwerger}} \bibnamefont{and}
  \bibinfo{author}{\bibfnamefont{E.}~\bibnamefont{M{\"u}ller}},
  \bibinfo{journal}{Astron. Astrophys.} \textbf{\bibinfo{volume}{{\bf 320}}},
  \bibinfo{pages}{209} (\bibinfo{year}{1997}).

\bibitem[{\citenamefont{Lax and Wendroff}(1960)}]{Lax60}
\bibinfo{author}{\bibfnamefont{P.~D.} \bibnamefont{Lax}} \bibnamefont{and}
  \bibinfo{author}{\bibfnamefont{B.}~\bibnamefont{Wendroff}},
  \bibinfo{journal}{Comm. Pure Appl. Math.} \textbf{\bibinfo{volume}{13}},
  \bibinfo{pages}{217} (\bibinfo{year}{1960}).

\bibitem[{\citenamefont{Hou and LeFloch}(1994)}]{Hou94}
\bibinfo{author}{\bibfnamefont{T.~Y.} \bibnamefont{Hou}} \bibnamefont{and}
  \bibinfo{author}{\bibfnamefont{P.~G.} \bibnamefont{LeFloch}},
  \bibinfo{journal}{Math. of Comp.} \textbf{\bibinfo{volume}{62}},
  \bibinfo{pages}{497} (\bibinfo{year}{1994}).

\bibitem[{\citenamefont{Laney}(1998)}]{Laney98}
\bibinfo{author}{\bibfnamefont{C.~B.} \bibnamefont{Laney}},
  \emph{\bibinfo{title}{Computational Gasdynamics}}
  (\bibinfo{publisher}{Cambridge University Press}, \bibinfo{year}{1998}).

\bibitem[{\citenamefont{Toro}(1999)}]{Toro99}
\bibinfo{author}{\bibfnamefont{E.~F.} \bibnamefont{Toro}},
  \emph{\bibinfo{title}{Riemann Solvers and Numerical Methods for Fluid
  Dynamics}} (\bibinfo{publisher}{Springer-Verlag}, \bibinfo{year}{1999}).

\bibitem[{\citenamefont{Leveque}(1998)}]{Leveque98}
\bibinfo{author}{\bibfnamefont{R.~J.} \bibnamefont{Leveque}}, in
  \emph{\bibinfo{booktitle}{Computational Methods for Astrophysical Fluid
  Flow}}, edited by \bibinfo{editor}{\bibfnamefont{O.}~\bibnamefont{Steiner}}
  \bibnamefont{and} \bibinfo{editor}{\bibfnamefont{A.}~\bibnamefont{Gautschy}}
  (\bibinfo{publisher}{Springer-Verlag}, \bibinfo{year}{1998}).

\bibitem[{\citenamefont{Thornburg}(2004)}]{Thornburg2003:AH-finding}
\bibinfo{author}{\bibfnamefont{J.}~\bibnamefont{Thornburg}},
  \bibinfo{journal}{Classical and Quantum Gravity}
  \textbf{\bibinfo{volume}{21}}, \bibinfo{pages}{743} (\bibinfo{year}{2004}).

\bibitem[{\citenamefont{Kidder et~al.}(2000{\natexlab{b}})\citenamefont{Kidder,
  Scheel, Teukolsky, and Cook}}]{Kidder00b}
\bibinfo{author}{\bibfnamefont{L.}~\bibnamefont{Kidder}},
  \bibinfo{author}{\bibfnamefont{M.}~\bibnamefont{Scheel}},
  \bibinfo{author}{\bibfnamefont{S.}~\bibnamefont{Teukolsky}},
  \bibnamefont{and} \bibinfo{author}{\bibfnamefont{G.}~\bibnamefont{Cook}}, in
  \emph{\bibinfo{booktitle}{Miniprogram on Colliding Black Holes: Mathematical
  Issues in Numerical Relativity}} (\bibinfo{publisher}{Institute for
  Theoretical Physics, UCSB}, \bibinfo{address}{Santa Barbara, CA},
  \bibinfo{year}{2000}{\natexlab{b}}).

\bibitem[{\citenamefont{Calabrese et~al.}(2003)\citenamefont{Calabrese, Lehner,
  Neilsen, Pullin, Reula, Sarbach, and Tiglio}}]{Calabrese:2003a}
\bibinfo{author}{\bibfnamefont{G.}~\bibnamefont{Calabrese}},
  \bibinfo{author}{\bibfnamefont{L.}~\bibnamefont{Lehner}},
  \bibinfo{author}{\bibfnamefont{D.}~\bibnamefont{Neilsen}},
  \bibinfo{author}{\bibfnamefont{J.}~\bibnamefont{Pullin}},
  \bibinfo{author}{\bibfnamefont{O.}~\bibnamefont{Reula}},
  \bibinfo{author}{\bibfnamefont{O.}~\bibnamefont{Sarbach}}, \bibnamefont{and}
  \bibinfo{author}{\bibfnamefont{M.}~\bibnamefont{Tiglio}},
  \bibinfo{journal}{Class. Quant. Grav} \textbf{\bibinfo{volume}{20}},
  \bibinfo{pages}{L245} (\bibinfo{year}{2003}).

\bibitem[{\citenamefont{Hawke et~al.}(2005)\citenamefont{Hawke, L{\"o}ffler,
  and Nerozzi}}]{Hawke04}
\bibinfo{author}{\bibfnamefont{I.}~\bibnamefont{Hawke}},
  \bibinfo{author}{\bibfnamefont{F.}~\bibnamefont{L{\"o}ffler}},
  \bibnamefont{and} \bibinfo{author}{\bibfnamefont{A.}~\bibnamefont{Nerozzi}}
  (\bibinfo{year}{2005}), \eprint{gr-qc/0501054}.

\bibitem[{\citenamefont{Friedman et~al.}(1988)\citenamefont{Friedman, Ipser,
  and Sorkin}}]{Friedman88}
\bibinfo{author}{\bibfnamefont{J.~L.} \bibnamefont{Friedman}},
  \bibinfo{author}{\bibfnamefont{J.~R.} \bibnamefont{Ipser}}, \bibnamefont{and}
  \bibinfo{author}{\bibfnamefont{R.~D.} \bibnamefont{Sorkin}},
  \bibinfo{journal}{Astrophys. J.} \textbf{\bibinfo{volume}{325}},
  \bibinfo{pages}{722} (\bibinfo{year}{1988}).

\bibitem[{\citenamefont{Stergioulas and Friedman}(1995)}]{Stergioulas95}
\bibinfo{author}{\bibfnamefont{N.}~\bibnamefont{Stergioulas}} \bibnamefont{and}
  \bibinfo{author}{\bibfnamefont{J.~L.} \bibnamefont{Friedman}},
  \bibinfo{journal}{Astrophys. J.} \textbf{\bibinfo{volume}{444}},
  \bibinfo{pages}{306} (\bibinfo{year}{1995}).

\bibitem[{\citenamefont{Stergioulas and Font}(2001)}]{Stergioulas01}
\bibinfo{author}{\bibfnamefont{N.}~\bibnamefont{Stergioulas}} \bibnamefont{and}
  \bibinfo{author}{\bibfnamefont{J.}~\bibnamefont{Font}},
  \bibinfo{journal}{Phys. Rev. Lett.} \textbf{\bibinfo{volume}{86}},
  \bibinfo{pages}{1148} (\bibinfo{year}{2001}).

\bibitem[{\citenamefont{Stergioulas}(2003)}]{Stergioulas03}
\bibinfo{author}{\bibfnamefont{N.}~\bibnamefont{Stergioulas}},
  \bibinfo{journal}{Living Rev. Relativity} \textbf{\bibinfo{volume}{6}},
  \bibinfo{pages}{3} (\bibinfo{year}{2003}).

\bibitem[{\citenamefont{Font et~al.}(2000{\natexlab{b}})\citenamefont{Font,
  Stergioulas, and Kokkotas}}]{Font99}
\bibinfo{author}{\bibfnamefont{J.~A.} \bibnamefont{Font}},
  \bibinfo{author}{\bibfnamefont{N.}~\bibnamefont{Stergioulas}},
  \bibnamefont{and} \bibinfo{author}{\bibfnamefont{K.~D.}
  \bibnamefont{Kokkotas}}, \bibinfo{journal}{Mon. Not. R. Astron. Soc.}
  \textbf{\bibinfo{volume}{313}}, \bibinfo{pages}{678}
  (\bibinfo{year}{2000}{\natexlab{b}}).

\bibitem[{\citenamefont{Stergioulas and Hawke}(2003)}]{Stergioulas03b}
\bibinfo{author}{\bibfnamefont{N.}~\bibnamefont{Stergioulas}} \bibnamefont{and}
  \bibinfo{author}{\bibfnamefont{I.}~\bibnamefont{Hawke}}, in
  \emph{\bibinfo{booktitle}{{\it Recent Developments in Gravity}, Proceedings
  of the 10th Hellenic Relativity Conference}}, edited by
  \bibinfo{editor}{\bibfnamefont{K.~D.} \bibnamefont{Kokkotas}}
  \bibnamefont{and}
  \bibinfo{editor}{\bibfnamefont{N.}~\bibnamefont{Stergioulas}}
  (\bibinfo{publisher}{World Scientific}, \bibinfo{address}{Singapore},
  \bibinfo{year}{2003}), p. \bibinfo{pages}{185}.

\bibitem[{\citenamefont{Cook}(2000)}]{Cook00a}
\bibinfo{author}{\bibfnamefont{G.~B.} \bibnamefont{Cook}},
  \bibinfo{journal}{Living Rev. Relativity} \textbf{\bibinfo{volume}{5}},
  \bibinfo{pages}{5} (\bibinfo{year}{2000}).

\bibitem[{\citenamefont{Woosley}(2000)}]{Woosley00}
\bibinfo{author}{\bibfnamefont{S.~E.} \bibnamefont{Woosley}}, in
  \emph{\bibinfo{booktitle}{Proceedings of the International Workshop held in
  Rome, CNR headquarters, October, 2000}}, edited by
  \bibinfo{editor}{\bibfnamefont{E.}~\bibnamefont{Costa}},
  \bibinfo{editor}{\bibfnamefont{F.}~\bibnamefont{Frontera}}, ,
  \bibnamefont{and} \bibinfo{editor}{\bibfnamefont{J.}~\bibnamefont{Hjorth}}
  (\bibinfo{publisher}{Springer}, \bibinfo{address}{Berlin Heidelberg},
  \bibinfo{year}{2000}), p. \bibinfo{pages}{257}.

\bibitem[{\citenamefont{{Zanotti} et~al.}(2003)\citenamefont{{Zanotti},
  {Rezzolla}, and {Font}}}]{Zanotti02}
\bibinfo{author}{\bibfnamefont{O.}~\bibnamefont{{Zanotti}}},
  \bibinfo{author}{\bibfnamefont{L.}~\bibnamefont{{Rezzolla}}},
  \bibnamefont{and} \bibinfo{author}{\bibfnamefont{J.~A.}
  \bibnamefont{{Font}}}, \bibinfo{journal}{MNRAS}
  \textbf{\bibinfo{volume}{341}}, \bibinfo{pages}{832} (\bibinfo{year}{2003}).

\bibitem[{\citenamefont{Shibata}(1999)}]{Shibata99c}
\bibinfo{author}{\bibfnamefont{M.}~\bibnamefont{Shibata}},
  \bibinfo{journal}{Phys. Rev. D} \textbf{\bibinfo{volume}{60}},
  \bibinfo{pages}{104052} (\bibinfo{year}{1999}).

\bibitem[{\citenamefont{Dimmelmeier et~al.}(2002)\citenamefont{Dimmelmeier,
  Font, and M{\"{u}}ller}}]{Dimmelmeier02b}
\bibinfo{author}{\bibfnamefont{H.}~\bibnamefont{Dimmelmeier}},
  \bibinfo{author}{\bibfnamefont{J.~A.} \bibnamefont{Font}}, \bibnamefont{and}
  \bibinfo{author}{\bibfnamefont{E.}~\bibnamefont{M{\"{u}}ller}},
  \bibinfo{journal}{Astron. and Astrophys.} \textbf{\bibinfo{volume}{393}},
  \bibinfo{pages}{523} (\bibinfo{year}{2002}).

\bibitem[{\citenamefont{Shibata and Uryu}(2002)}]{Shibata02a}
\bibinfo{author}{\bibfnamefont{M.}~\bibnamefont{Shibata}} \bibnamefont{and}
  \bibinfo{author}{\bibfnamefont{K.}~\bibnamefont{Uryu}},
  \bibinfo{journal}{Prog. Theor. Phys.} \textbf{\bibinfo{volume}{107}},
  \bibinfo{pages}{265} (\bibinfo{year}{2002}).

\bibitem[{\citenamefont{{Shapiro}}(2000)}]{Shapiro00}
\bibinfo{author}{\bibfnamefont{S.~L.} \bibnamefont{{Shapiro}}},
  \bibinfo{journal}{\apj} \textbf{\bibinfo{volume}{544}}, \bibinfo{pages}{397}
  (\bibinfo{year}{2000}).

\bibitem[{\citenamefont{{Cook} et~al.}(2003)\citenamefont{{Cook}, {Shapiro},
  and {Stephens}}}]{Shapiro03}
\bibinfo{author}{\bibfnamefont{J.~N.} \bibnamefont{{Cook}}},
  \bibinfo{author}{\bibfnamefont{S.~L.} \bibnamefont{{Shapiro}}},
  \bibnamefont{and} \bibinfo{author}{\bibfnamefont{B.~C.}
  \bibnamefont{{Stephens}}}, \bibinfo{journal}{\apj}
  \textbf{\bibinfo{volume}{599}}, \bibinfo{pages}{1272} (\bibinfo{year}{2003}).

\bibitem[{\citenamefont{{Liu} and {Shapiro}}(2004)}]{Shapiro04}
\bibinfo{author}{\bibfnamefont{Y.~T.} \bibnamefont{{Liu}}} \bibnamefont{and}
  \bibinfo{author}{\bibfnamefont{S.~L.} \bibnamefont{{Shapiro}}},
  \bibinfo{journal}{Phys. Rev. D} \textbf{\bibinfo{volume}{69}},
  \bibinfo{pages}{044009} (\bibinfo{year}{2004}).

\bibitem[{\citenamefont{Bardeen}(1970)}]{Bardeen70}
\bibinfo{author}{\bibfnamefont{J.}~\bibnamefont{Bardeen}},
  \bibinfo{journal}{Astrophys. J.} \textbf{\bibinfo{volume}{162}},
  \bibinfo{pages}{171} (\bibinfo{year}{1970}).

\bibitem[{\citenamefont{Cumming et~al.}(2000)\citenamefont{Cumming, Morsink,
  Bildsten, Friedman, and Holz}}]{Cumming00}
\bibinfo{author}{\bibfnamefont{A.}~\bibnamefont{Cumming}},
  \bibinfo{author}{\bibfnamefont{S.~M.} \bibnamefont{Morsink}},
  \bibinfo{author}{\bibfnamefont{L.}~\bibnamefont{Bildsten}},
  \bibinfo{author}{\bibfnamefont{J.~L.} \bibnamefont{Friedman}},
  \bibnamefont{and} \bibinfo{author}{\bibfnamefont{D.~E.} \bibnamefont{Holz}},
  \bibinfo{journal}{Astrophys. J.} \textbf{\bibinfo{volume}{564}},
  \bibinfo{pages}{343} (\bibinfo{year}{2000}).

\bibitem[{\citenamefont{Rezzolla et~al.}(2000)\citenamefont{Rezzolla, Lamb, and
  Shapiro}}]{Rezzolla00}
\bibinfo{author}{\bibfnamefont{L.}~\bibnamefont{Rezzolla}},
  \bibinfo{author}{\bibfnamefont{F.~K.} \bibnamefont{Lamb}}, \bibnamefont{and}
  \bibinfo{author}{\bibfnamefont{S.~L.} \bibnamefont{Shapiro}},
  \bibinfo{journal}{Astrophys. J.} \textbf{\bibinfo{volume}{531}},
  \bibinfo{pages}{L139} (\bibinfo{year}{2000}).

\bibitem[{\citenamefont{Spruit}(1999)}]{Spruit99}
\bibinfo{author}{\bibfnamefont{H.~C.} \bibnamefont{Spruit}},
  \bibinfo{journal}{Astron. and Astrophys.} \textbf{\bibinfo{volume}{341}},
  \bibinfo{pages}{L1} (\bibinfo{year}{1999}).

\bibitem[{\citenamefont{Diener}(2003)}]{Diener03a}
\bibinfo{author}{\bibfnamefont{P.}~\bibnamefont{Diener}},
  \bibinfo{journal}{Class. Quantum Grav.} \textbf{\bibinfo{volume}{20}},
  \bibinfo{pages}{4901} (\bibinfo{year}{2003}).

\bibitem[{\citenamefont{Anninos et~al.}(1994)\citenamefont{Anninos, Bernstein,
  Brandt, Hobill, Seidel, and Smarr}}]{Anninos93a}
\bibinfo{author}{\bibfnamefont{P.}~\bibnamefont{Anninos}},
  \bibinfo{author}{\bibfnamefont{D.}~\bibnamefont{Bernstein}},
  \bibinfo{author}{\bibfnamefont{S.~R.} \bibnamefont{Brandt}},
  \bibinfo{author}{\bibfnamefont{D.}~\bibnamefont{Hobill}},
  \bibinfo{author}{\bibfnamefont{E.}~\bibnamefont{Seidel}}, \bibnamefont{and}
  \bibinfo{author}{\bibfnamefont{L.}~\bibnamefont{Smarr}},
  \bibinfo{journal}{Phys. Rev. D} \textbf{\bibinfo{volume}{50}},
  \bibinfo{pages}{3801} (\bibinfo{year}{1994}).

\bibitem[{\citenamefont{Anninos
  et~al.}(1995{\natexlab{a}})\citenamefont{Anninos, Bernstein, Brandt, Libson,
  Mass{\'o}, Seidel, Smarr, Suen, and Walker}}]{Anninos94f}
\bibinfo{author}{\bibfnamefont{P.}~\bibnamefont{Anninos}},
  \bibinfo{author}{\bibfnamefont{D.}~\bibnamefont{Bernstein}},
  \bibinfo{author}{\bibfnamefont{S.}~\bibnamefont{Brandt}},
  \bibinfo{author}{\bibfnamefont{J.}~\bibnamefont{Libson}},
  \bibinfo{author}{\bibfnamefont{J.}~\bibnamefont{Mass{\'o}}},
  \bibinfo{author}{\bibfnamefont{E.}~\bibnamefont{Seidel}},
  \bibinfo{author}{\bibfnamefont{L.}~\bibnamefont{Smarr}},
  \bibinfo{author}{\bibfnamefont{W.-M.} \bibnamefont{Suen}}, \bibnamefont{and}
  \bibinfo{author}{\bibfnamefont{P.}~\bibnamefont{Walker}},
  \bibinfo{journal}{Phys. Rev. Lett.} \textbf{\bibinfo{volume}{74}},
  \bibinfo{pages}{630} (\bibinfo{year}{1995}{\natexlab{a}}).

\bibitem[{\citenamefont{Anninos
  et~al.}(1995{\natexlab{b}})\citenamefont{Anninos, Bernstein, Brandt, Hobill,
  Seidel, and Smarr}}]{Anninos95c}
\bibinfo{author}{\bibfnamefont{P.}~\bibnamefont{Anninos}},
  \bibinfo{author}{\bibfnamefont{D.}~\bibnamefont{Bernstein}},
  \bibinfo{author}{\bibfnamefont{S.}~\bibnamefont{Brandt}},
  \bibinfo{author}{\bibfnamefont{D.}~\bibnamefont{Hobill}},
  \bibinfo{author}{\bibfnamefont{E.}~\bibnamefont{Seidel}}, \bibnamefont{and}
  \bibinfo{author}{\bibfnamefont{L.}~\bibnamefont{Smarr}},
  \bibinfo{journal}{Austral. Journ. Phys.} \textbf{\bibinfo{volume}{48}},
  \bibinfo{pages}{1027} (\bibinfo{year}{1995}{\natexlab{b}}).

\bibitem[{\citenamefont{Brandt and Seidel}(1995{\natexlab{a}})}]{Brandt94c}
\bibinfo{author}{\bibfnamefont{S.}~\bibnamefont{Brandt}} \bibnamefont{and}
  \bibinfo{author}{\bibfnamefont{E.}~\bibnamefont{Seidel}},
  \bibinfo{journal}{Phys. Rev. D} \textbf{\bibinfo{volume}{52}},
  \bibinfo{pages}{870} (\bibinfo{year}{1995}{\natexlab{a}}).

\bibitem[{\citenamefont{Brandt et~al.}(2000{\natexlab{b}})\citenamefont{Brandt,
  Font, Ib{\'a}{\~n}ez, Mass{\'o}, and Seidel}}]{Brandt98}
\bibinfo{author}{\bibfnamefont{S.}~\bibnamefont{Brandt}},
  \bibinfo{author}{\bibfnamefont{J.~A.} \bibnamefont{Font}},
  \bibinfo{author}{\bibfnamefont{J.~M.} \bibnamefont{Ib{\'a}{\~n}ez}},
  \bibinfo{author}{\bibfnamefont{J.}~\bibnamefont{Mass{\'o}}},
  \bibnamefont{and} \bibinfo{author}{\bibfnamefont{E.}~\bibnamefont{Seidel}},
  \bibinfo{journal}{Comp. Phys. Comm.} \textbf{\bibinfo{volume}{124}},
  \bibinfo{pages}{169} (\bibinfo{year}{2000}{\natexlab{b}}).

\bibitem[{\citenamefont{Brandt and Seidel}(1995{\natexlab{b}})}]{Brandt94b}
\bibinfo{author}{\bibfnamefont{S.}~\bibnamefont{Brandt}} \bibnamefont{and}
  \bibinfo{author}{\bibfnamefont{E.}~\bibnamefont{Seidel}},
  \bibinfo{journal}{Phys. Rev. D} \textbf{\bibinfo{volume}{52}},
  \bibinfo{pages}{856} (\bibinfo{year}{1995}{\natexlab{b}}).

\bibitem[{\citenamefont{Ashtekar
  et~al.}(2000{\natexlab{a}})\citenamefont{Ashtekar, Beetle, and
  Fairhurst}}]{Ashtekar99a}
\bibinfo{author}{\bibfnamefont{A.}~\bibnamefont{Ashtekar}},
  \bibinfo{author}{\bibfnamefont{C.}~\bibnamefont{Beetle}}, \bibnamefont{and}
  \bibinfo{author}{\bibfnamefont{S.}~\bibnamefont{Fairhurst}},
  \bibinfo{journal}{Class. Quantum Grav.} \textbf{\bibinfo{volume}{17}},
  \bibinfo{pages}{253} (\bibinfo{year}{2000}{\natexlab{a}}).

\bibitem[{\citenamefont{Ashtekar
  et~al.}(2000{\natexlab{b}})\citenamefont{Ashtekar, Beetle, Dreyer, Fairhurst,
  Krishnan, Lewandowski, and Wisniewski}}]{Ashtekar00a}
\bibinfo{author}{\bibfnamefont{A.}~\bibnamefont{Ashtekar}},
  \bibinfo{author}{\bibfnamefont{C.}~\bibnamefont{Beetle}},
  \bibinfo{author}{\bibfnamefont{O.}~\bibnamefont{Dreyer}},
  \bibinfo{author}{\bibfnamefont{S.}~\bibnamefont{Fairhurst}},
  \bibinfo{author}{\bibfnamefont{B.}~\bibnamefont{Krishnan}},
  \bibinfo{author}{\bibfnamefont{J.}~\bibnamefont{Lewandowski}},
  \bibnamefont{and}
  \bibinfo{author}{\bibfnamefont{J.}~\bibnamefont{Wisniewski}},
  \bibinfo{journal}{Phys. Rev. Lett.} \textbf{\bibinfo{volume}{85}},
  \bibinfo{pages}{3564} (\bibinfo{year}{2000}{\natexlab{b}}).

\bibitem[{\citenamefont{Ashtekar et~al.}(2001)\citenamefont{Ashtekar, Beetle,
  and Lewandowski}}]{Ashtekar01a}
\bibinfo{author}{\bibfnamefont{A.}~\bibnamefont{Ashtekar}},
  \bibinfo{author}{\bibfnamefont{C.}~\bibnamefont{Beetle}}, \bibnamefont{and}
  \bibinfo{author}{\bibfnamefont{J.}~\bibnamefont{Lewandowski}},
  \bibinfo{journal}{Phys. Rev. D} \textbf{\bibinfo{volume}{64}},
  \bibinfo{pages}{044016} (\bibinfo{year}{2001}).

\bibitem[{\citenamefont{Ashtekar and
  Krishnan}(2002)}]{Ashtekar-etal-2002-dynamical-horizons}
\bibinfo{author}{\bibfnamefont{A.}~\bibnamefont{Ashtekar}} \bibnamefont{and}
  \bibinfo{author}{\bibfnamefont{B.}~\bibnamefont{Krishnan}},
  \bibinfo{journal}{Phys. Rev. Lett.} \textbf{\bibinfo{volume}{89}},
  \bibinfo{pages}{261101} (\bibinfo{year}{2002}).

\bibitem[{\citenamefont{Dreyer et~al.}(2002)\citenamefont{Dreyer, Krishnan,
  Shoemaker, and Schnetter}}]{Dreyer-etal-2002-isolated-horizons}
\bibinfo{author}{\bibfnamefont{O.}~\bibnamefont{Dreyer}},
  \bibinfo{author}{\bibfnamefont{B.}~\bibnamefont{Krishnan}},
  \bibinfo{author}{\bibfnamefont{D.}~\bibnamefont{Shoemaker}},
  \bibnamefont{and}
  \bibinfo{author}{\bibfnamefont{E.}~\bibnamefont{Schnetter}},
  \bibinfo{journal}{Phys. Rev. D} \textbf{\bibinfo{volume}{67}},
  \bibinfo{pages}{024018} (\bibinfo{year}{2002}).

\bibitem[{\citenamefont{Ashtekar and Krishnan}(2003)}]{Ashtekar03a}
\bibinfo{author}{\bibfnamefont{A.}~\bibnamefont{Ashtekar}} \bibnamefont{and}
  \bibinfo{author}{\bibfnamefont{B.}~\bibnamefont{Krishnan}},
  \bibinfo{journal}{Phys. Rev. D} \textbf{\bibinfo{volume}{{\bf 68}}},
  \bibinfo{pages}{104030} (\bibinfo{year}{2003}).

\bibitem[{\citenamefont{Christodoulou}(1970)}]{Christodoulou70}
\bibinfo{author}{\bibfnamefont{D.}~\bibnamefont{Christodoulou}},
  \bibinfo{journal}{Phys. Rev. Lett.} \textbf{\bibinfo{volume}{25}},
  \bibinfo{pages}{1596} (\bibinfo{year}{1970}).

\bibitem[{\citenamefont{Shu and Osher}(1988)}]{Shu88}
\bibinfo{author}{\bibfnamefont{C.~W.} \bibnamefont{Shu}} \bibnamefont{and}
  \bibinfo{author}{\bibfnamefont{S.~J.} \bibnamefont{Osher}},
  \bibinfo{journal}{Journ. Comput. Phys.} \textbf{\bibinfo{volume}{77}},
  \bibinfo{pages}{439} (\bibinfo{year}{1988}).

\bibitem[{\citenamefont{Gottlieb and Shu}(1998)}]{Gottlieb98}
\bibinfo{author}{\bibfnamefont{S.}~\bibnamefont{Gottlieb}} \bibnamefont{and}
  \bibinfo{author}{\bibfnamefont{C.}~\bibnamefont{Shu}},
  \bibinfo{journal}{Math. Comp.} \textbf{\bibinfo{volume}{67}},
  \bibinfo{pages}{73} (\bibinfo{year}{1998}).

\bibitem[{\citenamefont{Shu}(1999)}]{Shu99}
\bibinfo{author}{\bibfnamefont{C.~W.} \bibnamefont{Shu}}, in
  \emph{\bibinfo{booktitle}{High-{O}rder {M}ethods for {C}omputational
  {P}hysics}}, edited by \bibinfo{editor}{\bibfnamefont{T.~J.}
  \bibnamefont{Barth}} \bibnamefont{and}
  \bibinfo{editor}{\bibfnamefont{H.}~\bibnamefont{Deconinck}}
  (\bibinfo{publisher}{Springer}, \bibinfo{year}{1999}).

\bibitem[{\citenamefont{Pons et~al.}(2000)\citenamefont{Pons, Mart{\'\i}, and
  M{\"u}ller}}]{Pons00}
\bibinfo{author}{\bibfnamefont{J.~A.} \bibnamefont{Pons}},
  \bibinfo{author}{\bibfnamefont{J.~M.} \bibnamefont{Mart{\'\i}}},
  \bibnamefont{and}
  \bibinfo{author}{\bibfnamefont{E.}~\bibnamefont{M{\"u}ller}},
  \bibinfo{journal}{Journ. Fluid Mech.} \textbf{\bibinfo{volume}{422}},
  \bibinfo{pages}{125} (\bibinfo{year}{2000}).

\bibitem[{\citenamefont{Rezzolla and Zanotti}(2001)}]{Rezzolla01}
\bibinfo{author}{\bibfnamefont{L.}~\bibnamefont{Rezzolla}} \bibnamefont{and}
  \bibinfo{author}{\bibfnamefont{O.}~\bibnamefont{Zanotti}},
  \bibinfo{journal}{Journ. Fluid. Mech.} \textbf{\bibinfo{volume}{449}},
  \bibinfo{pages}{395} (\bibinfo{year}{2001}).

\bibitem[{\citenamefont{Rezzolla et~al.}(2003)\citenamefont{Rezzolla, Zanotti,
  and Pons}}]{Rezzolla03}
\bibinfo{author}{\bibfnamefont{L.}~\bibnamefont{Rezzolla}},
  \bibinfo{author}{\bibfnamefont{O.}~\bibnamefont{Zanotti}}, \bibnamefont{and}
  \bibinfo{author}{\bibfnamefont{J.~A.} \bibnamefont{Pons}},
  \bibinfo{journal}{Journ. Fluid. Mech.} \textbf{\bibinfo{volume}{479}},
  \bibinfo{pages}{199} (\bibinfo{year}{2003}).

\bibitem[{\citenamefont{Roe}(1981)}]{Roe81}
\bibinfo{author}{\bibfnamefont{P.~L.} \bibnamefont{Roe}}, \bibinfo{journal}{J.
  Comput. Phy.} \textbf{\bibinfo{volume}{43}}, \bibinfo{pages}{357}
  (\bibinfo{year}{1981}).

\bibitem[{\citenamefont{Donat and Marquina}(1996)}]{Donat96}
\bibinfo{author}{\bibfnamefont{R.}~\bibnamefont{Donat}} \bibnamefont{and}
  \bibinfo{author}{\bibfnamefont{A.}~\bibnamefont{Marquina}},
  \bibinfo{journal}{Journ. Comput. Phys.} \textbf{\bibinfo{volume}{125}},
  \bibinfo{pages}{42} (\bibinfo{year}{1996}).

\bibitem[{\citenamefont{Donat et~al.}(1998)\citenamefont{Donat, Font,
  Ib{\'a}nez, and Marquina}}]{Donat98}
\bibinfo{author}{\bibfnamefont{R.}~\bibnamefont{Donat}},
  \bibinfo{author}{\bibfnamefont{J.~A.} \bibnamefont{Font}},
  \bibinfo{author}{\bibfnamefont{J.~M.} \bibnamefont{Ib{\'a}nez}},
  \bibnamefont{and} \bibinfo{author}{\bibfnamefont{A.}~\bibnamefont{Marquina}},
  \bibinfo{journal}{Journ. Comput. Phys.} \textbf{\bibinfo{volume}{146}},
  \bibinfo{pages}{58} (\bibinfo{year}{1998}).

\bibitem[{\citenamefont{Duez et~al.}(2003)\citenamefont{Duez, Marronetti,
  Shapiro, and Baumgarte}}]{Duez:2002bn}
\bibinfo{author}{\bibfnamefont{M.~D.} \bibnamefont{Duez}},
  \bibinfo{author}{\bibfnamefont{P.}~\bibnamefont{Marronetti}},
  \bibinfo{author}{\bibfnamefont{S.~L.} \bibnamefont{Shapiro}},
  \bibnamefont{and} \bibinfo{author}{\bibfnamefont{T.~W.}
  \bibnamefont{Baumgarte}}, \bibinfo{journal}{Phys. Rev.}
  \textbf{\bibinfo{volume}{D67}}, \bibinfo{pages}{024004}
  (\bibinfo{year}{2003}).

\end{thebibliography}


\end{document}